\documentclass[11pt,a4paper]{article}
\usepackage[T1]{fontenc}
\usepackage[utf8]{inputenc}
\usepackage[margin=22mm]{geometry}
\usepackage{lmodern}
\usepackage{microtype}
\usepackage{graphicx}
\usepackage{bm}
\usepackage{amsmath,amsfonts,amssymb,amsthm}
\usepackage{braket}
\usepackage{subcaption}
\usepackage{multirow}
\usepackage{authblk}
\usepackage[numbers,sort&compress]{natbib}
\usepackage[hypertexnames=false,colorlinks=true,linkcolor=blue,citecolor=blue,urlcolor=blue]{hyperref}
\hypersetup{
  pdftitle={From Discrete to Continuous Variable Systems via Jordan--Schwinger Tomographic Transformation},
  pdfauthor={Liubov A. Markovich; Vladimir A. Orlov; Alexey N. Rubtsov; Vladimir I. Man'ko},
  pdfsubject={Annals of Physics 493 (2026) 170593},
  pdfkeywords={Tomography, Jordan--Schwinger map, Holstein--Primakoff map, CV--DV mappings}
}
\DeclareMathOperator{\tr}{tr}

\newtheorem{theorem}{Theorem}[section]

\newtheorem{definition}{Definition}[section]

\newtheorem{statement}{Statement}

\newtheorem{proposition}{Proposition}
\newtheorem{remark}{Remark}

\title{From Discrete to Continuous Variable Systems via Jordan--Schwinger Tomographic Transformation}
\author[1,2]{Liubov A. Markovich\thanks{Corresponding author: \href{mailto:markovich@mail.lorentz.leidenuniv.nl}{markovich@mail.lorentz.leidenuniv.nl}.}}
\author[3]{Vladimir A. Orlov}
\author[3,4]{Alexey N. Rubtsov}
\author[3,5]{Vladimir I. Man'ko}
\affil[1]{Instituut-Lorentz, Universiteit Leiden, P.O. Box 9506, 2300 RA Leiden, The Netherlands}
\affil[2]{Applied Quantum Algorithms Leiden, The Netherlands}
\affil[3]{Russian Quantum Center, Skolkovo, Moscow 121205, Russia}
\affil[4]{Department of Physics, Lomonosov Moscow State University, Moscow 119991, Russia}
\affil[5]{Lebedev Physical Institute, Russian Academy of Sciences, Leninskii Prospect 53, Moscow 119991, Russia}
\date{\small Published in \emph{Annals of Physics} \textbf{493} (2026), 170593\\
\href{https://doi.org/10.1016/j.aop.2026.170593}{doi:10.1016/j.aop.2026.170593}}
\begin{document}
\maketitle

\begin{abstract}
Hybrid quantum systems that combine discrete-variable (DV) and continuous-variable (CV) architectures represent a promising direction in quantum information science. However, transferring concepts, information, and states between such fundamentally different platforms entails both practical and theoretical challenges. The formalisms of these two ``universes'' differ significantly, and many notions, although sharing the same names, possess distinct properties and physical interpretations. In this work, we construct a sector wise bridge between DV spin systems and restricted CV bosonic systems by means of the tomographic probability representation of quantum states complemented by the Jordan--Schwinger and Holstein--Primakoff maps. While both maps are well known at the operator level, their action on the classical counterparts of quantum states, namely tomograms and other probability representations, has not been addressed in the literature. To the best of our knowledge, this work provides the first explicit demonstration of how the Jordan--Schwinger and Holstein--Primakoff maps act on tomographic probability distributions and Wigner functions, thereby establishing direct sector wise correspondences between the classical measurement statistical descriptions of DV systems and the corresponding restricted CV bosonic representations. Our tomographic mapping enables a direct transfer of measurement data between different quantum architectures by acting as an intrinsic data compression kernel that selects the Hilbert space sector relevant to the target representation. It allows one to obtain the tomogram of a target sector directly from experimentally acquired data in another representation, without reconstructing the density matrix. This provides a unified framework for transferring and comparing quantum information across heterogeneous quantum hardware platforms within the corresponding restricted sectors, facilitating hybrid protocols, device benchmarking, and the validation of error correction schemes that rely on mappings between finite-dimensional systems and sector restricted bosonic representations.
\end{abstract}

\section{Introduction}
Quantum information theory operates within two complementary paradigms: 
the discrete-variable (DV) framework, describing finite-dimensional systems such as qubits and qudits, and the continuous-variable (CV) framework, describing infinite-dimensional systems such as optical modes~\cite{PhysRevLett.112.120505}, microwave-cavity modes~\cite{rosenblum2018cnot}, and vibrational modes of trapped ions~\cite{fluhmann2019encoding}. 
Mapping between these  different types of quantum objects helps to unify different physical platforms and  enables \textit{hybrid continuous variable-discrete variable} (CV-DV) architectures, exploring the best from both CV and DV worlds  in one quantum device~\cite{takeda2015entanglement,andersen2015hybrid,he2022teleportation,lu2025quantumcoherentstatetransform}. Classic examples include the Jordan--Schwinger mapping~\cite{Jordan1935, Schwinger1952}  which connects Lie algebra generators to bilinear combinations of bosonic or fermionic operators, Jordan--Wigner transformation~\cite{jordan1928paulische}, which relates spin chains to fermionic systems, Holstein--Primakoff mapping~\cite{holstein1940field}  of spins to bosonic modes, and Dyson--Maleev transformation~\cite{dyson1956general, maleev1958scattering} providing a mapping of spin operators onto bosonic operators using a non-Hermitian representation. 
\par Since the density matrix of a given state cannot be measured directly, alternative but equivalent representations are required. For CV systems, this led to \textit{quasiprobability distribution functions} (QPDFs) such as the Wigner~\cite{wigner1932quantum}, Husimi 
$Q-$~\cite{1940264}, and Sudarshan--Glauber $P$-functions~\cite{sudarshan1963equivalence, PhysRev.177.1882}. 
All of these functions are QPDFs, since they can take negative values or the variables they depend on lack direct physical meaning, being not measurable experimentally. The development of \textit{symplectic tomography} resolved this by representing quantum states through positive probability distribution functions (PDFs) of quadratures~\cite{Mancini_1995}, a framework that generalizes optical homodyne tomography~\cite{vogel1989determination, bertrand1987tomographic, smithey1993measurement,Lvovsky_2004,lvovsky2002quantum, lvovsky2009continuous}.
The tomogram is informationally complete %
and, owing to the new nonparametric kernel quantum state reconstruction (KQSE) method~\cite{markovich2025nonparametric}, allows one to estimate any state (even non-Gaussian) while avoiding the drawbacks of reconstruction techniques based on QPDFs and maximum likelihood estimator~\cite{vogel1989determination, lvovsky2009continuous}. Similar to CV systems, DV spin states can be represented by probability functions (PF) known as  \textit{spin tomograms}~\cite{Dodonov1997}, obtained from projection measurements along different axes. In quantum optics,  \textit{photon-number tomography} provides an analogous description by measuring photon statistics, where Fock states correspond to photon-number eigenstates~\cite{banaszek1996direct, wallentowitz1996unbalanced, zucchetti1996quantum, mancini1997density}. 
\par
Tomograms are not auxiliary representations but probability density functions or probability functions of directly measurable quantum observables. Experimentally, one accesses numerical outcomes sampled from these distributions, making the tomographic description the natural level at which measured data are processed. Despite their positivity and classical statistical interpretation, tomograms are genuine quantum objects: their dynamics and composition are governed by star-product relations~\cite{man2000star,man2002alternative}, and they faithfully encode both classical and nonclassical properties of quantum states, including  entanglement and Bell-type correlations~\cite{andreev2001classification,lupo2005partial,chernega2022entangled,paul2023tomographic}. Beyond this, tomograms directly encode intrinsic state characteristics (purity, trace powers) and enable Gaussianity and non-Gaussianity tests, as well as distance-like metrics between quantum states, without full state reconstruction~\cite{markovich2024not}. Nevertheless, different tomographic representations are typically treated separately, reflecting differences in measurement schemes and underlying Hilbert-space structures.
\subsection{Contributions of this Paper}
\par {In this work, we employ the Jordan--Schwinger (JS) and Holstein--Primakoff (HP) operator mappings to establish explicit sector-wise correspondences between discrete-variable (DV) spin states and restricted continuous-variable (CV) bosonic states at the level of their tomograms}, thereby providing a unified probabilistic framework for comparing different quantum architectures for the first time. In a conventional CV-DV workflow, the conversion of CV optical data into a DV description typically proceeds via three steps. Homodyne detection yields an optical tomogram $W_{opt}(x|\theta)$, which is then used to numerically reconstruct the photonic density matrix $\hat{\rho}_{(1)}$ that is an intrinsically ill-posed inverse problem. Then the JS  type projection is applied to obtain the DV representation. The reconstruction stage is corrupted by noise, the regularization procedure is nearly unavoidable, and nonphysical artifacts may arise, ultimately limiting robustness and scalability of the data transferring.
\par In this work, we introduce \textit{tomographic JS and HP mappings} that bypass these problems entirely by fusing measurement, reconstruction, and mapping into a single integral transform acting directly on experimentally accessible tomograms,
\begin{align}
    \omega_{spin}(m,\Omega)=\int W_{opt}(x|\theta) K(m,\Omega;x,\theta) dx d\theta.  
\end{align}
Our kernel-based formulation enables a direct transfer of information between CV and DV architectures, while simultaneously performing an intrinsic data compression by filtering only the Hilbert space sector relevant to the target representation. As a result, tomograms corresponding to one physical architecture can be obtained directly from measurement data acquired in another (see Sec.~\ref{sim} for the real homodyne data based example), providing a unified framework for comparing, transferring, and benchmarking quantum information across heterogeneous hardware platforms.
\par A further key advantage of the proposed method is that the transformation kernels admit fully analytical closed form expressions. Rather than relying on numerical manipulation of density matrices, which in the CV setting are infinite dimensional, require truncation at high photon numbers, and are sensitive to cutoff artifacts, our kernel formulation resums the relevant Fock space contributions into compact expressions involving known special functions. As a result, large scale numerical linear algebra is replaced by direct evaluation of analytical kernels,  which can reduce computational cost and improved numerical stability.
\par
\begin{figure*}
    \centering
    \includegraphics[width=0.8\linewidth]{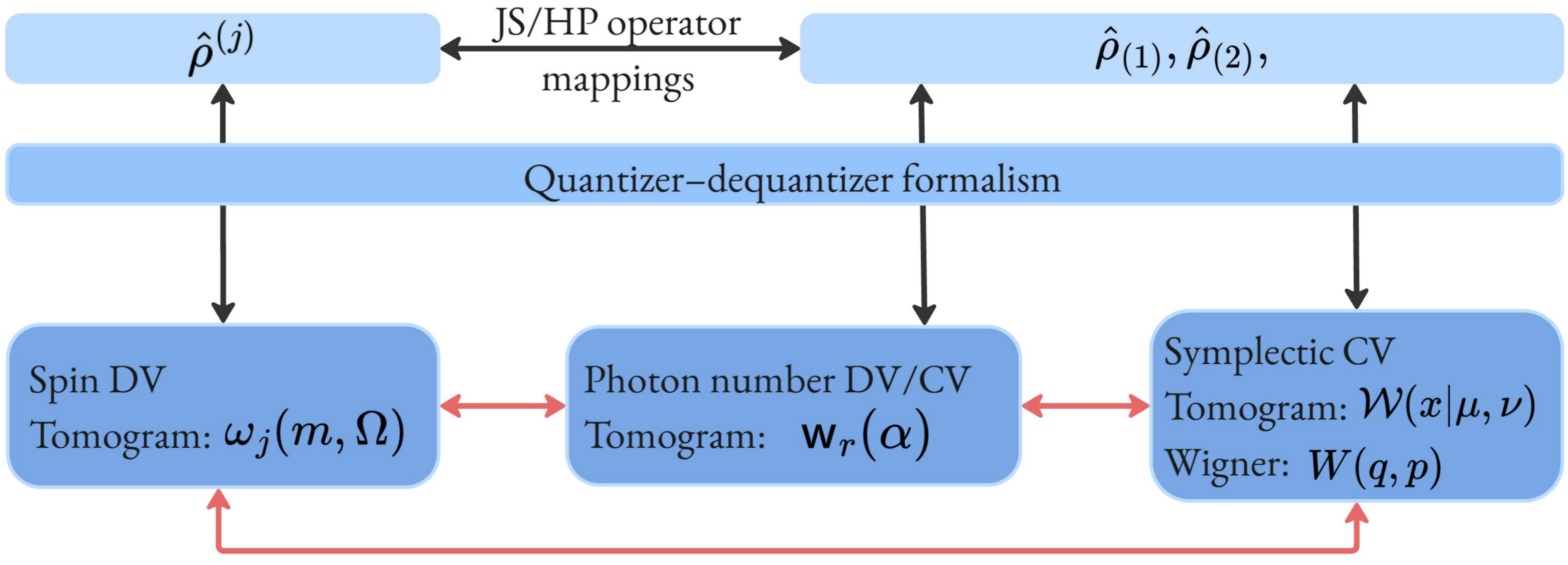}
   \caption{Unified tomographic framework connecting DV and CV quantum systems. The spin tomogram $\omega_j(m,\Omega)$, the photon-number tomogram $\mathsf{w}_r(\alpha)$, and the symplectic tomogram $\mathcal{W}(x|\mu,\nu)$ are linked by explicit quantizer-dequantizer kernel transformations, {establishing invertible correspondences among tomographic representations within the appropriate restricted sector and, separately, between informationally complete tomograms and the Wigner function $\mathbb{W}(q,p)$ in their common domain of definition.} Each tomogram is associated with a measurable random variable, spin projection $m$, photon-number $r$ (after displacement $\alpha$), and quadrature $x_{\mu,\nu}$, yielding experimentally accessible probability distributions across DV and CV regimes. The Jordan--Schwinger and Holstein--Primakoff mappings provide the unifying link between the density operators $\hat{\rho}^{(j)}$ and $\hat{\rho}_{(2)}, \hat{\rho}_{(1)}$, {enabling direct translation of tomographic data between finite spin representations and the corresponding restricted bosonic sectors.}}
   \label{fig:omega-diagram}
\end{figure*}

Both the JS and HP mappings are exact only after restriction to a fixed total-occupation sector of the bosonic Fock space. In particular, the sector with total occupation number $n=2j$ is isomorphic to the spin-$j$ Hilbert space and carries a single irreducible $\mathrm{SU}(2)$ representation.
We introduce the concept of a \emph{restricted Fock state}, namely a bosonic state supported on a fixed-$n$ subspace, together with the associated \emph{restricted tomogram}. {The normalized restricted state is defined only for sectors with nonzero probability weight.} In contrast to standard optical tomograms, which in general encode contributions from all photon-number sectors $n=0,1,2,\dots$, the correspondence with a fixed spin-$j$ system involves only the invariant bosonic sector with total excitation number $n=2j$. Thus, the JS and HP mappings should be understood as sector-wise correspondences: they relate the spin-$j$ Hilbert space to the restricted bosonic subspace $\mathcal H_{2j}$ and, at the tomographic level, induce kernels acting only on the associated restricted tomograms. Within this fixed sector, both mappings are invertible at the level of states, yielding a one-to-one correspondence between the discrete-variable and continuous-variable tomographic descriptions (see Fig.~\ref{fig:omega-diagram}). This provides the first explicit mapping between spin and bosonic tomographic probability distributions, extending the JS and HP frameworks from operator level correspondences to experimentally accessible probability representations.
The restricted density operators in our construction have direct operational meanings. In the JS realization, the spin-$j$ Hilbert space is identified with the two-mode sector of fixed total excitation number $n_1+n_2=2j$. Operationally, this corresponds to preparing or postselecting states within this exact manifold. Examples include heralded photonic qudits~\cite{yoshikawa2018heralded}, excitation-resolved quantum polarization tomography where $SU(2)$ observables act blockwise~\cite{kothe2013experimental,soderholm2012quantum}, and two-mode Bose--Einstein condensates with conserved atom number~\cite{dalton2012two,gross2012spin}.
In the HP realization, the restriction has a different interpretation: it corresponds to a finite truncation of a single-mode Fock space ($k=0,\ldots,2j$). This defines the exact physical Hilbert space for finite spin systems and magnons~\cite{holstein1940field}, as well as collective atomic ensembles~\cite{RevModPhys.82.1041,kurucz2010multilevel}. When implemented in an actual oscillator, this restriction implies preparing finite Fock superpositions, which can be achieved via quantum scissors and optical truncation~\cite{pegg1998optical,babichev2003quantum,miranowicz2005optical}, conditional few-photon engineering~\cite{bimbard2010quantum,yukawa2013generating}, or deterministic state synthesis in circuit QED and trapped ions~\cite{hofheinz2009synthesizing,benkish2003experimental}.
\par Importantly, the JS and HP mappings remain exact at the operator and density-matrix levels. The HP loss of exactness commonly encountered in practice originates from replacing operator-valued functions with truncated Taylor expansions, which are inappropriate for operators with discrete spectra~\cite{10.21468/SciPostPhys.10.1.007}. By contrast, our construction preserves exactness within each fixed-$n$ sector and remains practically exact when combined with an appropriate discrete (Newton) expansion. 
\par We emphasize that the space restriction is not a practical limitation. Recent results show that the commonly used ``CV limit'' should be understood as an approximation arising from a (large) fixed-$n$ sector, thereby providing a controlled bridge between discrete and continuous descriptions \cite{descamps2024superselection,DescampsUnified2025,SaharyanMetrology2025}.
\par A further contribution of this work is a unified analysis of the large-$j$ limit of spin tomograms within both the HP and JS frameworks. We show that the spin tomogram admits an exact finite-$j$ representation in terms of a deformed quadrature observable, without invoking the linearized HP approximation. This reveals that, in general, the large-$j$ limit leads to a  deformed quadrature, and that the standard symplectic tomogram is recovered only under additional scaling or localization conditions. 
We then formulate the CV limit at the level of density operators by introducing HP-lifted sector restricted states and proving that trace norm convergence of these states implies convergence of the associated CFs and distributions. This provides a rigorous and physically transparent criterion for when a sequence of finite-dimensional spin states admits a genuine CV limit. 
Finally, we clarify the conceptual difference between the HP and JS constructions: while the HP correspondence realizes the CV limit through an increasing sequence of truncations, the JS construction involves fixed excitation sectors and requires an additional identification to define a consistent single-mode limit. Together, these results establish a coherent framework connecting DV spin tomography with CV quantum optics.
\subsection{Practical Motivation}
\par Hybrid quantum architectures combining DV and CV components are already realized across several leading platforms. In superconducting circuits, advances in quantum control~\cite{descamps2024superselection,Liu_2021} and bosonic error correction~\cite{Ma2020ErrorTransparent,Gertler2021ProtectingBosonicQubit} have enabled early qubit--oscillator processors, where quantum information is jointly encoded in qubits and microwave resonator modes. These systems constitute small but fully functional hybrid devices supporting coherent operations across both subsystems.
Trapped-ion platforms exhibit a similar structure. Ion-based devices~\cite{Bruzewicz2019TrappedIonReview,PhysRevLett.131.033604,PhysRevLett.124.170502} naturally integrate DV qubits with CV motional modes that mediate interactions and provide access to non-Gaussian resources. While traditionally qubit centric, there is growing activity in bosonic operations and states~\cite{bazavan2024squeezing,kbv4-jj51}, including GKP state preparation~\cite{PhysRevLett.133.050602}, bosonic quantum simulations~\cite{Kang2024QuantumAdvantageChemistry}, and machine-learning applications using motional modes~\cite{nguyen2021quantum}.
Neutral-atom platforms, typically operated in a qubit regime, are also beginning to exploit bosonic degrees of freedom. Experiments~\cite{PhysRevA.75.013406, shaw2025erasure} have demonstrated hybrid qubit--phonon operations, long-lived motional modes, and hyper-entanglement across internal and motional states, while theory proposals point toward bosonic error-correcting codes~\cite{PhysRevA.111.022432} and the use of intrinsic trap nonlinearities~\cite{Grochowski_2025}.
\par Despite rapid progress~\cite{lu2025quantumcoherentstatetransform,Liu_2021, PRXQuantum.5.030322, Hastrup_2022}, theoretical and algorithmic tools for comparing, characterizing, and controlling hybrid CV--DV systems remain limited. Different platforms rely on incompatible mathematical descriptions. Finite dimensional Hilbert spaces for DV systems and continuous phase-space representations for CV systems~\cite{Arienzo2025BosonicRB} are making it difficult to define unified performance metrics, quantify nonclassical correlations, or design interoperable error-correction protocols~\cite{liu2025hybridoscillatorqubitquantumprocessors}.
General-purpose certification methods~\cite{Kliesch2021QuantumCertification,Eisert2020QuantumCertification} and device-independent schemes~\cite{Supic2020SelfTestingReview} can, in principle, be adapted to CV systems, while CV-specific benchmarking based on Gaussian measurements such as heterodyne detection~\cite{Chabaud2020BuildingTrustCV} has enabled verification of Gaussian boson sampling~\cite{Chabaud2021EfficientVerification}. In circuit-QED and trapped-ion platforms, multimode Wigner tomography using joint photon-number parity measurements further provides access to non-Gaussian phase-space representations.
\par Nevertheless, the development of a unified benchmarking protocol for hybrid architectures remains an open problem. The proposed research takes a first step toward this goal.  It enables a direct, operational procedure for the  hybrid device verification. For example, the experimentally measured photonic tomograms can be mapped via the kernel into a virtual spin tomogram, which can then be directly compared with the tomogram obtained from the DV systems. Agreement between the two tomograms provides an immediate and experimentally meaningful benchmark for hybrid quantum interfaces, avoiding ill-posed reconstructions and representation dependent metrics.
\par The paper is organized as follows. In Sec.~\ref{sec_1}, we review the quantizer-- dequantizer formalism and introduce the symplectic, spin, and photon-number tomograms. Sections~\ref{sec:JS} and~\ref{hp_sec} recall the JS and HP mappings and define single- and two-mode restricted states together with their tomograms, which naturally lead to the direct and inverse transformations between spin and bosonic density matrices. {In Sec.~\ref{sec_4}, these results are employed to establish sector-wise mappings between spin and bosonic tomographic representations, together with their relation to the corresponding Wigner QPDF.} In Sec.~\ref{sec_6}, we formulate the large-$j$ CV limit at the level of sector restricted density operators and tomograms. Section~\ref{sim} presents numerical simulations illustrating experimentally relevant CV--DV transitions, using real homodyne data from~\cite{lvovsky2002quantum}. The paper concludes in Sec.~\ref{sec_5}.
\section{Probability Picture of Quantum Mechanics}\label{sec_1}
The quantum state is fully characterized by the density matrix $\hat{\rho}$, a positive semi-definite, Hermitian operator with unit trace acting on the Hilbert space $\mathcal{H}$. While $\hat{\rho}$ provides a complete description of the state, its matrix representation depends on the chosen basis and thus on the observables relevant to a given experimental setup. In the literature, the same symbol $\hat{\rho}$ may therefore represent structurally different objects; to avoid ambiguity, we use dimensional or contextual labels when needed, particularly in hybrid quantum settings.
\par We consider a family of Hermitian operators $\hat U(z)$ acting on $\mathcal H$, called the \textit{dequantizer}~\cite{andreev2020quantizer}.
The dequantizer  defines the \textit{symbol} of an operator $\hat{\rho}$ as a function:  
\begin{eqnarray}\label{01}
     f_{\rho}(z) = \tr{\hat{\rho}\hat{U}(z)}.
\end{eqnarray}
Here $z$ is a vector of parameters. This symbol establishes a linear mapping from the operator space to the functional space. Conversely, one can reconstruct 
$\hat{\rho}$ from its symbol $f_{\rho}(z)$, using a set of \textit{quantizer} operators $ \hat{D}(z) \in \mathcal{B}(\mathcal{H})$. The reconstruction transformation is
\begin{eqnarray}\label{2}
\hat{\rho}=
\begin{cases}
\displaystyle \int f_{\rho}(z)\,\hat{D}(z)\,dz, & z \ \text{continuous},\\[0.4em]
\displaystyle \sum\limits_{k} f_{\rho}(k)\,\hat{D}(k), & z=k \ \text{discrete}.
\end{cases}
\end{eqnarray}
In the continuous and discrete cases, quantizers and dequantizers are connected by the relations:
\begin{eqnarray}\label{1727}
   \!\!\!\!\! \tr{\hat{U}(z')^\dagger\hat{D}(z)}\! = \!\delta(z - z'),\! \!\!\!\quad 
   \tr{ \hat{U}(k)^\dagger \hat{D}(k')} = \delta_{k, k'}.
\end{eqnarray}
The quantizer--dequantizer pair associated with an operator $\hat{\rho}$ is not unique. In general, the dequantizer is only required to be Hermitian, in which case the resulting symbol is not necessarily a PDF (for CV states) or a probability function (for DV states). Requiring the dequantizer $\hat{U}(z)$ to be positive semi-definite and complete on $\mathcal{H}$ makes it a POVM, and Eq.~\eqref{01} then defines a valid PDF or probability function over~$z$.
\subsection{Quantum States and Tomograms Across Architectures}
\par We briefly collect the definitions of quantum states and their tomographic representations in the CV and DV architectures considered in this work, fixing notation and emphasizing their probabilistic interpretation. 
\par For a finite Hilbert space $\mathcal{H}\cong\mathbb{C}^d$, a quantum state is described by a density matrix operator 
\begin{eqnarray}
\hat{\rho}^{(d)}=\sum_{i,j=1}^{d}\rho_{ij}\ket{i}\bra{j}, \qquad 
\rho_{ij}=\langle i|\hat{\rho}^{(d)}|j\rangle,
\end{eqnarray}
with respect to an orthonormal basis $\{\ket{i}\}_{i=1}^{d}$. A paradigmatic example is a spin-$j$ system, where $d=2j+1$ and the natural basis is given by the joint eigenstates $\{\ket{j,m}\}_{m=-j}^{j}$ of the spin operators $\hat S^2$ and $\hat S_z$,
\begin{eqnarray}
\hat{\rho}^{(j)}=\sum_{m,m'=-j}^{j}\rho^{(j)}_{mm'}\ket{j,m}\bra{j,m'},\quad \rho_{mm'}^{(j)}=\langle j,m|\hat{\rho}^{(j)}|j,m'\rangle.
\end{eqnarray}
The experimentally accessible description of such states is provided by the \emph{spin tomogram}, defined as the probability of obtaining outcome $m$ when measuring the spin projection along a given direction.
\begin{definition}Let $\hat{\rho}^{(j)}$ be the density operator of a spin-$j$ system.  
The \emph{spin tomogram} of the state $\hat{\rho}^{(j)}$ is defined by \begin{eqnarray}\label{spin_deqaunt}
    \omega_j(m,  \Omega)&=&
    \tr{\hat{\rho}^{(j)} \hat{\mathcal{U}}(m, \Omega)}\equiv \mathcal{S}[\hat{\rho}^{(j)}], 
\end{eqnarray}
where  $ \hat{\mathcal{U}}(m, \Omega)\equiv \hat{R}^{\dagger}(\Omega) |j,m\rangle \langle j,m| \hat{R}(\Omega)$ is the corresponding dequantizer dependent on a rotation operator $\hat{R}(\Omega)=e^{-i\alpha \hat{S}_{z}} \, e^{-i\beta \hat{S}_{y}} \, e^{-i\gamma \hat{S}_{z}}$ in terms of Euler angles $\Omega\equiv (\alpha,\beta,\gamma)$.
\end{definition}
Using the quantizer operator $\hat{\mathcal{D}}(m,\Omega)$ the state is reconstructed as follows:
\begin{eqnarray}\label{1234567}
\hat{\rho}^{(j)} = \sum_{m=-j}^{j} \int \omega_j(m,\Omega) \hat{\mathcal{D}}(m,\Omega)d\Omega\equiv \mathcal{S}^{-1}[\omega_j(m,  \Omega)].
\end{eqnarray}
The dequantizer and quantizer operator can be
written in the basis of irreducible tensor operators (see~\ref{ap_1}). 
Although for each fixed $m$ the spin tomogram $\omega_j(m,\Omega)$ is a PDF over $\Omega$ (since $\hat{\mathcal{U}}(m, \Omega)$ is a projection-valued measure), not every PDF corresponds to a physical quantum state. \begin{theorem}\label{thm_1}
A real-valued nonnegative normalized function 
$\omega_j(m,\Omega)$ is the spin tomogram of a spin-$j$ state if and only if 
\begin{equation}
\sum_{m=-j}^{j}\int \omega_j(m,\Omega)\,
\langle \psi|\hat{\mathcal D}(m,\Omega)|\psi\rangle\,d\Omega
\ge 0,
\qquad
\forall\,|\psi\rangle\in\mathcal H_j .
\end{equation}
\end{theorem}
Indeed, the $\hat{\rho}$ trace-one condition is automatically satisfied due to the normalization of the spin-tomogram
together with $\tr{\hat{\mathcal D}(m,\Omega)}=\tfrac{1}{8\pi^2}$,
while hermiticity holds automatically since $\omega_j(m,\Omega)$ is real-valued and the quantizer $\hat{\mathcal D}(m,\Omega)$ is hermitian (see~\ref{ap_1}).
\par If the Hilbert space is infinite, then $\hat{\rho}$ is a trace-class, compact operator, and its spectrum consists of infinitely many eigenvalues. This density operator may be expressed in a CV basis (e.g., position or momentum) or in the DV Fock basis. For a single mode Fock basis $\{\ket{n}\}_{n=0}^{\infty}$, the density matrix operator is
\begin{eqnarray}\label{1939}
\hat{\rho}_{(1)}=\sum\limits_{n,m=0}^{\infty}\rho_{nm}\ket{n}\bra{m}, \quad \rho_{nm} = \bra{n} \hat{\rho}_{(1)} \ket{m}.
\end{eqnarray}
This matrix has a countably infinite number of entries. The diagonal elements $\rho_{nn}$ are probabilities of measuring 
$n$ photons in the system.  A natural tomographic description in this setting is provided by the \emph{photon-number tomogram}~\cite{10.1063/1.4773139}:
\begin{definition}
Let $\hat{\rho}_{(1)}$ be the density operator of a single mode state. The photon-number tomogram of the state  $\hat{\rho}_{(1)}$ is defined by
\begin{eqnarray}\label{dequantizer-ph-number}
\mathsf{w}_r(\alpha)=\tr{\hat{\rho}_{(1)} \hat{\mathcal{U}}_r(\alpha)}\equiv \mathcal{P}[\hat{\rho}_{(1)}],
\end{eqnarray}with the dequantizer $\hat{\mathcal{U}}_r(\alpha) =  \hat{D}^\dagger(\alpha) \ket{r}\bra{r}\hat{D}(\alpha)$ dependent on the Weyl displacement operator $
\hat{D}(\alpha) = \exp\left(\alpha\,\hat{a}^\dagger - \alpha^{*}\,\hat{a}\right)$,  $\alpha\in \mathbb{C}$ is a  field amplitude, while $r$ is the non-negative integer photon-number. 
\end{definition}
The function $\mathsf{w}_r(\alpha)$ gives the probability of detecting $r$ photons after a phase-space displacement by $\alpha$ and  $\sum_{r=0}^{\infty}  \mathsf{w}_r(\alpha) = 1$. This comes  from the resolution of the identity $\sum_{r=0}^{\infty}\hat{\mathcal{U}}_r(\alpha)=\hat{1}$ and the fact that the dequantizer $\hat{\mathcal{U}}_r(\alpha)$ is a  POVM. 
The inverted transformation is
\begin{eqnarray}\label{1355}
     \hat{\rho}_{(1)} = \sum_{r=0}^{\infty} \int\limits_{\mathbb{C}} d^2\alpha \, \mathsf{w}_r(\alpha) \hat{\mathcal{D}}_r(\alpha)\equiv  \mathcal{P}^{-1}[\mathsf{w}_r(\alpha)],
\end{eqnarray}
where the quantizer operator  is given by:
\begin{equation}\label{quantizer-ph-number}
\hat{\mathcal{D}}_r(\alpha) = 4(\pi (1-s^2))^{-1} g(s)^{(\hat{a}^\dagger+\alpha^{*})(\hat{a}+\alpha)-r},
\end{equation}
where $g(s)\equiv (s-1)/(s+1)$,
with the ordering parameter $s\in(-1,1)$, where the limits $s\to-1$ and $s\to+1$ correspond to
antinormal/normal ordering of operators $\hat{a}$ and $\hat{a}^{\dagger}$, respectively.  One may note that the photon--number tomogram can also be viewed as a PDF of the CV variable $\alpha$ for a fixed DV index $r$, normalized as $\int \mathsf{w}_r(\alpha)\tfrac{d^2\alpha}{\pi} = 1$. This observation reveals a direct connection to the Husimi $Q$-function, which is likewise nonnegative and admits a POVM dequantizer $\hat{E}(\alpha)=\pi^{-1}\ket{\alpha}\bra{\alpha}$, where $\ket{\alpha}=\hat{D}(\alpha)\ket{0}$ is a coherent state. For $r=0$, the correspondence reads $\mathsf{w}_0(-\alpha)=\pi Q(\alpha)$, reflecting a displacement opposite to that of the coherent-state projection. Experimentally, the $Q$-function is accessed via heterodyne detection, whose outcomes directly sample the POVM and thus the $Q$-function over phase space, rather than the random variable $\alpha$ itself. For this reason, the $Q$-function should be classified as a QPDF. The tomogram \eqref{dequantizer-ph-number} may also be interpreted as a conditional PDF of photon-number for a fixed field amplitude $\alpha$. In particular, the photon--number distribution $P(r)\equiv\langle r|\hat{\rho}_{(1)}|r\rangle=\mathsf{w}_r(\alpha=0)$ gives the probability of finding $r$ photons in the state $\hat{\rho}_{(1)}$.
\\Similarly to the spin tomographic case (Theorem~\ref{thm_1}), not every PDF $\mathsf w_r(\alpha)$ defines a physical state $\hat{\rho}_{(1)}$:
\begin{theorem}\label{thm_2}A real-valued nonnegative normalized function 
$\mathsf w_r(\alpha)$ is the photon-number tomogram of a one-mode CV-state $\hat{\rho}_{(1)}$ if and only if 
\begin{equation}
\sum_{r=0}^{\infty}
\int_{\mathbb C}
\mathsf w_r(\alpha)
\langle\psi|\hat{\mathcal D}_r(\alpha)|\psi\rangle\,d^2\alpha
\ge0
\qquad
\forall\,|\psi\rangle\in\mathcal H.
\label{eq:photon_direct_positive}
\end{equation}
\end{theorem}
The proof follows the same steps as in the spin-tomographic case and is therefore omitted. In particular, one verifies normalization ($\tr{\hat\rho_{(1)}}=1$), Hermiticity, and positivity of the reconstructed density operator, which together ensure that $\mathsf w_r(\alpha)$ corresponds to a physical quantum state.
\par In a continuous basis, such as position, the density operator admits
\begin{eqnarray}
\hat{\rho}_{(1)}=\iint_{\mathbb{R}^2}\rho(y,y')\ket{y}\bra{y'},dy,dy',
\end{eqnarray}
where $\rho(y,y')=\langle y|\hat{\rho}_{(1)}|y'\rangle$ is an uncountable integral kernel, whose diagonal elements define the position PDF. The most general tomographic description in this setting is provided by the \emph{symplectic tomogram}~\cite{Mancini_1995}, defined as the probability density of the generalized quadrature operator $ \hat{X}_{\mu,\nu}=\mu \hat{q}+\nu \hat{p}$:
\begin{definition}
    Let $\hat{\rho}_{(1)}$ be the density operator of a single mode state. The symplectic  tomogram of the state  $\hat{\rho}_{(1)}$ is defined by \begin{eqnarray}\label{eq:deq-simpl}
\mathcal{W}(x|\mu,\nu)=\tr{\hat{\rho}_{(1)} \hat{\mathcal{U}}(x, \mu, \nu)}\equiv \mathcal{R}[\hat{\rho}_{(1)}],\quad \mu, \nu \in \mathbb{R},
\end{eqnarray}
where $\hat{\mathcal{U}}(x, \mu, \nu) = \delta(x\hat{1} - \mu \hat{q} - \nu \hat{p})$ is the dequantizer operator.
\end{definition}
For fixed $\mu,\nu$, $\mathcal W(x|\mu,\nu)$ is a normalized PDF in $x$, and the transformation represents a quantum analogue of the classical Radon transform. The inverse transformation is given by:
\begin{eqnarray}\label{sympl_tomography_inverse}
\hat{\rho}_{(1)}&=&\iiint\limits_{\mathbb{R}^3}\mathcal{W}(x|\mu,\nu) \hat{\mathcal{D}}(x, \mu, \nu) dx\, d\mu\, d\nu\equiv \mathcal{R}^{-1}[\mathcal{W}(x|\mu,\nu)],
\end{eqnarray}
where the corresponding quantizer operator in the symplectic-tomography framework reads as
\begin{eqnarray}\label{symplectic_quantizer}
\hat{\mathcal{D}}(x, \mu, \nu) =(2\pi)^{-1} \exp{ i x \hat{1} - i \nu \hat{p} - i \mu \hat{q}}. 
\end{eqnarray}
Optical tomograms, directly sampled in homodyne detection, correspond to the special case $(\mu,\nu)=(\cos\theta,\sin\theta)$. Closely related phase-space representations, such as the Wigner function, are obtained from symplectic tomograms via integral transforms. In Ref.~\cite{markovich2024not} we introduced the notion of the characteristic function (CF) of the symplectic tomogram as
\begin{align}\label{characteristic_function}
    \phi(t;\mu,\nu)=\int\limits_{\mathbb{R}}\mathcal{W}(x|\mu,\nu)e^{itx}dx,
\end{align}
that is the core of the CVs state reconstruction KQSE method ~\cite{markovich2025nonparametric} providing the nearly optimal $\tilde{O}(1/T)$ rate of convergence in $L_2$ norm, where $T$ is the amount of homodyne measurements. 
The density matrix operator can be reconstructed from the CF of the symplectic tomogram as
\begin{align}\label{1_1}
\hat{\rho}_{(1)}=\frac{1}{2\pi}\!\iint
  \phi(1;\mu,\nu)
\exp{-i(\mu\hat{q}+\nu\hat{p})}\,d\mu\,d\nu,
\end{align}
In Ref.~\cite{markovich2024not}, we formulated conditions on the CF (symplectic tomogram) ensuring $\boldsymbol{\rho}$ reconstruction. Here, we present its equivalent integral formulation:
\begin{theorem}{(Hermiticity,  Normalization, Positivity)}\label{thm_3_0}
The integral \eqref{1_1}  defines the density operator corresponding to a quantum state if and only if the CF of the tomogram satisfies:
 \begin{eqnarray}\label{thm_3}
&&\phi(1;\mu,\nu)= \phi(-1;-\mu,-\nu) \quad \forall \mu,\nu,\quad \phi(1;0,0)=1,\nonumber\\
&& 0\leq  \frac{1}{2\pi}
    \iint \phi(1;\mu,\nu)\phi_{{\chi}}^{\star}(1;\mu,\nu)d\mu d\nu\leq 1, \quad \forall \ket{\chi}.
\end{eqnarray}
\end{theorem}
Further we use the tomogram and CF of the excited Harmonic oscillator state that are the following
\begin{eqnarray}\label{excited_HO}
     \mathcal{W}_m(x|\mu,\nu)
    &=& \mathcal{W}_0(x|\mu,\nu) \frac{1}{2^m m!} H^2_m \left(\frac{x}{\sqrt{\mu^2+\nu^2}}\right),\\\nonumber
    {\phi}_m(1;\mu,\nu)
    &=& {\phi}_0(1;\mu,\nu) L_m\left(\frac{\mu^2+\nu^2}{2}\right),
\end{eqnarray}
where the tomogram and CF of the ground state  are 
\begin{eqnarray}\label{ground_HO}
\mathcal{W}_0(x|\mu,\nu) = \frac{1}{\sqrt{\pi(\mu^2+\nu^2)}} \exp\!\Bigl[-\tfrac{x^2}{\mu^2+\nu^2}\Bigr],\!\! \quad 
{\phi}_0(t;\mu,\nu) = \exp\!\Bigl(-\tfrac{t^2(\mu^2+\nu^2)}{4}\Bigr),
\end{eqnarray}
being both Gaussian. 
\par The Wigner function $\mathbb{W}(q,p)$ is a QPDF widely used in quantum optics and statistical physics. It is defined via the Wigner--Weyl transform
\begin{eqnarray}
\mathbb{W}(q, p) = \tr{\hat{\rho}_{(1)}\hat{\mathbb{U}}(q,p)}\equiv \mathcal{F}[\hat{\rho}_{(1)}],
\end{eqnarray}
with dequantizer
\begin{eqnarray}
 \hat{\mathbb{U}}(q,p)=\frac{1}{2\pi} \int\limits_{\mathbb{R}} \ket{q - \frac{\tau}{2}}\bra{q + \frac{\tau}{2}} e^{-i\tau p} d\tau.
\end{eqnarray}
The inverse transform reads
\begin{eqnarray} \hat{\rho}_{(1)} = \iint\limits_{\mathbb{R}^2} \mathbb{W}(q, p) \hat{\mathbb{D}}(q,p) dq dp\equiv \mathcal{F}^{-1}[\mathbb{W}(q,p)], \end{eqnarray}
where the quantizer is $\hat{\mathbb{D}}(q,p)=2\pi \hat{\mathbb{U}}(q,p)$.
It is often convenient to express the Wigner function in terms of the complex phase-space variable $\alpha=(q+ip)/\sqrt{2}$. In this representation, the dequantizer and quantizer of $\mathbb{W}(\alpha)$ are given by Glauber's displacement--parity operators~\cite{man2020integral}:
\begin{eqnarray}\label{quant_and_dequant_wigner}
\hat{\mathbb{U}}(\alpha) =(\pi)^{-1} \hat{D}(2\alpha)\hat{P},\quad \hat{\mathbb{D}}(\alpha)=2\hat{D}(2\alpha)\hat{P},
\end{eqnarray}
where $\hat{P}$ denotes the parity operator.
Similarly to the tomograms above not every real normalized phase-space function
$\mathbb W(q,p)$ is the Wigner function of a physical one-mode quantum
state. Since the Wigner function is a quasiprobability distribution, its
pointwise nonnegativity is not required. The admissibility condition is
instead imposed on the operator reconstructed by the inverse Wigner--Weyl
transform.
\begin{theorem}
\label{thm:wigner_admissibility}
Let \(\mathbb W(q,p)\) be a real-valued phase-space function such that $\iint_{\mathbb R^2}\mathbb W(q,p)\,dq\,dp=1$.
Then \(\mathbb W(q,p)\) is the
Wigner function of a one-mode CV quantum state if and only if
\begin{align}
\iint_{\mathbb R^2}
\mathbb W(q,p)\,
\langle\psi|\hat{\mathbb D}(q,p)|\psi\rangle\,dq\,dp
\ge 0
\qquad
\forall\,|\psi\rangle\in\mathcal H .
\label{eq:wigner_direct_positive}
\end{align}
\end{theorem}
The normalization condition is satisfied, since $\operatorname{Tr}\hat{\mathbb D}(q,p)=1$. Hermiticity follows from the reality of $\mathbb W(q,p)$ together with the Hermiticity of the Wigner quantizer $\hat{\mathbb D}(q,p)$. Therefore, the only nontrivial admissibility requirement is the positivity of the reconstructed operator, as expressed in Eq.~\eqref{eq:wigner_direct_positive}.
{Taken together, spin, photon-number, and symplectic tomograms provide a unified probabilistic language for describing quantum states across discrete- and continuous-variable architectures. Despite arising from different measurement schemes and Hilbert-space structures, all tomograms are bona fide probability distributions that fully encode the underlying quantum state within their respective domains of definition. This shared probabilistic structure underlies the sector-wise mappings between tomograms associated with the corresponding restricted quantum architectures developed below.}
\section{Jordan--Schwinger Isomorphism and Restricted State}\label{sec:JS}
\par In 1935, Jordan introduced a correspondence between irreducible representations of the symmetric group $\mathrm{S}_n$ and the general linear group $\mathrm{GL}(n)$ via bilinear operators obeying bosonic or fermionic (anti)commutation relations of $\mathfrak{gl}(n)$~\cite{Jordan1935}. This was motivated by clarifying the relation between the wave-function and operator (second-quantization) formulations of many-particle quantum mechanics, whose equivalence he had previously established at the matrix level~\cite{jordan1932methode}. Schwinger later proposed a unified oscillator-based approach to angular momentum, reducing spin commutation relations to those of harmonic oscillators~\cite{Schwinger1952}.
\par We consider a set of $n\times n$ matrices $\{A_k\}_{k=1}^N$ forming a representation of a Lie algebra $\mathfrak{g}$ with commutation relations $[A_i,A_j]=\sum_{k=1}^N c_{ij}^k A_k$,
where $c_{ij}^k$ are the structure constants of $\mathfrak{g}$. Denoting by $(A_i)_{\alpha\beta}$, with $\alpha,\beta\in[1,n]$, the $(\alpha,\beta)$-th entry of the matrix $A_i$, we define 
$ \hat{A}_i = \sum_{\alpha,\beta=1}^n \hat{a}_{\alpha}^{\dagger} (A_i)_{\alpha\beta}\hat{a}_{\beta}$.
These operators also form a representation of  $\mathfrak{g}$, preserving the commutation relations $[\hat{A}_i,\hat{A}_j]=\sum_{k=1}^N c_{ij}^k \hat{A}_k$.
The homomorphism $A_i \mapsto \hat{A}_i$ from matrices to operators on a bosonic Fock space is known as the JS map.
\par We focus on the angular momentum algebra $\mathfrak{su}(2)$, whose fundamental representation is generated by $S_i=\sigma_i/2$, $i\in\{x,y,z\}$, satisfying $[S_i,S_j]=i\varepsilon_{ijk}S_k$, with $\sigma_i$ the Pauli matrices.
\par
\begin{figure}
    \centering
\includegraphics[width=0.45\linewidth]{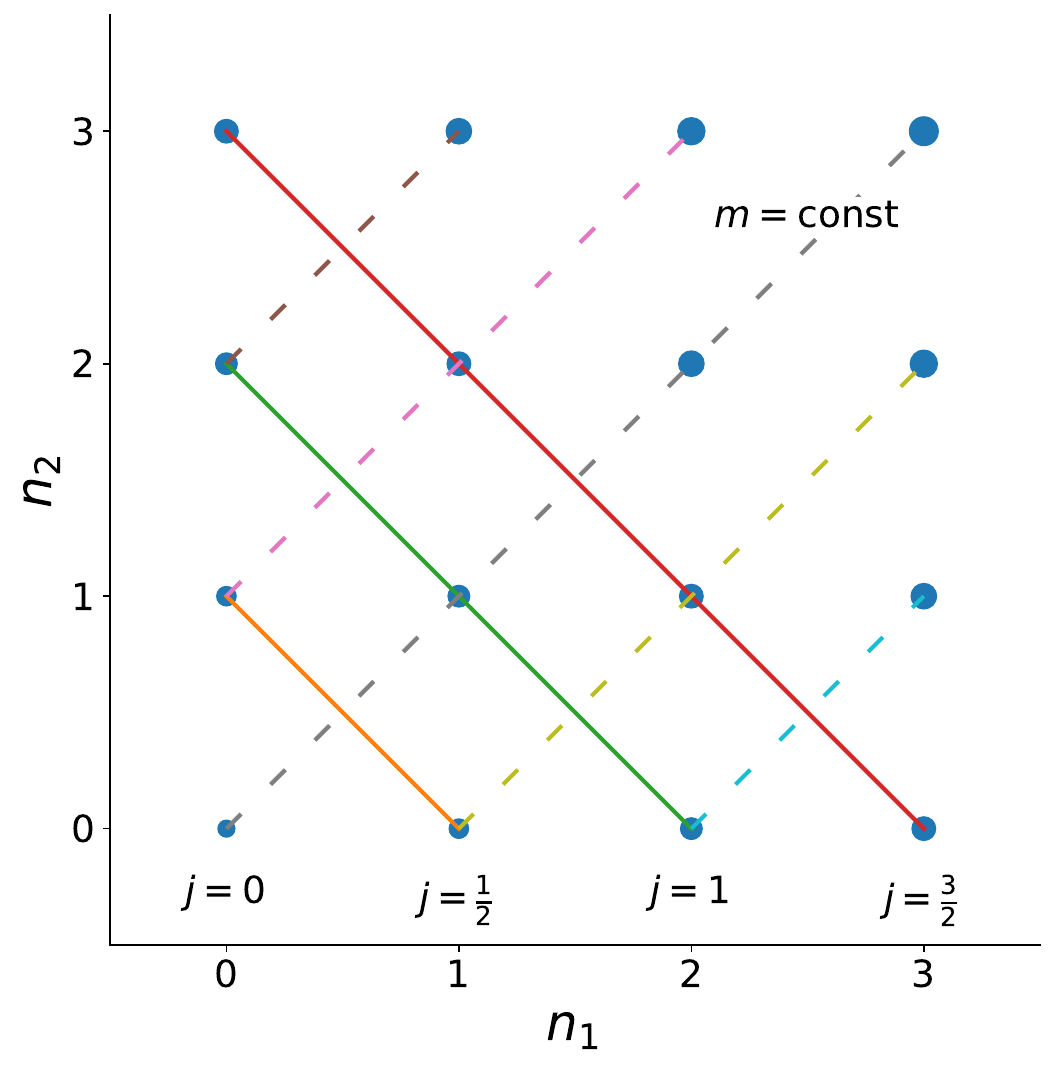}
    \caption{A schematic of the two-mode bosonic lattice with total particle number $\hat{n} =\hat{n}_1 + \hat{n}_2$, where $\hat{n}_1$ and  $\hat{n}_2$ denote the number of particles in each mode. Bold lines connect states with fixed $n = 2j$, corresponding to spin-$j$ representations. Dashed lines connect states with constant $\hat{S}_z = (\hat{n}_1 - \hat{n}_2)/2$, where $m$ is the magnetic quantum number.}
    \label{fig:2}
\end{figure}

 \begin{definition}
The JS map is the representation of the generators of $\mathfrak{su}(2)$ in terms of two bosonic modes given by
\begin{eqnarray}\label{eq:su2_schwinger}
\hat{S}_x=\tfrac{1}{2}\left(\hat{a}_1^{\dagger}\hat{a}_2+\hat{a}_2^{\dagger}\hat{a}_1\right),
\hat{S}_y=\tfrac{i}{2}\left(\hat{a}_2^{\dagger}\hat{a}_1-\hat{a}_1^{\dagger}\hat{a}_2\right),
\hat{S}_z=\tfrac{1}{2}\left(\hat{a}_1^{\dagger}\hat{a}_1-\hat{a}_2^{\dagger}\hat{a}_2\right).
\end{eqnarray}
\end{definition}
Equivalently, introducing the number operators $\hat n_i=\hat a_i^\dagger \hat a_i$, one has $\hat{S}_z=(\hat{n}_1-\hat{n}_2)/2$. The associated Casimir operator takes the form
$\hat{S}^2=\hat{S}_x^2+\hat{S}_y^2+\hat{S}_z^2
=((\hat{n}_1+\hat{n}_2)/2)((\hat{n}_1+\hat{n}_2)/2+1)$. Hence the total boson number $\hat n=\hat n_1+\hat n_2$
is conserved, and each subspace of fixed $\hat n=2j$ carries an irreducible spin-$j$ representation of $\mathfrak{su}(2)$, with magnetic quantum number $m=(n_1-n_2)/2$ (see Fig.~\ref{fig:2}).
\par Without fixing $n=2j$, the operators act on the full Fock space, which decomposes into a direct sum of all spin representations. 
The normalized two-mode Fock states are
\begin{equation}
\ket{n_1,n_2}_2=\frac{(\hat{a}_{1}^{\dagger})^{n_1}}{\sqrt{n_1!}}\,
\frac{(\hat{a}_{2}^{\dagger})^{n_2}}{\sqrt{n_2!}}\ket{0,0}, \qquad 
n_1,n_2\in\mathbb{N}_0,
\end{equation}
defined on the Hilbert space $\mathcal{H}_{\mathrm{Fock}}=\operatorname{span}\{\ket{n_1,n_2}_2:n_1,n_2\in\mathbb{N}_0\}$.
This space decomposes into subspaces of fixed total excitation:
\begin{equation}
\mathcal{H}_{\mathrm{Fock}}=\bigoplus_{n=0}^\infty S_n,\qquad 
S_n=\operatorname{span}\{\ket{n_1,n_2}_2:n_1+n_2=n\},
\end{equation}
with $\dim S_n=n+1$. Setting $n=2j$, the corresponding subspace is
\begin{equation}\label{s2j}
\!\!\!S_{2j}=\operatorname{span}\{\ket{j+m,\,j-m}_2\!\!:m=\!-j,-j+1,\ldots,j\},
\end{equation}
which the Schwinger construction identifies with angular momentum eigenstates~\cite{Schwinger1952},
\begin{equation}
\ket{j,m}=\frac{(\hat{a}_{1}^{\dagger})^{j+m}}{\sqrt{(j+m)!}}\,
\frac{(\hat{a}_{2}^{\dagger})^{j-m}}{\sqrt{(j-m)!}}\ket{0,0}.
\end{equation}
The parametrization $n=2j$, $m=(n_1-n_2)/2$
establishes the isomorphism
\begin{equation}\label{1139}
\ket{n_1,n_2}_2\;\longleftrightarrow\;
\ket{j=\tfrac{n_1+n_2}{2},\,m=\tfrac{n_1-n_2}{2}},
\end{equation}
between the Fock subspace $S_{2j}$ and the spin-$j$ Hilbert space $\mathcal{H}_{2j}$ of dimension $2j+1$. 
This correspondence is precisely the JS map and is one-to-one within each fixed sector $n=2j$.
\begin{definition}\label{def_3_2}
    The associated unitary relabeling is
\begin{eqnarray}\label{1256}
{\mathcal{E}}_j:S_{2j}\rightarrow\mathcal{H}_{2j},\quad {\mathcal{E}}_j\ket{j+m,j-m}_2=\ket{j,m}.
\end{eqnarray}
\end{definition}
This is not a physical operator associated with an observable, but a unitary identification of the bosonic sector
\eqref{s2j} with the spin-$j$ Hilbert space $\mathcal{H}_{2j}$.
The map is exact and  bijective within each block $S_{2j}$.
\par Using \eqref{1139}, we reparametrize the two-mode state $\hat{\rho}_{(2)}$, the analogue of \eqref{1939}, by setting $n_1=j+m$, $n_2=j-m$, $n'_1=j+m'$, $n'_2=j-m'$.  The resulting operator is referred to as the \textit{restricted two-mode Fock state}:
\begin{definition}
Let $\hat{\Pi}_{2j}=\sum_{m=-j}^{j}\ket{j+m,j-m}_2\bra{j+m,j-m}_2$ be an orthogonal  projector onto $n=2j$ subspace.
Then, for a two-mode bosonic state $\hat\rho_{(2)}$, we define its restriction to the fixed total occupation sector $n=2j$ by
\begin{equation}\label{1042}
\hat\rho_{(2)}^{(2j)}:=\Pi_{2j}\hat\rho_{(2)}\Pi_{2j}.
\end{equation}
\end{definition}
{Equation \eqref{1042} is a standard example of a linear completely positive trace-nonincreasing map, since it has the single Kraus form with Kraus operator $\Pi_{2j}$. The restricted operator remains Hermitian and positive semidefinite. However, its trace is 
$p_j\equiv \tr{\hat{\Pi}_{2j}\hat{\rho}_{(2)}}$, with $0\leq p_j\leq 1$. 
The normalized restriction is defined only for sectors with nonzero probability weight, $p_j>0$, and is given by
\begin{align}
 \hat\rho_{r(2)}^{(2j)}\equiv    
 \frac{\hat{\Pi}_{2j}\hat{\rho}_{(2)}\hat{\Pi}_{2j}}
 {\tr{\hat{\Pi}_{2j}\hat{\rho}_{(2)}}}.
\end{align}} 
\begin{proposition}\label{prop1}
  At the level of density operators, the JS map takes the form
\begin{eqnarray}\label{1401}
    \hat{\rho}^{(j)}=\mathcal{E}_j  \hat\rho_{r(2)}^{(2j)}\mathcal{E}_j^{\dagger}\equiv \mathcal{JS}[\hat{\rho}_{(2)}].
\end{eqnarray} Its inverse on the spin-$j$ state space is
\begin{eqnarray}\label{inv_1409}
   \hat{\rho}_{r(2)}^{(2j)}={\mathcal{E}}_j^{\dagger}\hat{\rho}^{(j)}{\mathcal{E}}_j\equiv  \mathcal{JS}^{-1}[\hat{\rho}^{(j)}].
\end{eqnarray}
\end{proposition}
The map from the definition~\eqref{def_3_2} is unitary on the fixed sector and therefore preserves Hermiticity, positivity, and trace
exactly. Thus, at the state level, the sector-restricted JS correspondence maps physical states to physical states.
\begin{remark}
The JS map is one-to-one when restricted to the fixed-excitation sector $\mathcal H_{2j}$. In particular, if $\hat\rho_{(2)}$ is already supported on $\mathcal H_{2j}$, then $\hat\rho_{r(2)}^{(2j)}=\hat\rho_{(2)}$ and the map reduces to
\begin{equation}
\hat\rho^{(j)}=\mathcal E_j \hat\rho_{(2)} \mathcal E_j^\dagger,
\end{equation}
which is a unitary equivalence between the bosonic sector $\mathcal H_{2j}$ and the spin-$j$ Hilbert space. In this case, the inverse $\mathcal{JS}^{-1}$ is exact and the correspondence is one-to-one at the level of states.
\end{remark}
Restricting to a single excitation number $n=2j$ amounts to selecting a $(2j+1)\times(2j+1)$ block of the infinite dimensional density matrix, which is isomorphic to the spin-$j$ density matrix on the finite dimensional Hilbert space $\mathcal{H}_{2j}$. The full two-mode Hilbert space decomposes as $\mathcal{H}_{\mathrm{Fock}}=\bigoplus_{j\in\{0,\tfrac12,1,\tfrac32,\ldots\}}\mathcal{H}_{2j}$, and reconstructing the full operator $\hat{\rho}_{(2)}$ therefore requires all blocks $\{\hat{\rho}^{(j)}\}_{j\in\{0,\tfrac12,1,\tfrac32,\ldots\}}$. A single $j$ block only captures the state restricted to that excitation sector.
If the total number of excitations is fixed, the corresponding block contains all relevant information: occupations as well as correlations and interference effects are encoded in its off-diagonal elements. In this case, the two-mode density matrix is block diagonal with a single nonzero block, while all elements $\rho_{n_1,n_2;n'_1,n'_2}$ vanish unless $n_1+n_2=2j$ and $n'_1+n'_2=2j$. By contrast, a generic Fock state involves superpositions across different excitation numbers. For example, $\tfrac{1}{\sqrt{2}}(\ket{0,0}+\ket{1,1})$ contains contributions from $n=0$ and $n=2$, with $\ket{0,0}\in S_0$ and $\ket{1,1}\in S_2$, which are orthogonal subspaces. Thus, when the excitation number is not fixed, the full direct sum structure is required.
Extending the JS mapping beyond two modes is nontrivial, as spin quantum numbers no longer provide a complete labeling of states. A bijective multimode construction was recently proposed in Ref.~\cite{dubus2024bosons}. Here, we focus on tomographic representations and restrict attention to the fundamental two-mode case.
\section{Holstein--Primakoff Isomorphism}\label{hp_sec}
\par In the previous section, we discussed a representation of the Lie algebra $\mathfrak{su}(2)$ in terms of two bosonic modes. It is therefore natural to ask whether a single-mode representation of $\mathfrak{su}(2)$ exists. Such a transformation is known as the Holstein--Primakoff (HP) map~\cite{holstein1940field} and can be obtained from the JS construction by imposing additional restrictions. 
The HP map is particularly useful for describing collective spin excitations, such as in atomic ensembles or magnetic systems, as bosonic quasiparticles (magnons), while the square root nonlinearity accounts for saturation effects at large excitation numbers~\cite{RevModPhys.82.1041,PhysRevLett.93.237204,MA201189}.
\par We start from the two-mode JS map \eqref{eq:su2_schwinger}. In the standard HP construction, the bosonic vacuum is identified with the spin state of maximal projection $m=+j$. From \eqref{1256}, this corresponds to $n_1=2j$ and $n_2=0$. It is therefore natural to define the boson number of the single-mode HP representation as the occupation of the second JS mode, $\hat{k}\equiv\hat{n}_2$. The conservation law $\hat{n}_1+\hat{n}_2=2j$ then implies $\hat{n}_1=2j-\hat{k}$. Accordingly, the basis vectors of the $(2j+1)$-dimensional subspace can be relabeled as
\begin{equation}\label{1146}
|k\rangle_{\rm HP} \equiv |2j-k,k\rangle_2,
\qquad k = 0,1,\dots ,2j.
\end{equation}
The HP map is defined on the finite dimensional Hilbert subspace
$S_{2j}={\rm span}\{|0\rangle_{\rm HP},\allowbreak|1\rangle_{\rm HP},\allowbreak\dots ,\allowbreak|2j\rangle_{\rm HP}\}$ and the latter relabeling means that within the fixed $n=2j$ sector the two-mode Fock space is put into one-to-one correspondence with a single bosonic mode whose occupation number is $k$. However, this does not mean that the physical system has become a genuine single-mode bosonic system. The state \eqref{1146} behaves like single-mode number states only inside this finite dimensional space. Outside $S_{2j}$ the mapping does not exist, the HP representation is not defined globally and cannot describe superpositions of different total excitation numbers.
\par Consider the raising and lowering operators $\hat{S}_+ = \hat{S}_x+i\hat{S}_y = \hat{a}_1^\dagger \hat{a}_2$, $\hat{S}_- = \hat{S}_x-i\hat{S}_y = \hat{a}_2^\dagger \hat{a}_1$, whose action on a two-mode Fock state is
\begin{align}
    &\hat{S}_+ |n_1, n_2\rangle_2 = \sqrt{n_1+1}\sqrt{n_2}\ket{n_1+1, n_2-1}_2,\\\nonumber
    &\hat{S}_- |n_1, n_2\rangle_2
= \sqrt{n_2+1}\sqrt{n_1}  \ket{n_1-1, n_2+1}_2.
\end{align}
Using the HP parametrization $n_1 = 2j-k$, $n_2 = k$, we obtain
\begin{align}
    &\hat{S}_+ \ket{k}_{\rm HP}
    = \sqrt{2j-k+1}\sqrt{k}\ket{k-1}_{\rm HP},\\\nonumber
    & \hat{S}_- \ket{k}_{\rm HP}
    = \sqrt{2j-k}\sqrt{k+1} \ket{k+1}_{\rm HP}
\end{align}
Since the single-mode ladder operators act as $\hat{a}|k\rangle_{\rm HP} = \sqrt{k}|k-1\rangle_{\rm HP}$, $\hat{a}^{\dagger}|k\rangle_{\rm HP} = \sqrt{k+1} |k+1\rangle_{\rm HP}$ comparison yields
\begin{equation}
    \hat{S}_+ = \hat{a}\,\sqrt{2j-\hat{k}+1}, \quad \hat{S}_- = \hat{a}^\dagger\sqrt{2j-\hat{k}}.
\end{equation}
Using the identity $\hat{a}f(\hat{k}) = f(\hat{k}+1)\hat{a}$, where the function $f(\cdot)$ is defined on the integers, one can reorder the square roots as $ \hat{S}_+ = \sqrt{2j-\hat{k}}\hat{a}$. 
Finally, the $z$-component follows directly from the JS expression $\hat{S}_z = \tfrac12(\hat{n}_1-\hat{n}_2)$.
Thus we obtain the standard HP representation
\begin{equation}\label{eq:HP_standard}
    \hat{S}_{+} = \sqrt{2j-\hat{k}}\hat{a},\qquad
    \hat{S}_{-} = \hat{a}^{\dagger}\sqrt{2j-\hat{k}},\qquad
    \hat{S}_z = j - \hat{k}.
\end{equation}
The HP map provides an exact isomorphism between the truncated bosonic space $S_{2j}$ and the spin-$j$ Hilbert space $\mathcal{H}_{2j}$. \begin{definition} The associated unitary relabeling is
\begin{equation}\label{1249}
\mathsf{E}_j: S_{2j}\rightarrow\mathcal{H}_{2j},\qquad
\mathsf{E}_j|k\rangle_{\rm HP} = {\mathsf{E}}_j\ket{2j-k,k}_{2}=\ket{j,m}, \quad m=j-k.
\end{equation}
\end{definition}
\par  {The \textit{restricted single-mode Fock state}, that is, the restriction of the single-mode bosonic density operator \eqref{1939} to the physical subspace $S_{2j}$, is obtained by projection. 
The normalized restriction is defined only when the selected subspace has nonzero probability weight,}
\begin{equation}
p_j^{\rm HP}\equiv \tr\!\bigl(\hat{P}_{2j}\hat{\rho}_{(1)}\bigr)>0.
\end{equation}
In this case,
\begin{equation}\label{1352}
\hat{\rho}_{r(1)}^{(2j)} = 
\frac{\hat{P}_{2j}\,\hat{\rho}_{(1)}\,\hat{P}_{2j}}
{\tr\!\bigl(\hat{P}_{2j}\hat{\rho}_{(1)}\bigr)},\qquad
\hat{P}_{2j} = \sum_{k=0}^{2j} |k\rangle_{\rm HP}\langle k|_{\rm HP}.
\end{equation}
\begin{proposition}\label{prop2}
At the level of density operators, the Holstein--Primakoff map is given by
\begin{equation}\label{eq:HP_state_map}
\hat{\rho}^{(j)}
=
\mathsf{E}_j\,\hat{\rho}_{r(1)}^{(2j)}\,\mathsf{E}_j^{\dagger}
\equiv
\mathcal{HP}_j\bigl[\hat{\rho}_{(1)}\bigr].
\end{equation}
This correspondence is invertible on the fixed sector $\mathcal H_{2j}$, with inverse
\begin{equation}
\hat{\rho}_{r(1)}^{(2j)}
=
\mathsf{E}_j^{\dagger}\,\hat{\rho}^{(j)}\,\mathsf{E}_j
\equiv
\mathcal{HP}_j^{-1}\bigl[\hat{\rho}^{(j)}\bigr].
\end{equation}
\end{proposition}
\par The square root function  $\sqrt{2j-\hat{k}}$  introduces an essential nonlinearity in \eqref{eq:HP_standard}. In the limit $j \gg 1$ with $\langle \hat{k} \rangle \ll 2j$, the square root can be expanded, yielding the linearised approximation $\hat{S}_+ \approx \sqrt{2j}\hat{a}$, which describes non-interacting spin waves (magnons). The full nonlinear HP map captures the saturation effects that become significant when the excitation number approaches the maximal value $2j$.
\section{Connecting DV and CV States via the Tomographic JS and HP Maps}\label{sec_4}
\par {In this section we demonstrate how JS and HP maps induce sector-wise connections between DV spin tomograms and the corresponding restricted CV bosonic tomograms.}
\par As noted in proposition~\ref{prop1} the inversion of the JS map proceeds through the restricted state \eqref{inv_1409}. 
\begin{definition}
    The \textit{restricted symplectic tomogram} of the two-mode bosonic system $ \boldsymbol{x}=(x_1, x_2)$, $\boldsymbol{\mu}=(\mu_1, \mu_2)$,  $\boldsymbol{\nu}=(\nu_1, \nu_2)$ is
\begin{eqnarray}\label{1035}
\mathcal{W}^{(2j)}_{\mathrm{JS}}(\boldsymbol{x}|\boldsymbol{\mu},\boldsymbol{\nu}) &=& \tr{\hat{\rho}_{r(2)}^{(2j)} \hat{\mathcal{U}}(\boldsymbol{x}, \boldsymbol{\mu}, \boldsymbol{\nu})},
\end{eqnarray}
which can be viewed as the projection of the full symplectic tomogram \eqref{eq:deq-simpl} onto the $2j$ excitation subspace.
\end{definition}
Similarly, we introduce:
\begin{definition} The
 \textit{restricted photon-number tomogram} and \textit{restricted Wigner function} are
\begin{eqnarray}\label{1085}
     \mathsf{w}_{\vec r,\mathrm{JS}}^{(2j)}(\boldsymbol{\alpha}) = \tr{\hat{\rho}_{r(2)}^{(2j)} \hat{\mathcal{U}}_{\vec r}(\boldsymbol{\alpha})},\quad
 \mathbb{W}^{(2j)}_{\mathrm{JS}}(\boldsymbol{\alpha}) = \tr{\hat{\rho}_{r(2)}^{(2j)} \hat{\mathbb{U}}(\boldsymbol{\alpha})},
\end{eqnarray}
where $\vec{r}=(r_1,r_2)$, $\boldsymbol{\alpha}=(\alpha_1,\alpha_2)$.
\end{definition}
The restricted tomograms and the Wigner function associated with the single-mode bosonic state are constructed in the same way, using \eqref{1352}:
\begin{eqnarray}
&\mathcal{W}^{(2j)}_{\mathrm{HP}}(x|\mu,\nu) = \tr{\hat{\rho}_{r(1)}^{(2j)}\,\hat{\mathcal{U}}(x,\mu,\nu)} \\\nonumber\label{eq:restricted_symplectic} 
&\mathsf{w}_{r,\mathrm{HP}}^{(2j)}(\alpha)= \tr{\hat{\rho}_{r(1)}^{(2j)}\,\hat{\mathcal{U}}_r(\alpha)},\quad 
\mathbb{W}^{(2j)}_{\mathrm{HP}}(\alpha) = \tr{\hat{\rho}_{r(1)}^{(2j)}\,\hat{\mathbb{U}}(\alpha)}. 
\end{eqnarray}
Although the spin-$j$ systems density operator $\hat{\rho}^{(2j)}$ is supported on the finite subspace $S_{2j}$,
the photon-number tomogram $\mathsf{w}^{(2j)}_{r}(\alpha)$
is defined with the displaced number POVM on the full Fock space~\eqref{dequantizer-ph-number}.
Since the displacement operator mixes number sectors, $\mathsf{w}^{(2j)}_{r}(\alpha)$ generally has support for an arbitrarily large $r$ when $\alpha\neq 0$ and the normalization $\sum_{r=0}^{\infty}\mathsf{w}^{(2j)}_{r}(\alpha)=1$ holds. 
\par
\begin{figure}
    \centering
\includegraphics[width=0.9\linewidth]{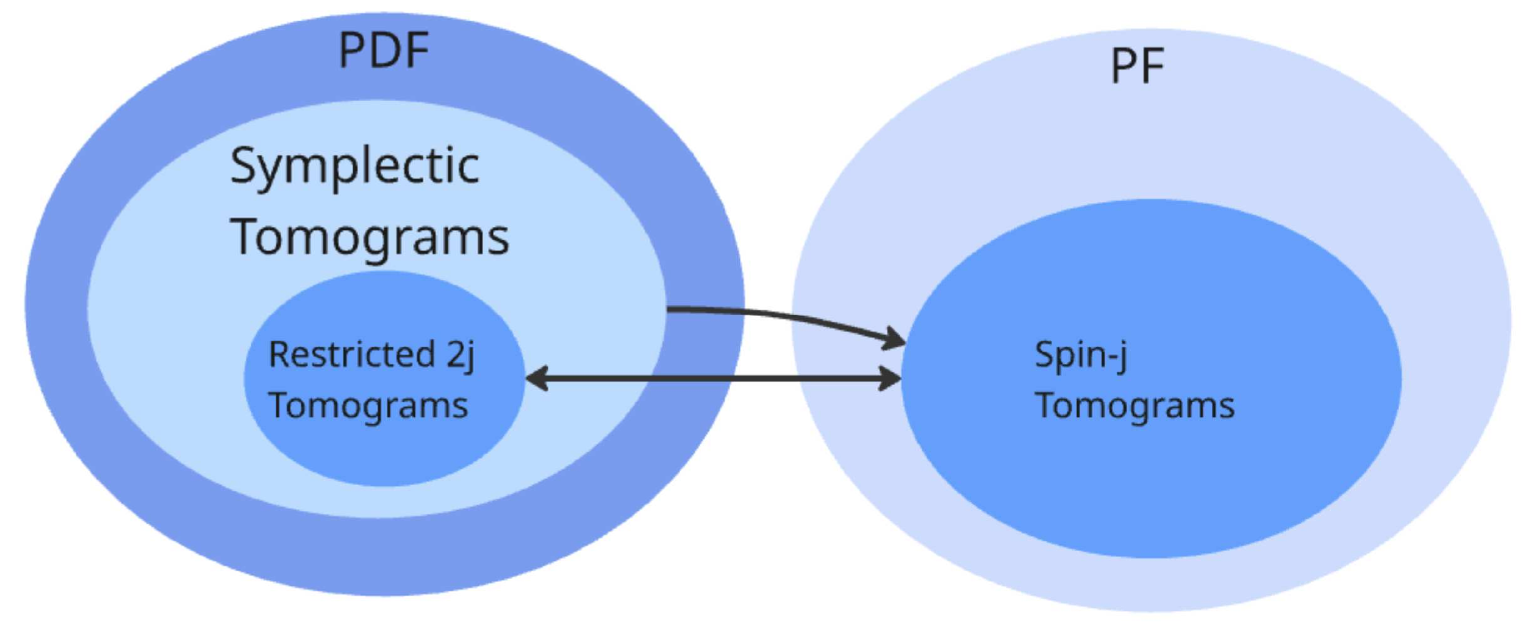}
   \caption{
Relation between CV and DV tomographic descriptions. Every physical CV quantum state admits a symplectic tomogram (that is a PDF), providing a complete representation in the CV setting. However, not every PDF corresponds to a valid quantum state (Theorem~\ref{thm_2}-\ref{thm_3_0}). {The JS and HP maps implement a normalized restriction of the underlying bosonic state to the fixed excitation sector $n=2j$, yielding the corresponding spin-$j$ tomogram.} The inverse transformations reconstruct only the restricted $2j$ tomogram, so the correspondence is invertible only within this sector. In particular, states with support on multiple excitation sectors (e.g., $\tfrac{1}{\sqrt{2}}(\ket{0,0}+\ket{1,1})$) cannot be represented by a single restricted tomogram.
}
    \label{fig:3_0}
\end{figure}

 The transformation from the single- and two-mode bosonic system symplectic tomograms  to the spin-$j$ tomogram $\omega_j(m,\Omega)$ can be written as a composition of three maps:
\begin{eqnarray}\label{1033}
&&\omega_j(m,\Omega) =
\mathcal{S} \left[ \mathcal{JS} \left[ \mathcal{R}^{-1} \left[
\mathcal{W}(\boldsymbol{x}| \boldsymbol{\mu}, \boldsymbol{\nu}) \right] \right] \right],\\\nonumber
&&\omega_j(m,\Omega) = \mathcal{S}\left[\,\mathcal{HP}\!\left[\,
\mathcal{R}^{-1}[\mathcal{W}(x|\mu,\nu)]\right]\right].
\end{eqnarray}
The first step reconstructs the one- and two-mode density operators $\hat{\rho}_{(1)}$ and $\hat{\rho}_{(2)}$ from the symplectic tomogram via the inverse Radon transform \eqref{sympl_tomography_inverse}. The second step applies the JS \eqref{1401} or HP \eqref{1249} mappings, establishing an isomorphism between the fixed $n=2j$ Fock subspace and the spin-$j$ Hilbert space. The final step uses the relation \eqref{spin_deqaunt} between the spin density operator and the spin tomogram. For states supported on $S_{2j}$, the reverse transformation is given by
\begin{eqnarray}\label{1159}
&&\mathcal{W}^{(2j)}_{\mathrm{JS}}(\boldsymbol{x}|\boldsymbol{\mu},\boldsymbol{\nu})=\mathcal{R}[\mathcal{JS}^{-1}[\mathcal{S}^{-1}[\omega_j(m,\Omega)]]],\\\nonumber
&&\mathcal{W}^{(2j)}_{\mathrm{HP}}(x|\mu,\nu) = \mathcal{R}\!\bigl[\,
\mathcal{HP}^{-1}\!\bigl[\,
\mathcal{S}^{-1}\bigl[\omega_j(m,\Omega)\bigr]
\bigr]\bigr].
\end{eqnarray}
Figure~\ref{fig:3_0} illustrates the later relations.
We define \textit{lifted spin dequantizer and quantizer} on Fock space as
\begin{align}
&\hat{\mathcal{U}}^{\mathrm{JS}}_j(m,\Omega)
\equiv \mathcal{E}_j^{\dagger}\hat{\mathcal{U}}(m,\Omega)\mathcal{E}_j,
\label{lifted_spin_dequantizer} \quad 
\hat{\mathcal{D}}^{\mathrm{JS}}_j(m,\Omega)
\equiv \mathcal{E}_j^{\dagger}\hat{\mathcal{D}}(m,\Omega)\mathcal{E}_j,\\\nonumber
&\hat{\mathcal{U}}^{\mathrm{HP}}_j(m,\Omega) = \mathsf{E}_j^{\dagger}\,\hat{\mathcal{U}}(m,\Omega)\,\mathsf{E}_j,
\qquad
\hat{\mathcal{D}}_j^{\mathrm{HP}}(m,\Omega) = \mathsf{E}_j^{\dagger}\,\hat{\mathcal{D}}(m,\Omega)\,\mathsf{E}_j,
\end{align}
They act as $\hat{\mathcal{U}}(m,\Omega)$, $\hat{\mathcal{D}}(m,\Omega)$  on $S_{2j}$ and zero otherwise.
\begin{proposition}
    The  transition  from the full  symplectic tomogram of the two-mode bosonic state to the spin tomogram of the spin-$j$ state and its partial inverted transformation to the restricted symplectic tomogram are
\begin{align}
&\omega_j(m,\Omega)
=\frac{\displaystyle \int_{\mathbb{R}^6} d^2\boldsymbol{x}
d^2\boldsymbol{\mu}
 d^2\boldsymbol{\nu}
\mathcal{W}(\boldsymbol{x}\mid \boldsymbol{\mu},\boldsymbol{\nu})\,
K^{(j,\mathrm{JS})}_{\mathcal{W}\to\omega}(m,\Omega;\boldsymbol{x},\boldsymbol{\mu},\boldsymbol{\nu})}
{\displaystyle \int_{\mathbb{R}^6}\! d^2\boldsymbol{x}d^2\boldsymbol{\mu}
 d^2\boldsymbol{\nu}\;
\mathcal{W}(\boldsymbol{x}\mid \boldsymbol{\mu},\boldsymbol{\nu})\tr{\hat{\mathcal{D}}(\boldsymbol{x}, \boldsymbol{\mu}, \boldsymbol{\nu})\hat{\Pi}_{2j}}} ,
\label{eq:omega_from_W} \\
&\mathcal{W}^{(2j)}_{\mathrm{JS}}(\boldsymbol{x}\mid\boldsymbol{\mu},\boldsymbol{\nu})
=\int d\Omega\sum_{m=-j}^j \omega_j(m,\Omega)\,
K^{(j,\mathrm{JS})}_{\omega\to \mathcal{W}}(m,\Omega;\boldsymbol{x},\boldsymbol{\mu},\boldsymbol{\nu}),
\label{eq:W_from_omega}
\end{align}
where the transformation kernels are the following
\begin{align}
\!\!\!\!K^{(j,\mathrm{JS})}_{\mathcal{W}\to\omega}(m,\Omega;\boldsymbol{x},\boldsymbol{\mu},\boldsymbol{\nu})
&\equiv \tr{\hat{\mathcal{D}}(\boldsymbol{x},\boldsymbol{\mu},\boldsymbol{\nu})\,\hat{\mathcal{U}}^{\mathrm{JS}}_j(m,\Omega)},
\label{952} \\
\!\!\!\!K^{(j,\mathrm{JS})}_{\omega\to \mathcal{W}}(m,\Omega;\boldsymbol{x},\boldsymbol{\mu},\boldsymbol{\nu})
&\equiv \tr{\hat{\mathcal{U}}(\boldsymbol{x},\boldsymbol{\mu},\boldsymbol{\nu})\,\hat{\mathcal{D}}^{\mathrm{JS}}_j(m,\Omega)}.
\label{953}
\end{align}
\end{proposition}
The spin tomogram is obtained from the full bosonic tomogram through the normalized sector restriction to $n=2j$, followed by the JS correspondence on the restricted sector.
The transition  from the restricted symplectic tomogram to a spin tomogram can be deduced in the similar fashion, using the  kernel \eqref{952}. The explicit view of the latter kernels in terms of the Laguerre polynomials is  provided in~\ref{app_2}.
\par We can simplify the latter transformations, using the CF of the two mode  tomogram, that is defined similarly to its single mode version \eqref{characteristic_function} as
\begin{align}
\phi(t_1,t_2;\boldsymbol{\mu},\boldsymbol{\nu})=\int\limits_{\mathbb{R}}\mathcal{W}(\boldsymbol{x}|\boldsymbol{\mu},\boldsymbol{\nu})e^{i(t_1x_1+t_2x_2)}dx_1dx_2.
\end{align}
\begin{proposition}\label{prop_4}
The  transition  from the CF of the  symplectic tomogram of the two-mode bosonic state to the spin tomogram of the spin-$j$ state %
is
\begin{equation}
\omega_j(m,\Omega)
=
\frac{\displaystyle\sum_{m_1,m_2=-j}^j  {\Phi}^{(j)}_{m_1,m_2}(m,\Omega)
\int_{\mathbb R^4}
d^2\boldsymbol\mu\,d^2\boldsymbol\nu\;
\phi(1,1;\boldsymbol{\mu},\boldsymbol{\nu})\,
\mathcal{A}^{(j)}_{m_1m_2}(\boldsymbol\mu,\boldsymbol\nu)
}{
\displaystyle \sum_{k=0}^{2j}
\int_{\mathbb R^4}
d^2\boldsymbol\mu\,d^2\boldsymbol\nu\;
\phi(1,1;\boldsymbol{\mu},\boldsymbol{\nu}) {\phi}_k(1;\mu_1,\nu_1) {\phi}_{2j-k}(1;\mu_2,\nu_2)
}
\end{equation}
where ${\phi}_k(1;\mu,\nu)$  is the CF corresponding to the excited state of the HO \eqref{excited_HO}. We used the notations
\begin{equation}
r\equiv|m_1-m_2|,
\qquad
\eta_1\equiv j+\min\{m_1,m_2\},
\qquad
\eta_2\equiv j-\max\{m_1,m_2\},
\end{equation}
\begin{equation}
a\equiv\max(0,m_2-m_1),
\qquad
b\equiv\max(0,m_1-m_2),
\end{equation}
\begin{align}
 {\Phi}^{(j)}_{m_1,m_2}(m,\Omega)\equiv    D_{m,m_2}^{(j)*}(\Omega)\,
D_{m,m_1}^{(j)}(\Omega)(-1)^{r} \sqrt{
\frac{\eta_1!\,\eta_2!}
{(2j-\eta_2)!\,(2j-\eta_1)!}
}
\end{align}
where $D^{(j)}_{m,k}(\Omega)$ is the Wigner D-matrix.
If $m_1=m_2=l$, then $\mathcal{A}^{(j)}_{ll}(\boldsymbol\mu,\boldsymbol\nu)=\allowbreak\phi_{j+l}(1;\mu_1,\nu_1)
\allowbreak\phi_{j-l}(1;\mu_2,\nu_2)$. If $m_1\neq m_2$, then
\begin{align}\label{amm}
 \mathcal{A}^{(j)}_{m_1m_2}(\boldsymbol\mu,\boldsymbol\nu)&=   (\xi_1^{*}\xi_2)^a
(\xi_1\xi_2^{*})^b
\sum_{s_1=0}^{\eta_1}
\sum_{s_2=0}^{\eta_2}
\binom{r+s_1-1}{s_1}
\binom{r+s_2-1}{s_2}\\\nonumber
&\times
\phi_{\eta_1-s_1}(1;\mu_1,\nu_1)
\phi_{\eta_2-s_2}(1;\mu_2,\nu_2)
\end{align}
where  $\xi_{k}=-\tfrac{(\nu_{k}-i\mu_{k})}{\sqrt{2}}$, $k=1,2$.
\end{proposition}
The integrals appearing in the above expression are well defined and absolutely convergent for any fixed $j$. Indeed, the CFs of the excited harmonic oscillator states have the form \eqref{excited_HO} and are therefore given by a Gaussian factor multiplied by a polynomial of finite degree. The prefactors $(\xi_1^*\xi_2)^a(\xi_1\xi_2^*)^b$ in \eqref{amm} are also polynomial in $(\boldsymbol\mu,\boldsymbol\nu)$. Since any CF satisfies $|\phi(1,1;\boldsymbol\mu,\boldsymbol\nu)|\le 1$, each integrand is bounded by a Gaussian times a polynomial. Hence the integrals over $\mathbb R^4$ are absolutely convergent. The denominator coincides with the weight of the fixed excitation sector $n=2j$. Thus the expression for $\omega_j(m,\Omega)$ is well defined whenever this weight is nonzero, that is, when the initial bosonic state has a nonvanishing projection onto the sector $\mathcal H_{2j}$.
For the inverse transformation we can write
\begin{proposition}
The  transition  from the spin tomogram of the spin-$j$ state to the  restricted two-mode symplectic tomogram supported on the fixed sector $\mathcal H_{2j}$ is
\begin{eqnarray}\label{1448_10}
    \mathcal{W}^{(2j)}_{\mathrm{JS}}(\boldsymbol{x}\mid\boldsymbol{\mu},\boldsymbol{\nu})
&=&\!\! \mathcal W_0(x_1|\mu_1,\nu_1)
\mathcal W_0(x_2|\mu_2,\nu_2)\!\!\!\! \sum\limits_ {m_1,m_2=-j}^j \!\!\!\!  K^{(j)}_{m_1,m_2}(\boldsymbol{x}, \boldsymbol{\mu}, \boldsymbol{\nu})\\\nonumber
&\times&\sum_{m=-j}^j \int d\Omega \omega_j(m,\Omega)
 \mathcal{D}_{m_2 m_1}^{(j)}(m,\Omega),
\end{eqnarray}
where we used the notations
\begin{align}
K^{(j)}_{m_1,m_2}(\boldsymbol{x},\boldsymbol{\mu},\boldsymbol{\nu})
&\equiv
\mathcal{P}^{(j)}_{m_1,m_2}(x_1,\xi_1)
\mathcal{P}^{(j)}_{m_1,m_2}(x_2,\xi_2)
\mathcal{K}^{(j)}_{m_1,m_2}(\xi_1,\xi_2).
\end{align}
where  we used the notations $\xi_{k}=-\tfrac{(\nu_{k}-i\mu_{k})}{\sqrt{2}}$, $k=1,2$ and 
\begin{align}\label{1426_1}
\mathcal{P}^{(j)}_{m_1,m_2}(x_k,\xi_k)
&=
\sum_{s=0}^{\eta_k}
\binom{\eta_k+r}{\,\eta_k-s\,}\,
\frac{1}{s!\,2^{s}}\,
H_{r+2s}\!\Big(\tfrac{x_k}{\sqrt{2}|\xi_k|}\Big),\quad k=1,2,
\end{align}
\begin{align}
  \mathcal{K}^{(j)}_{m_1,m_2}(\xi_1,\xi_2)
&=
\sqrt{
\frac{\eta_1!\,\eta_2!}
{(2j-\eta_2)!\,(2j-\eta_1)!}
}(2|\xi_1\xi_2|)^{-r}
(\xi_1^{*}\xi_2)^{a}
(\xi_1\xi_2^{*})^{b},
\end{align}
\begin{eqnarray}\label{1250_1}
\mathcal{D}_{m_1 m_2}^{(j)}(m,\Omega) =\sum_{L=0}^{2j}c_L\!\!\!\!\sum_{M=-L}^{L} C_{L,M}(m,\Omega) (-1)^{j-m_2} \langle j,m_1; j,-m_2 | L,M \rangle,
\end{eqnarray}
\end{proposition}
The sums in \eqref{1448_10} over $m,m_1,m_2,L,M$ are finite, and the integration over $\Omega$ is over the compact rotation group. Since the spin tomogram $\omega_j(m,\Omega)$ is a probability distribution in $m$ for each $\Omega$, and the spin dequantizer matrix elements $\mathcal D^{(j)}_{m_2m_1}(m,\Omega)$ are finite combinations of Wigner functions and Clebsch--Gordan coefficients, the integral in \eqref{1448_10}
is finite. The apparent singularities in the factors involving $|\xi_k|$ are the usual tomographic singularities at $(\mu_k,\nu_k)=(0,0)$, where the quadrature variable becomes degenerate. Thus the formula should be understood for $(\mu_k,\nu_k)\neq(0,0)$, or distributionally at the singular point.
\begin{proposition} The  transition  from the  symplectic tomogram of the single-mode bosonic state to the spin tomogram of the spin-$j$ state and its partial inverted transformation to the restricted symplectic tomogram are
\begin{align}
\omega_j(m,\Omega) &=
\frac{\displaystyle\int_{\mathbb{R}^3} \,dx\,d\mu\,d\nu\mathcal{W}(x|\mu,\nu)
K^{(j,\mathrm{HP})}_{\mathcal{W}\to\omega}(m,\Omega;x,\mu,\nu)}
{\displaystyle\int_{\mathbb{R}^3} \,dx\,d\mu\,d\nu \mathcal{W}(x|\mu,\nu)\,
\tr{\hat{\mathcal{D}}(x,\mu,\nu)\,\hat{P}_{2j}}},
\label{eq:W_to_omega_int} \\
\mathcal{W}^{(2j)}_{\mathrm{HP}}(x|\mu,\nu) &=
\int d\Omega \sum_{m=-j}^{j} \omega_j(m,\Omega)\,
K^{(j,\mathrm{HP})}_{\omega\to\mathcal{W}}(x,\mu,\nu;m,\Omega),
\label{eq:omega_to_W_int}
\end{align}
where the transformation kernels are given by
\begin{align}
K^{(j,\mathrm{HP})}_{\mathcal{W}\to\omega}(m,\Omega;x,\mu,\nu) &\equiv
\tr{\hat{\mathcal{D}}(x,\mu,\nu)\,\hat{\mathcal{U}}^{\mathrm{HP}}_j(m,\Omega)}, \label{eq:kernel-W-to-omega-HP} \\
K^{(j,\mathrm{HP})}_{\omega\to\mathcal{W}}(x,\mu,\nu;m,\Omega) &\equiv
\tr{\hat{\mathcal{U}}(x,\mu,\nu)\,\hat{\mathcal{D}}_j^{\mathrm{HP}}(m,\Omega)}. \label{eq:kernel-omega-to-W-HP}
\end{align}
\end{proposition}
The explicit view of the latter kernels is given in~\ref{app:HP_symplectic_spin}.  
Similarly to the JS case, we can reformulate  \eqref{eq:W_to_omega_int}, using the CF of the single-mode symplectic tomogram \eqref{characteristic_function}. 
\begin{proposition}
   The  transition  from the CF of the  symplectic tomogram of the single-mode bosonic state to the spin tomogram of the spin-$j$ state is
   \begin{align}
\omega_j(m,\Omega)
&=
\frac{
\iint_{\mathbb{R}^2}d\mu\, d\nu
\phi(1;\mu,\nu) A^{(j,\mathrm{HP})}_{\phi\to\omega}(m,\Omega,\mu,\nu)\,
}{\sum_{k=0}^{2j}
\iint_{\mathbb{R}^2}d\mu\, d\nu
\phi(1;\mu,\nu)\phi_k(1;\mu,\nu)
}\label{eq:W_to_omega_int_12} 
\end{align}
where we used the notation
\begin{align}
    A^{(j,\mathrm{HP})}_{\phi\to\omega}(m,\Omega,\mu,\nu)&\equiv
\sum_{m_1,m_2=-j}^{j}
D^{(j)*}_{m,m_2}(\Omega)\,
D^{(j)}_{m,m_1}(\Omega)\,
M(j-m_1,\,j-m_2;\xi).
\end{align}
Here $M(a,b;\xi)=\phi_a(1;\mu,\nu)$ if $a=b$. Else 
\begin{equation}
M(a,b;\xi)=
\displaystyle
\sqrt{\frac{n!}{(n+r)!}}\,
(-\xi^*)^{\max(0,b-a)}\,
\xi^{\max(0,a-b)}
\sum_{s=0}^{n}
\binom{r+s-1}{s}\,
\phi_{n-s}(1;\mu,\nu),
\end{equation}
where we used the notations  $\xi=\frac{\nu-i\mu}{\sqrt2}$, %
\end{proposition}
The integrals in Eq.~\eqref{eq:W_to_omega_int_12} are absolutely convergent for every fixed $j$ for essentially the same reason as in the JS case above. 
The denominator is the weight of the truncated HP sector with $0\leq n\leq 2j$. Therefore, the normalized spin tomogram is well defined provided this weight is nonzero.
\begin{proposition}
The  transition  from the spin tomogram of the spin-$j$ state to the  restricted single-mode symplectic tomogram supported on the fixed sector $\mathcal H_{2j}$ is
\begin{eqnarray}
    \mathcal{W}^{(2j)}_{\mathrm{HP}}(x|\mu,\nu) &=&
\mathcal W_0(x|\mu,\nu) \sum_{m_1,m_2=-j}^{j}K^{(j)}_{m_1,m_2}(x,\mu,\nu)\\\nonumber&\times&
\sum_{m=-j}^{j}\int d\Omega \omega_j(m,\Omega)\,
\mathcal D^{(j)}_{m_2m_1}(m,\Omega)\,
\end{eqnarray}
where 
\begin{align}
    K^{(j)}_{m_1,m_2}(x,\mu,\nu)\equiv \mathcal{P}^{(j)}_{m_1,m_2}(x,\xi)
\sqrt{\frac{\eta_2!}{(\eta_2+r)!}}
\left(\frac{i\,\xi'}{\sqrt{2}|\xi|}\right)^{r},
\end{align}
and \begin{equation}
\xi'\equiv 
\begin{cases}
\xi, & m_1\le m_2,\\
-\xi^{*}, & m_1>m_2.
\end{cases}
\end{equation}
The $\mathcal{P}^{(j)}_{m_1,m_2}(x,\xi)$ is given in \eqref{1426_1}, and $\mathcal D^{(j)}_{m_2m_1}(m,\Omega)$ in \eqref{1250_1}.
\end{proposition} The integral is well defined: for fixed $j$ the sums are finite, the $\Omega$-domain is compact, and the integrand is bounded. The same convergence arguments as above apply.
\par The transformations from  the photon-number tomograms of the single- and two-mode bosonic states to  the spin-$j$ tomogram are
 \begin{eqnarray}
   \!\!\!\!\!\!\!\!  && \omega_{j}(m,\Omega) = \mathcal{S} [ \mathcal{JS} [ \mathcal{P}^{-1} [\mathsf{w}_{\vec r}( \boldsymbol{\alpha})]]],\!\!\quad  \mathsf{w}^{(2j)}_{\vec r, \mathrm{JS}}( \boldsymbol{\alpha}) =  \mathcal{P} [ \mathcal{JS}^{-1} [ \mathcal{S}^{-1}  [\omega_{j}(m,\Omega)]]]\\\nonumber
    \!\!\!\!\!\!\!\! && \omega_j(m,\Omega)= \mathcal{S}\!\bigl[\,\mathcal{HP}\!\bigl[\,
\mathcal{P}^{-1}\bigl[\mathsf{w}_r(\alpha)\bigr]\,\bigr]\bigr],\!\!\!\!\quad
      \mathsf{w}_{r, \mathrm{HP}}^{(2j)}(\alpha)= \mathcal{P}\!\bigl[\,
\mathcal{HP}^{-1}\!\bigl[\,
\mathcal{S}^{-1}\bigl[\omega_j(m,\Omega)\bigr]
\bigr]\bigr].
\end{eqnarray}
where $\mathcal{P}$ denotes the map \eqref{dequantizer-ph-number} from $\hat{\rho}_{(1)}$ and  $\hat{\rho}_{(2)}$  to the photon-number tomograms.
\begin{proposition} The connection of the spin and the photon-number tomograms via the JS transformation is:
\begin{eqnarray}\label{1055_2}
\!\!\!\!\!\!\!\!\omega_j(m,\Omega)
&=&\!\frac{\displaystyle\int_{\mathbb{C}^2}d^2\boldsymbol{\alpha} \sum\limits_{\vec r=0}^{\infty}\mathsf{w}_{\vec r}(\boldsymbol{\alpha}) K^{(j,\mathrm{JS})}_{\mathsf{w}\rightarrow \omega}(m,\Omega; \vec r, \boldsymbol{\alpha})  }{\displaystyle \int_{\mathbb{C}^2}d^2\boldsymbol{\alpha} \sum\limits_{\vec r=0}^{\infty}\mathsf{w}_{\vec r}(\boldsymbol{\alpha})\tr{\hat{\Pi}_{2j}\hat{\mathcal{D}}_{\vec r}(\boldsymbol{\alpha})}},\\\nonumber
  \mathsf{w}^{(2j)}_{\vec r,\mathrm{JS}}( \boldsymbol{\alpha})   & =& \displaystyle\int d\Omega\sum\limits_{m=-j}^j\omega_{j}(m,\Omega) K^{(j,\mathrm{JS})}_{\omega\rightarrow \mathsf{w}}(m,\Omega; \vec r, \boldsymbol{\alpha}),
\end{eqnarray}
and via the HP transformation is:
\begin{align}
\omega_j(m,\Omega) &=
\frac{\displaystyle\int_{\mathbb{C}} d\alpha\;\sum_{r=0}^{\infty} \mathsf{w}_r(\alpha)\;
K^{(j,\mathrm{HP})}_{\mathsf{w}\to\omega}(m,\Omega;\,r,\alpha)}
{\displaystyle\int_{\mathbb{C}} d\alpha\;\sum_{r=0}^{\infty} \mathsf{w}_r(\alpha)\;
\mathrm{tr}\!\bigl\{\hat{\mathcal{D}}_r(\alpha)\,\hat{P}_{2j}\bigr\}}, \label{eq:photon-to-omega} \\[4pt]
\mathsf{w}_{r, \mathrm{HP}}^{(2j)}(\alpha) &=
\int d\Omega\sum_{m=-j}^{j} \omega_j(m,\Omega)\;
K^{(j,\mathrm{HP})}_{\omega\to\mathsf{w}}(m,\Omega;\,r,\alpha), \label{eq:omega-to-photon}
\end{align}
The transition kernels are the following
\begin{align}
    K^{(j,\mathrm{JS})}_{\mathsf{w}\rightarrow \omega}(m,\Omega; \vec r, \boldsymbol{\alpha}) &\equiv\tr{ \hat{\mathcal{D}}_{\vec r}(\boldsymbol{\alpha}) \hat{\mathcal{U}}^{\mathrm{JS}}_{j}(m,\Omega) },\label{1554_1}\\
    K^{(j,\mathrm{JS})}_{\omega\rightarrow \mathsf{w}}( m,\Omega; \vec r, \boldsymbol{\alpha}) &\equiv  \tr{  \hat{\mathcal{U}}_{\vec r}(\boldsymbol{\alpha})\hat{\mathcal{D}}^{\mathrm{JS}}_j(m,\Omega)}.\label{1554_2}\\
K^{(j,\mathrm{HP})}_{\mathsf{w}\to\omega}(m,\Omega;\,r,\alpha)
&\equiv \mathrm{tr}\!\bigl\{\hat{\mathcal{D}}_r(\alpha)\,\hat{\mathcal{U}}^{\mathrm{HP}}_j(m,\Omega)\bigr\}, \label{eq:kernel-w-to-omega-HP}\\
K^{(j,\mathrm{HP})}_{\omega\to\mathsf{w}}(m,\Omega;\,r,\alpha)
&\equiv \mathrm{tr}\!\bigl\{\hat{\mathcal{U}}_r(\alpha)\,\hat{\mathcal{D}}^{\mathrm{HP}}_j(m,\Omega)\bigr\}. \label{eq:kernel-omega-to-w-HP}
\end{align}
\end{proposition}
The explicit view of the latter kernels is given in \ref{app_3} and in \ref{app_7}.
We do not present the explicit substitution of the kernels in this case, as it proceeds in complete analogy with the JS and HP transformations considered above and does not introduce any new technical features.
The expressions are well defined for photon-number tomograms
$\mathsf w_r(\alpha)$ which, by definition, satisfy
$\int_{\mathbb{C}^2} d^2\boldsymbol{\alpha}
\sum_{\vec r=0}^{\infty}\mathsf{w}_{\vec r}(\boldsymbol{\alpha})=1$.
The kernels \eqref{1554_2}--\eqref{eq:kernel-omega-to-w-HP} are constructed from Gaussian factors multiplied by associated Laguerre polynomials and finite powers of $\alpha$ and $\alpha^*$ (see \ref{app_3} and in \ref{app_7}). Hence, for fixed $j$, they exhibit at most polynomial growth in the indices and variables.
The Gaussian factors provide exponential decay in $\alpha$, which dominates the polynomial growth. Consequently, the integrands are controlled by an integrable function, and the sum integral expressions converge absolutely. 
\par The transformations from Wigner functions of the single- and two-mode bosonic states to the spin tomogram  are:
 \begin{align}\label{eq:Wig_to_omega_HP}
    &  \omega_{j}(m,\Omega) = \mathcal{S} [ \mathcal{JS} [ \mathcal{F}^{-1} [\mathbb{W}( \boldsymbol{\alpha})]]], 
     \mathbb{W}^{(2j)}_{\mathrm{JS}}( \boldsymbol{\alpha}) = \mathcal{F} [  \mathcal{JS}^{-1} [ \mathcal{S}^{-1} [\omega_{j}(m,\Omega)]]],\\\nonumber
&\omega_j(m,\Omega) = \mathcal{S}\!\bigl[\,\mathcal{HP}\!\bigl[\mathcal{F}^{-1}\bigl[\mathbb{W}(\alpha)\bigr]\bigr]\bigr], 
\mathbb{W}^{(2j)}_{\mathrm{HP}}(\alpha) = \mathcal{F}\!\bigl[\,
\mathcal{HP}^{-1}\!\bigl[\,
\mathcal{S}^{-1}\!\bigl[\,\omega_j(m,\Omega) \bigr]
\bigr]\bigr].
\end{align}
\begin{proposition}The transformation between the spin tomogram and the Wigner function is given by JS transformation:
\begin{eqnarray}\label{1248_4}
\omega_j(m,\Omega) &=& 
\frac{\displaystyle\int_{\mathbb{C}^2} d^2 \boldsymbol{\alpha}\mathbb{W}( \boldsymbol{\alpha})  K^{(j, \mathrm{JS})}_{\mathbb{W}\rightarrow \omega}(m,\Omega; \boldsymbol{\alpha})}{\displaystyle\int_{\mathbb{C}^2}  d^2 \boldsymbol{\alpha}\mathbb{W}( \boldsymbol{\alpha})\tr{\hat{\Pi}_{2j} \hat{\mathbb{D}}( \boldsymbol{\alpha})}},\\\nonumber
\mathbb{W}^{(2j)}_{\mathrm{JS}}( \boldsymbol{\alpha})&=&\int d\Omega\sum\limits_{m=-j}^j \omega_{j}(m,\Omega) K^{(j, \mathrm{JS})}_{\omega\rightarrow \mathbb{W}}(m,\Omega; \boldsymbol{\alpha}),
\end{eqnarray}
and HP transformation:
\begin{align}
\omega_j(m,\Omega) &=
\frac{\displaystyle\int_{\mathbb{C}} d\alpha\;\mathbb{W}(\alpha)\;
K^{(j,\mathrm{HP})}_{\mathbb{W}\to\omega}(m,\Omega;\,\alpha)}
{\displaystyle\int_{\mathbb{C}} d\alpha\;\mathbb{W}(\alpha)\;
\mathrm{tr}\!\bigl\{\hat{\mathbb{D}}(\alpha)\,\hat{P}_{2j}\bigr\}}, \label{eq:Wigner-to-omega} \\[4pt]
\mathbb{W}^{(2j)}_{\mathrm{HP}}(\alpha) &=
\int d\Omega\sum_{m=-j}^{j} \omega_j(m,\Omega)\;
K^{(j,\mathrm{HP})}_{\omega\to\mathbb{W}}(m,\Omega;\,\alpha), \label{eq:omega-to-Wigner}
\end{align}
where the kernels are
\begin{align}\label{1713}
    K^{(j,\mathrm{JS})}_{\mathbb{W}\rightarrow \omega}(m,\Omega; \boldsymbol{\alpha})&\equiv \tr{ \hat{\mathbb{D}}( \boldsymbol{\alpha})   \hat{\mathcal{U}}^{\mathrm{JS}}_j(m,\Omega)},\\\nonumber
    K^{(j,\mathrm{JS})}_{\omega\rightarrow \mathbb{W}}(m,\Omega; \boldsymbol{\alpha})&\equiv\tr{ \hat{\mathbb{U}}( \boldsymbol{\alpha})   \hat{\mathcal{D}}^{\mathrm{JS}}_j(m,\Omega)},
    \\\nonumber
K^{(j,\mathrm{HP})}_{\mathbb{W}\to\omega}(m,\Omega;\,\alpha)
& \equiv \mathrm{tr}\!\bigl\{\hat{\mathbb{D}}(\alpha)\,\hat{\mathcal{U}}^{\mathrm{HP}}_j(m,\Omega)\bigr\},\\\nonumber
K^{(j,\mathrm{HP})}_{\omega\to\mathbb{W}}(m,\Omega;\,\alpha)
& \equiv \mathrm{tr}\!\bigl\{\hat{\mathbb{U}}(\alpha)\,\hat{\mathcal{D}}^{\mathrm{HP}}_j(m,\Omega)\bigr\}. 
\end{align}
\end{proposition}
The explicit view of the latter kernels is provided in~\ref{app_4} and~\ref{app_8}.
The integrals in the latter proposition are well defined. The Wigner function is integrable, while the
kernels are Gaussian weighted Laguerre functions with at most polynomial
growth. Hence the integrands are dominated by an integrable function, and the
same convergence arguments as above apply.
\par The connection between the symplectic $\mathcal{W}(x|\mu,\nu)$ and photon-number $\mathsf{w}_r(\alpha)$ tomograms is
\begin{eqnarray}
\mathsf{w}_{r}( \alpha) = \mathcal{P}\!\left[\mathcal{R}^{-1}\!\left[\mathcal{W}(x|\mu,\nu)\right]\right], \quad 
\mathcal{W}(x|\mu,\nu) = \mathcal{R}\!\left[\mathcal{P}^{-1}\!\left[\mathsf{w}_{r}( \alpha)\right]\right].
\end{eqnarray}
The forward transformation was previously reported in the literature~\cite{man2010photon}, while the corresponding inverse transformation had not been derived. 
{Unlike the transformations to the spin tomogram, the transformation derived here is one-to-one on the physical one-mode state space, since no sector restriction is imposed.}
\begin{proposition} The integral form of this transformation can be written as:
\begin{align}
&\mathsf{w}_r(\alpha) = \int_{\mathbb{R}^3} \mathcal{W}(x|\mu,\nu) K_{\mathcal{W}\to\mathsf{w}}(x, \mu, \nu, r, \alpha) dx d\mu d\nu, \label{1414_1}\\
&\mathcal{W}(x|\mu,\nu) = \sum_{r=0}^{\infty} \int_{\mathbb{C}} \mathsf{w}_r(\alpha) K_{\mathsf{w}\to\mathcal{W}}(x, \mu, \nu, r, \alpha) d^2\alpha,\label{1414_2} 
\end{align}
where the kernels are
\begin{align}\label{636}
K_{\mathcal{W}\to\mathsf{w}}(x, \mu, \nu, r, \alpha) &\equiv \tr{\hat{\mathcal{U}}_{ r}(\alpha) \hat{\mathcal{D}}(x, \mu, \nu)}, \\
K_{\mathsf{w}\to\mathcal{W}}(x, \mu, \nu, r, \alpha)& \equiv \tr{\hat{\mathcal{D}}_{ r}(\alpha) \hat{\mathcal{U}}(x, \mu, \nu)}.
\label{636_1}
\end{align}
\end{proposition}
The explicit view of these kernels is provided in~\ref{app_5}. The latter one-to-one transformation can be easily written for a multi-mode state.
Using the fact that the Wigner function is a Radon transform of the symplectic tomogram, the connection of the photon-number tomogram to the Wigner function is straightforward. 
\par Using \eqref{1414_1}, with the exact view of the kernel given in~\ref{app_5}, we get the transformation from the CF of the symplectic tomogram to the photon-count tomogram:
\begin{align}\label{1350_1}
\mathsf{w}_r(\alpha)
&=
\frac{1}{2\pi}
\int_{\mathbb R^2}
\phi(1;\mu,\nu)\,
\phi_r(1;\mu,\nu)\,
e^{ \frac{\nu - i\mu}{\sqrt{2}}\alpha^*
- \frac{\nu + i\mu}{\sqrt{2}}\alpha}
\,d\mu\,d\nu .
\end{align}
Due to the properties of the CF, the integral defining \eqref{1350_1} is absolutely convergent for every fixed $r$ and every finite $\alpha$.
This integral transform is the most convenient for KQSE method application, where we start from the homodyne data, estimate the CF of the symplectic tomogram in a nearly optimal way and reconstruct the photon-count tomogram by simply substituting the CFs in \eqref{1350_1}. We show how this works on the real experimental data in the section~\ref{sim}.
\section{Large spin tomographic limit}\label{sec_6}
The status of coherent states has long been discussed in connection with phase references and particle-number superselection rules. In particular, Mølmer argued that optical coherence can be regarded as a ``convenient fiction'', which led to subsequent discussions and clarifications by Gea-Banacloche and by Bartlett, Rudolph, and Spekkens~\cite{molmer1997optical, gea1998comment, molmer1998reply, bartlett2006dialogue, bartlett2007reference}. More recently, Descamps \textit{et al.} reformulated this problem within a superselection-rule-compliant (SSRC) framework, where the phase reference is treated as a physical quantum system and the CV description emerges in the large, strongly imbalanced particle number limit~\cite{DescampsUnified2025, SaharyanMetrology2025, descamps2024superselection}. In this setting, the finite dimensional description is not merely an approximation to CV quantum optics, but rather provides a physically transparent framework in which bosonic resources, mode structure, and computational or metrological advantages can be consistently characterized.
This perspective is closely related to the sector wise JS construction employed in the present work. The SSRC state
\begin{align}
   \ket{\Psi_n}=\sum_{k=0}^{n}c_k\,|k\rangle_A|n-k\rangle_R 
\end{align}
is a two-mode state with fixed total occupation number. Identifying $n=2j$, it belongs to the same spin-$j$ sector that appears in the JS map. The CV regime is recovered when one mode acts as a macroscopic reference, i.e., when $\sum_k |c_k|^2 k \ll n$, so that an effective single-mode CV description emerges only in the large-$n$ limit.
Thus, a continuous description can arise from a discrete one not only via the isomorphisms discussed above, but also through an appropriate limiting procedure. This naturally raises the question of whether our discrete tomographic description admits a well defined continuous limit as the total occupation number becomes large. As before, we do not impose a global identification between a single spin sector and the full bosonic Fock space. Instead, we consider a controlled large $j$ limit within a fixed sector $n=2j$, with low excitation relative to $j$, in which the HP realization approaches an effective bosonic mode.
\subsection{Large spin tomographic contraction with linear HP approximation}\label{sec_6_0}
Using the HP convention \eqref{eq:HP_standard}, the low-excitation regime is characterized by $\langle \hat k\rangle=o(j)$. It means that the HP bosonic mode carries only a small number of excitations compared with the large reference size. In this regime, $\sqrt{2j-\hat{k}}\approx \sqrt{2j}$ and we can write
\begin{equation}
\frac{\hat S_+}{\sqrt{2j}}\simeq \hat a,
\qquad
\frac{\hat S_-}{\sqrt{2j}}\simeq \hat a^\dagger.
\end{equation}
Moreover,
\begin{equation}
\left[
\frac{\hat S_+}{\sqrt{2j}},
\frac{\hat S_-}{\sqrt{2j}}
\right]
=
\frac{1}{2j}[\hat S_+,\hat S_-]
=
\frac{\hat S_z}{j}
=
1-\frac{\hat k}{j},
\end{equation}
which converges to the canonical commutation relation $[\hat a,\hat a^\dagger]\longrightarrow 1$ on states for which $\hat k/j$ becomes negligible.
Then we have\begin{equation}
\hat S_x \simeq \sqrt{j}\,\hat q,
\qquad
\hat S_y \simeq \sqrt{j}\,\hat p,
\qquad
\hat S_z = j-\hat k,
\end{equation}
where $\hat q=\frac{\hat a+\hat a^\dagger}{\sqrt2},$ $\hat p=\frac{\hat a-\hat a^\dagger}{i\sqrt2}.$
Then for 
\begin{equation}
\hat J_{\mathbf n}
=
\hat S_x\sin\beta\cos\gamma
+
\hat S_y\sin\beta\sin\gamma
+
\hat S_z\cos\beta
\end{equation}
that is the rotated spin projection measured in the spin tomogram \cite{man2006probabilistic}
\begin{eqnarray}\label{15444}
    \omega_j(m,\beta,\gamma)=\tr{\hat \rho^{2j}_{r(1)}\delta(m-\hat J_{\mathbf n})},
\end{eqnarray} we have 
\begin{equation}\label{1516}
\hat J_{\mathbf n}-j\cos\beta
\simeq 
\tilde{\mu}\hat q+\tilde{\nu} \hat p
-\hat k\cos\beta,
\end{equation}
where we introduce the notations  $\tilde{\mu} \equiv \sqrt{j}\sin\beta\cos\gamma,$ $\tilde{\nu} \equiv \sqrt{j}\sin\beta\sin\gamma$.
We can write
\begin{eqnarray}
&& \hat{X}^{HP}_{\tilde{\mu},\tilde{\nu},\beta}= \hat{X}_{\tilde{\mu},\tilde{\nu}}-\hat k\cos\beta,
\end{eqnarray}
where $\hat{X}_{\tilde{\mu},\tilde{\nu}}=\tilde{\mu}\hat q+\tilde{\nu} \hat p$ is the ordinary CV quadrature.
\par If $\langle \hat k\rangle=o(\sqrt{j}\sin\beta)$ then an additional number operator correction term $\hat k\cos\beta$  is negligible compared with the quadrature scale. Then 
\begin{eqnarray}
    \hat J_{\mathbf n}-j\cos\beta\rightarrow  \hat{X}_{\tilde{\mu},\tilde{\nu}}.
\end{eqnarray} In this case $\tilde{\mu}$, $\tilde{\nu}$ parameters are not finite and grow with $j$. 
Thus, in the quadrature-dominated regime, the spin tomogram approaches the symplectic tomogram of the effective CV mode,
\begin{eqnarray}
\omega_j(m,\beta,\gamma)
\;\longrightarrow\;
\mathcal{W}(x\,|\,\tilde{\mu},\tilde{\nu}),
\end{eqnarray}
where $x=m-j\cos\beta$.
\par However, if we demand the $\tilde{\mu}$, $\tilde{\nu}$ parameters to be finite ($\tilde{\mu},\tilde{\nu}=O(1)$) then $\sqrt{j}\sin\beta=O(1)$. Then $\sin{\beta}=O(1/\sqrt{j})$ and $\cos{\beta}\rightarrow 1$. But then $\hat k\cos\beta\rightarrow \hat k$, and this term cannot be neglected. Then
\begin{eqnarray}
    \hat J_{\mathbf n}-j\cos\beta\rightarrow  \hat{X}_{\tilde{\mu},\tilde{\nu}}-\hat{k}.
\end{eqnarray}
{Thus, in the scaling with finite $(\tilde{\mu},\tilde{\nu})$, the rescaled spin-tomographic measures converge, in the weak sense, to the distribution of the number-corrected observable,}
\begin{eqnarray}
\omega_j(m,\beta,\gamma)
\;\longrightarrow\;
\mathcal{W}_{\mathrm{HP}}(x\,|\,\tilde{\mu},\tilde{\nu}),
\end{eqnarray}
with
\begin{eqnarray}
\mathcal{W}_{\mathrm{HP}}(x\,|\,\tilde{\mu},\tilde{\nu})
=
\mathrm{Tr}\!\left[
\hat\rho\,
\delta\!\left(x-\hat X_{\tilde{\mu},\tilde{\nu}}+\hat k\right)
\right].
\end{eqnarray}
This conclusion is consistent with the earlier tomographic analysis~\cite{castanos2003kernel}. Unlike derivations based on asymptotics of Wigner $d$-functions, the large-$j$ limit is obtained here directly from the sector-wise HP construction. This approach reveals that the centered spin projection does not, in general, reduce to the symplectic quadrature, but rather to a number-corrected quadrature observable, with the standard CV limit recovered only under an additional quadrature-dominated scaling.
\subsection{Large spin tomographic contraction beyond the linear HP approximation}
We now analyze the large-$j$ tomographic limit without invoking the linear approximation of the HP map. Instead, we retain the full square root structure, which leads to a finite-$j$ deformation of the quadrature observable. In this formulation, the spin projection does not directly reduce to the standard symplectic quadrature, but rather to a number corrected quadrature operator. The usual CV quadrature is recovered only after an additional contraction limit, in which the excitation number becomes negligible relative to the quadrature scale.
On the HP subspace $S_{2j}=\operatorname{span}\{|k\rangle\}_{k=0}^{2j}$, the
spin generators are given by \eqref{eq:HP_standard}.
We introduce the finite-$j$ HP deformation factor
\begin{equation}
\hat F_j
=
\left(1-\frac{\hat k}{2j}\right)^{1/2}
\end{equation}
on $S_{2j}$. Then \eqref{eq:HP_standard} can be rewritten as
\begin{equation}
\hat S_+
=
\sqrt{2j}\,\hat F_j\hat a,
\qquad
\hat S_-
=
\sqrt{2j}\,\hat a^\dagger \hat F_j .
\end{equation}
For a phase $\gamma$, define the exact HP-deformed quadrature
\begin{equation}
\hat Q_{j,\gamma}
=
\frac{1}{\sqrt2}
\left(
e^{-i\gamma}\hat F_j\hat a
+
e^{i\gamma}\hat a^\dagger \hat F_j
\right).
\label{eq:finite_j_deformed_quadrature}
\end{equation}
This operator is self-adjoint on the finite-dimensional HP subspace. On states satisfying $\hat k/j\to 0$, one has $\hat F_j\to \hat 1$ in the strong sense, and the operator $\hat Q_{j,\gamma}$ reduces to the standard quadrature:
\begin{equation}
\hat Q_\gamma
=
\hat q\cos\gamma+\hat p\sin\gamma
=
\frac{1}{\sqrt2}
\left(
e^{-i\gamma}\hat a
+
e^{i\gamma}\hat a^\dagger
\right).
\end{equation}
 Using
\[
\hat S_x\cos\gamma+\hat S_y\sin\gamma
=
\frac12
\left(
e^{-i\gamma}\hat S_+
+
e^{i\gamma}\hat S_-
\right),
\]
we obtain the identity \eqref{1516} that is now exact
\begin{equation}
\hat J_{\mathbf n}
-
j\cos\beta
=
\sqrt j\,\sin\beta\,\hat Q_{j,\gamma}
-
\hat k\cos\beta,
\label{eq:exact_HP_spin_projection}
\end{equation}
where $\hat{X}_{\tilde{\mu},\tilde{\nu}}=\sqrt j\,\sin\beta\,\hat Q_{j,\gamma}$.
\par For any fixed $\lambda>0$ and $\sin\beta\neq0$, we can write
\begin{equation}
\frac{\lambda}{\sqrt j\,\sin\beta}
\left(
\hat J_{\mathbf n}-j\cos\beta
\right)
=
\lambda\hat Q_{j,\gamma}
-
\frac{\lambda\cot\beta}{\sqrt j}\hat k .
\label{eq:scaled_centered_spin_projection}
\end{equation}
This shows that, at finite $j$, the spin projection does not correspond to a standard quadrature, but rather to a deformed quadrature operator $\hat Q_{j,\gamma}$ together with a longitudinal correction. The usual CV quadrature emerges only after an additional contraction in which both $\hat F_j\to \hat 1$ and the number correction becomes negligible.
This gives an exact finite-$j$ tomographic object. For a spin tomogram
$\omega_j(m,\beta,\gamma)$, define the rescaled tomographic measure
\begin{equation}
\mathcal T_{j,\beta,\gamma}^{(\lambda)}(x)
=
\sum_{m=-j}^{j}
\omega_j(m,\beta,\gamma)\,
\delta\!\left(
x-
\frac{\lambda}{\sqrt j\,\sin\beta}
(m-j\cos\beta)
\right),
\label{eq:finite_j_rescaled_tomogram}
\end{equation}
that is the same probability distribution, but rewritten as a measure on the real line. Each discrete value of $m$ becomes a delta peak at the corresponding rescaled value of $x=\tfrac{\lambda}{\sqrt j\,\sin\beta}
(m-j\cos\beta)$.
The characteristic function is the  Fourier transform of this measure, that is equal to
\begin{align}
\widehat{\mathcal T}_{j,\beta,\gamma}^{(\lambda)}(t)
&=
\sum_{m=-j}^{j}
\omega_j(m,\beta,\gamma)
\exp\!\left[
it\frac{\lambda}{\sqrt j\,\sin\beta}
(m-j\cos\beta)
\right]
\end{align}
Using \eqref{15444}, we get
\begin{align}
\widehat{\mathcal T}_{j,\beta,\gamma}^{(\lambda)}(t)
&=\tr\!\left[
\hat\rho_{r(1)}^{(2j)}
\exp\!\left\{
it\left(
\lambda\hat Q_{j,\gamma}
-
\frac{\lambda\cot\beta}{\sqrt j}\hat k
\right)
\right\}
\right].
\label{eq:finite_j_CF_exact}
\end{align}
The function $\mathcal T_{j,\beta,\gamma}^{(\lambda)}(x)$ is the probability measure of the centered and rescaled spin projection. Its CF \eqref{eq:finite_j_CF_exact} is exactly the expectation value of the finite $j$ HP-deformed quadrature observable, including the longitudinal number correction. Thus, Eq.~\eqref{eq:finite_j_CF_exact} is exact on the HP sector and does not rely on the linearized HP approximation.
The most transparent contraction is obtained for equatorial spin
directions, $\beta=\pi/2$. Then the longitudinal correction vanishes
identically and
\begin{equation}
\frac{\lambda}{\sqrt j}\hat J_{\mathbf n}
=
\lambda \hat Q_{j,\gamma}.
\label{eq:equatorial_exact_contraction}
\end{equation}
Consequently,
\begin{equation}
\widehat{\mathcal T}_{j,\pi/2,\gamma}^{(\lambda)}(t)
=
\tr\!\left[
\hat\rho_{r(1)}^{(2j)}
e^{it\lambda\hat Q_{j,\gamma}}
\right].
\label{eq:equatorial_CF_exact}
\end{equation}
If the sequence of HP states is localized in the low excitation region, in the sense that the probability weight of $k$ comparable to $j$ vanishes as $j\to\infty$, then $\hat F_j\to \hat 1$ on the relevant part of the state and
$\hat Q_{j,\gamma}
\longrightarrow
\hat Q_\gamma$.
In this case,
one obtains the CF of the symplectic tomogram:
\begin{equation}
\lim_{j\to\infty}
\widehat{\mathcal T}_{j,\pi/2,\gamma}^{(\lambda)}(t)
=
\phi(t;\lambda\cos\gamma,\lambda\sin\gamma).
\label{eq:spin_CF_to_symplectic_CF}
\end{equation}
Equivalently, at the level of probability measures, the rescaled spin tomogram converges weakly to the symplectic tomogram. More precisely, defining the rescaled variable $x=\frac{\lambda}{\sqrt j}m$,
the corresponding probability measures converge weakly to
$\mathcal W(x|\lambda\cos\gamma,\lambda\sin\gamma)$. Formally, this can be expressed at the level of densities as
\begin{equation}
\frac{\sqrt j}{\lambda}\,
\omega_j\!\left(
m_j(x),\frac{\pi}{2},\gamma
\right)
\;\approx\;
\mathcal W(x|\lambda\cos\gamma,\lambda\sin\gamma),
\qquad
m_j(x)\simeq \frac{\sqrt j}{\lambda}x ,
\label{eq:spin_tomogram_to_symplectic_tomogram_equator}
\end{equation}
where the approximation reflects the discrete-to-continuous transition.
\par More generally, for fixed $\beta\in(0,\pi)$, one obtains the same limiting behavior under the additional assumption that the longitudinal correction term
$\lambda\cot\beta\,\hat k/\sqrt j$ in \eqref{eq:scaled_centered_spin_projection} vanishes on the considered sequence of states. In this case,
\begin{equation}
\frac{\sqrt j\,\sin\beta}{\lambda}\,
\omega_j\!\left(
m_j(x),\beta,\gamma
\right)
\;\approx\;
\mathcal W(x|\lambda\cos\gamma,\lambda\sin\gamma),
\label{eq:spin_tomogram_to_symplectic_tomogram_general_beta}
\end{equation}
with
\begin{equation}
m_j(x)
\simeq
j\cos\beta
+
\frac{\sqrt j\,\sin\beta}{\lambda}x.
\end{equation}
Thus the large-$j$ CV limit should be understood as a contraction of the exact finite-$j$ HP-deformed quadrature tomography. The symplectic tomogram emerges only after two ingredients: the deformation factor $\hat F_j$ approaches the identity and the longitudinal number correction becomes negligible.
Without a localization condition ensuring that the occupation number remains small compared to $j$, the limiting object is not the standard symplectic tomogram of a harmonic oscillator, but rather a nonlinear finite-$j$ tomogram associated with the deformed quadrature $\hat Q_{j,\gamma}$.
\subsection{Large spin limit at the level of density operators}
The large-$j$ interpretation of our construction can be formulated directly at the level of density operators and their tomographic representations. In the present framework, the fundamental objects are sector restricted density operators and their associated tomograms. Accordingly, the large-$j$ limit is understood as a limit of such sector-restricted states, together with the corresponding tomographic distributions.
For each fixed $j$, the HP correspondence identifies the spin-$j$ Hilbert
space with the truncated oscillator subspace
$S_{2j}$.
Thus, if $\hat\rho^{(j)}$ is a spin-$j$ density operator, its HP inverse is
the finite-rank oscillator state
\begin{equation}
\hat\sigma_j
=
\mathsf E_j^\dagger \hat\rho^{(j)}\mathsf E_j,
\qquad
\operatorname{supp}\hat\sigma_j\subset S_{2j}.
\label{eq:large_j_sigma_j}
\end{equation}
The corresponding restricted symplectic tomogram is
\begin{equation}
\mathcal W^{(j)}(x|\mu,\nu)
=
\tr{
\hat\sigma_j\,\delta(x\hat 1-\mu\hat q-\nu\hat p)}.
\label{eq:large_j_restricted_sympl_tomogram}
\end{equation}
Equivalently, in terms of the CF 
\begin{equation}
\phi_{j}(t;\mu,\nu)
=
\tr{
\hat\sigma_j
e^{it(\mu\hat q+\nu\hat p)}}.
\label{eq:large_j_restricted_CF}
\end{equation}
We say that a sequence of spin tomograms $\omega_j(m,\Omega)$ has a CV limit
in the HP sense if the corresponding HP-lifted density operators
$\hat\sigma_j=\mathsf E_j^\dagger\mathcal S^{-1}[\omega_j]\mathsf E_j$
converge in trace norm to a trace-class density operator $\hat\rho_{(1)}$:
\begin{equation}
\lim_{j\to\infty}
\left\|
\hat\sigma_j-\hat\rho_{(1)}
\right\|_1
=0.
\label{eq:trace_norm_HP_limit}
\end{equation}
In this case the characteristic functions converge pointwise for all $t\in\mathbb R$:
\begin{equation}
\lim_{j\to\infty}
\phi_j(t;\mu,\nu)
=
\phi(t;\mu,\nu),
\qquad
\phi(t;\mu,\nu)
=
\operatorname{Tr}\!\left[
\hat\rho_{(1)}
e^{it(\mu\hat q+\nu\hat p)}
\right].
\label{eq:CF_convergence_HP}
\end{equation}
Indeed, since \(e^{it(\mu\hat q+\nu\hat p)}\) is unitary, one has
\begin{equation}
\left|
\phi_j(t;\mu,\nu)-\phi(t;\mu,\nu)
\right|
\le
\left\|
\hat\sigma_j-\hat\rho_{(1)}
\right\|_1.
\label{eq:CF_trace_norm_bound}
\end{equation}
Therefore the convergence of density operators implies the convergence of all
CFs. In the weak sense of
probability  measures,
\begin{equation}
\mathcal W^{(j)}(x|\mu,\nu)
\longrightarrow
\mathcal W(x|\mu,\nu),
\qquad
j\to\infty,
\label{eq:sympl_tomogram_weak_limit}
\end{equation}
where $\mathcal W(x|\mu,\nu)$ is defined in \eqref{eq:deq-simpl}.
{Conversely, starting from a fixed single-mode CV state $\hat\rho_{(1)}$, one
may define its finite-$j$ HP restrictions by \eqref{1352} whenever
$\operatorname{Tr}(\hat P_{2j}\hat\rho_{(1)})>0$.}
Since $\hat P_{2j}\to \hat 1$ strongly and $\hat\rho_{(1)}$ is trace class,
one has $\hat\rho_{r(1)}^{(2j)}\longrightarrow\hat\rho_{(1)}$
in trace norm whenever
$ \tr{\hat P_{2j}\hat\rho_{(1)}}\to1$. This condition is satisfied, for example, for states with finite mean photon number or, more generally, whenever the occupation of high Fock levels is sufficiently suppressed.
Thus the associated spin states
$\hat\rho^{(j)}=\mathsf E_j\hat\rho_{r(1)}^{(2j)}\mathsf E_j^\dagger$
provide finite-dimensional spin representatives of the same CV state, and the
corresponding restricted symplectic tomograms converge to the full CV
symplectic tomogram.
\par For the JS correspondence the situation is fundamentally different. The fixed total occupation sectors $S_{2j}$ are mutually orthogonal and do not form an increasing sequence approximating the full two-mode Fock space. Consequently, the limit $2j\to\infty$ of normalized states supported on $S_{2j}$ cannot, in general, be interpreted as a trace norm approximation to a generic two-mode CV state.
Rather, it is a large dimensional fixed number, or relational, limit. A genuine
single-mode CV limit is obtained only after choosing an identification of each
sector $S_{2j}$ with a common oscillator space, for example through the HP
relabeling
$|2j-k,k\rangle_2\longleftrightarrow |k\rangle$, $k=0,\ldots,2j$.
\subsubsection{Jordan--Schwinger origin of the large spin limit.}
The large-$j$ CV interpretation is most naturally formulated at the level of fixed JS sectors. For each $j$, the subspace \eqref{s2j}
is unitarily equivalent to the spin-$j$ Hilbert space. Consequently, the spin tomogram and the restricted two-mode symplectic tomogram are exactly related at finite $j$ through the tomographic JS kernels derived above.
The HP description arises only after selecting one of the JS modes as a reference and introducing the relabeling \eqref{1146}.
In this representation, the large-$j$ limit corresponds to a strongly imbalanced regime characterized by $\langle \hat k\rangle=o(j)$
in which the first mode carries $2j-\hat k$ quanta and acts as a macroscopic reference, while the second mode supports the effective bosonic excitation.
Thus, the CV tomographic limit does not rely on a global identification of the spin system with the full Fock space. Rather, it emerges from a sequence of exact finite-dimensional JS sectors $S_{2j}$, together with an appropriate asymptotic regime selecting low excitations within each sector.
\section{Experimental Data Study}\label{sim}
\par To illustrate the analytically derived tomographic transformations presented in Sec.~\ref{sec_4}, we use experimental homodyne-tomography data reported in Refs.~\cite{lvovsky2002quantum,Lvovsky_2004}. In these experiments, nonclassical single-mode optical states were generated by conditional measurements at a highly unbalanced beam splitter. A weak coherent state of amplitude $\alpha$ and a conditionally prepared single-photon Fock state were injected into the two input ports, with reflectivity $r^2=0.92$ ($t=\sqrt{1-r^2}$). A single-photon detector in one output arm provided the conditioning: homodyne data in the other arm were acquired only upon a detector click. The resulting quantum interference removes which-path information and prepares a nonclassical state in the signal mode. In the ideal limit, the conditional state takes the form
\begin{eqnarray}
\ket{\psi_s}=c_0\ket{0}+c_1\ket{1},\qquad
c_0=\frac{t}{\sqrt{t^2+\lambda^2}},\quad
c_1=\frac{\lambda}{\sqrt{t^2+\lambda^2}},
\end{eqnarray}
which is commonly referred to as a Schrödinger kitten state.
The generated states were analyzed by balanced homodyne detection with a phase-scanned local oscillator, yielding quadrature probability distributions at multiple phases. The experimental dataset comprises
$T=14153$ homodyne samples $\{x_{\theta_j},\theta_j\}_{j=1}^{T}$,
where $x_{\theta_j}$ is the measured quadrature corresponding to the local-oscillator phase $\theta_j$.
\par Owing to experimental imperfections, such as finite single-photon preparation efficiency, optical losses, and detector dark counts, the reconstructed state is not pure. It can be described by the mixed state
\begin{eqnarray}\label{eq:rho_mixed}
\rho=(1-p)\ket{0}\bra{0}+p\ket{\psi_s}\bra{\psi_s},
\end{eqnarray}
where for $\lambda=0.3$ one can estimate $p$ from the reported purity to be equal to $p=0.6810463$. 
In \cite{markovich2025nonparametric} we show that the symplectic tomogram of the latter state is
\begin{eqnarray}\label{1449}
    \mathcal{W}(x|\chi)
    =
    \mathcal{W}_0(x|\chi)
    \left(
    1-pc_1^2
    +2pc_1^2\frac{x^2}{\chi^2}
    +2\sqrt{2}pc_0c_1\frac{\mu x}{\chi^2}
    \right),\!\!\!\quad \chi=\sqrt{\mu^2+\nu^2}, 
\end{eqnarray}
where the tomogram  of the ground state of the harmonic oscillator is 
\begin{eqnarray}\label{1530}
\mathcal{W}_0(x|\chi) = \frac{1}{\sqrt{\pi}\chi} \exp\!\Bigl[-\tfrac{x^2}{\chi^2}\Bigr].
\end{eqnarray}
Using \eqref{1350_1}, the corresponding photon-count tomogram is  
\begin{eqnarray}\label{1417}
w_r(\alpha)
\!=\!
\mathsf{w}_{0r}(\alpha)\!\left[\!
\left(1-p|c_1|^2\right)
+p|c_1|^2\,\frac{(r-|\alpha|^2)^2}{|\alpha|^2}
+2p\,\frac{(r-|\alpha|^2)}{|\alpha|^2}\,
\mathrm{Re}\!\left(c_0 c_1^* \alpha\right)
\right]\!,
\!\end{eqnarray}
where the  photon-count tomogram corresponding to \eqref{1530} is
\begin{align}
    \mathsf{w}_{0r}(\alpha)\equiv 
e^{- |\alpha|^2}\frac{ |\alpha|^{2r}}{r!},
\qquad r=0,1,2,\dots
\end{align}
Similar result can be achieved using the definition of the photon-count tomogram \eqref{dequantizer-ph-number}, that verifies our calculations. 
\par The next step is to obtain the corresponding spin tomograms associated with the selected HP sectors. To this end we must substitute the symplectic tomogram estimated from the homodyne data into the transformation \eqref{eq:omega_to_W_int}. For our example we use the transformation \eqref{eq:W_to_omega_int_12} to achieve the spin-$1/2$ tomogram corresponding to the  \eqref{eq:rho_mixed} state:
\begin{align}\label{1222}
   \omega_{1/2}(m,\beta,\gamma)=  \frac{1}{2}+2m[(\frac{1}{2}-pc_1^2)\cos{\beta}-pc_0c_1\sin{\beta}\cos{\gamma}], \quad m=\pm \frac{1}{2}.
\end{align}
Similarly, we get the spin tomogram for the spin-$1$ as
\begin{align}\label{1223}
\omega_{1}(m,\beta,\gamma)&=
\frac{1}{3}
+\frac{1-pc_{1}^{2}}{2}\,m\cos\beta
+\frac{1-3pc_{1}^{2}}{12}\,(3m^{2}-2)\,\bigl(3\cos^{2}\beta-1\bigr)\\\nonumber
&-\frac{pc_{0}c_{1}}{\sqrt2}\,\sin\beta\;\Bigl[m+(3m^{2}-2)\cos\beta\Bigr]\cos\gamma,
\qquad m\in\{-1,0,1\}.
\end{align}
\begin{figure}[t]
\centering
\begin{subfigure}{0.32\textwidth}
    \centering
    \includegraphics[width=\linewidth]{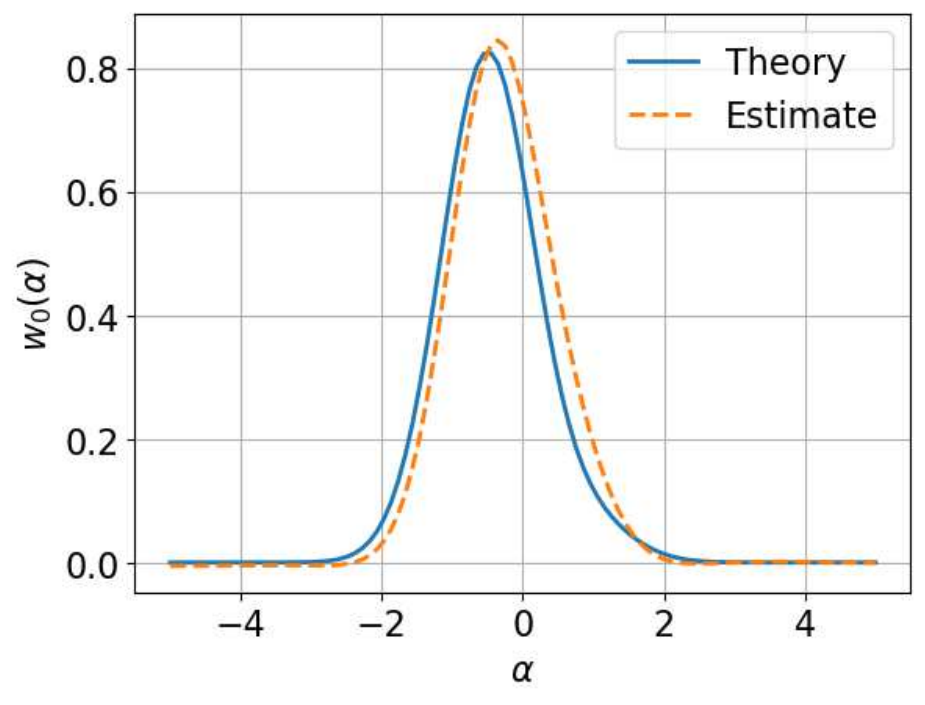}
\end{subfigure}
\hfill
\begin{subfigure}{0.32\textwidth}
    \centering
    \includegraphics[width=\linewidth]{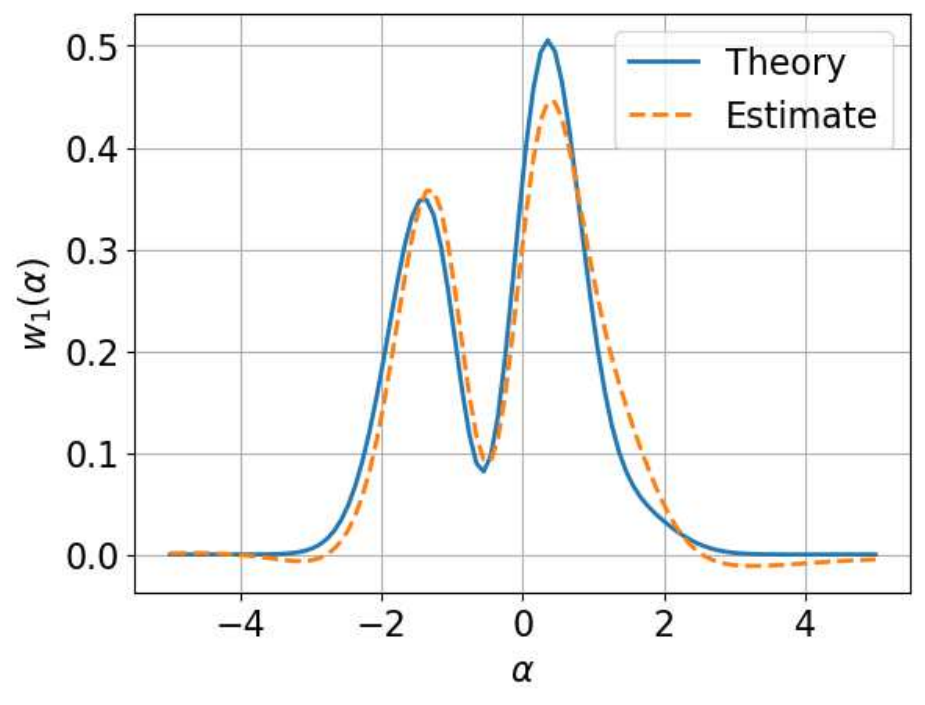}
\end{subfigure}
\hfill
\begin{subfigure}{0.32\textwidth}
    \centering
    \includegraphics[width=\linewidth]{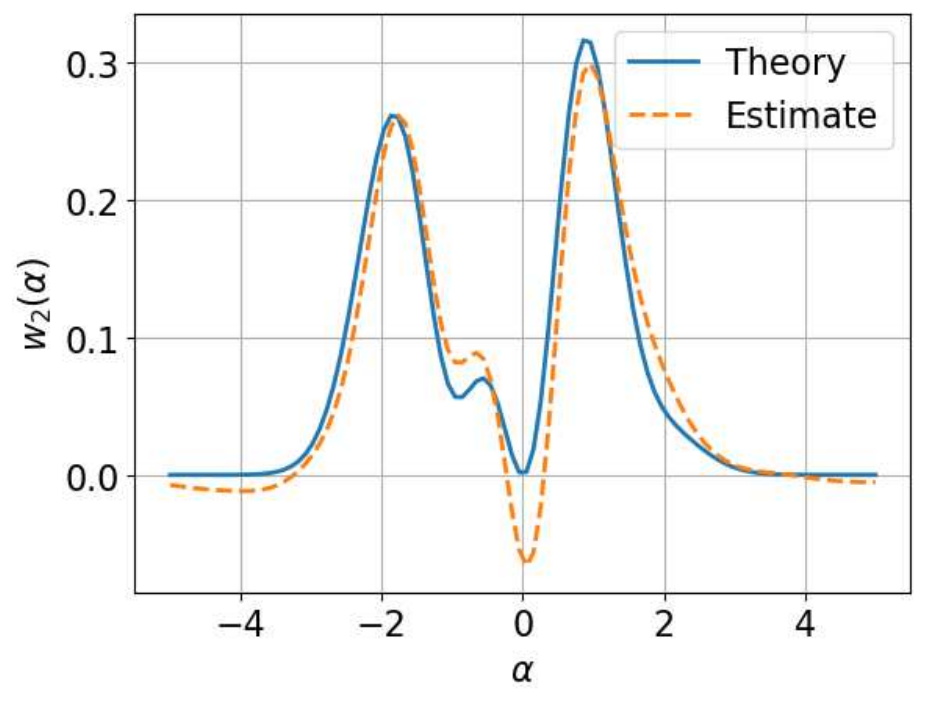}
\end{subfigure}
\caption{Photon-number tomograms $w_r(\alpha)$ for fixed $r=0,1,2$ as functions of $\alpha\in\mathbb{R}$. The solid blue curves represent the theoretical photon-number tomograms \eqref{1417}, while the orange dashed curves show the corresponding tomograms reconstructed from the homodyne data using the KQSE method and the HP tomographic mapping.}
\label{fig:w1r}
\end{figure}
\begin{figure}[t]
\centering
    \begin{subfigure}{0.49\linewidth}
        \centering
        \includegraphics[width=\linewidth]{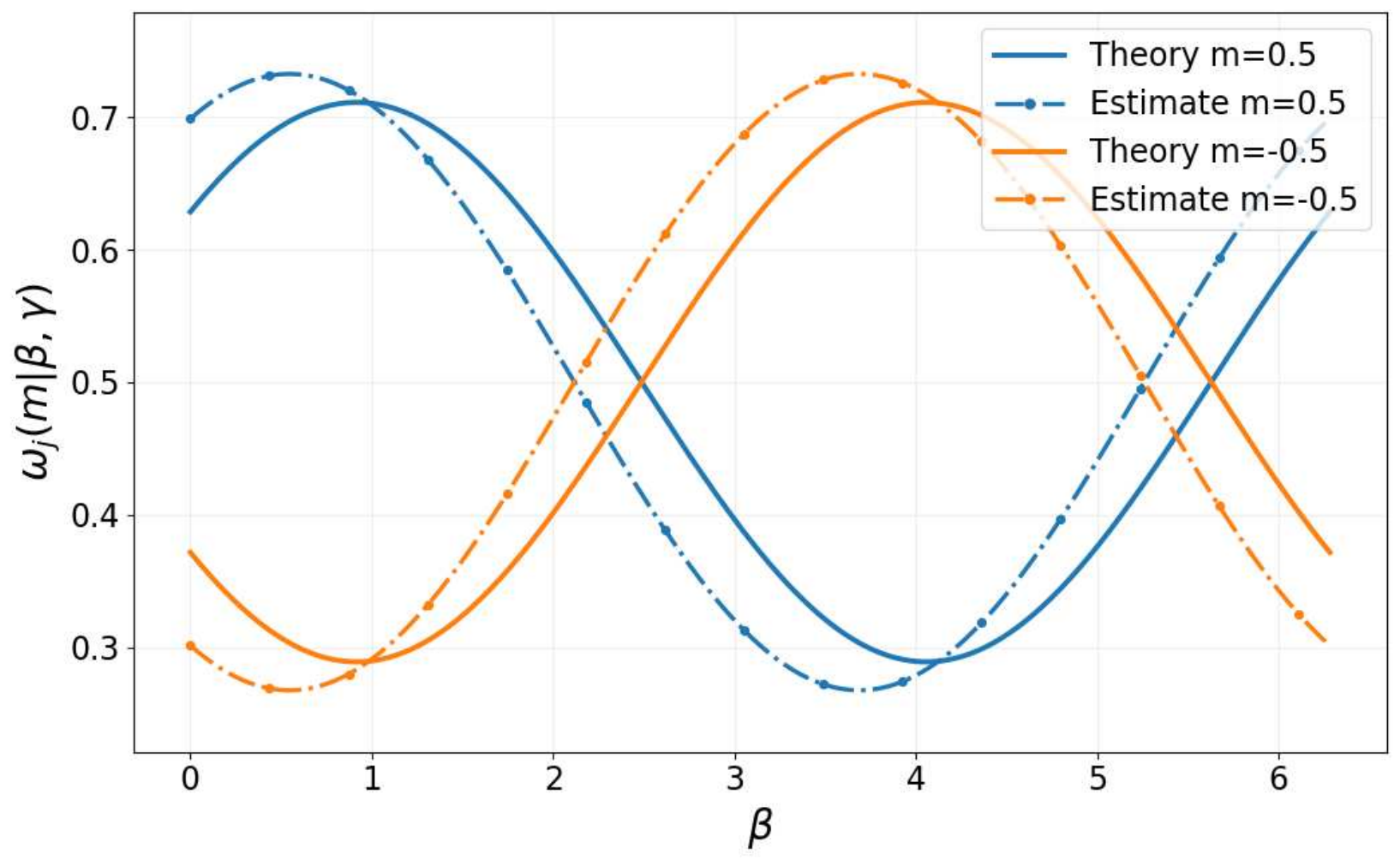}
        \caption{Fixed $\gamma$}
        \label{fig:spin_jhalf_fixed_gamma}
    \end{subfigure}\hfill
    \begin{subfigure}{0.49\linewidth}
        \centering
        \includegraphics[width=\linewidth]{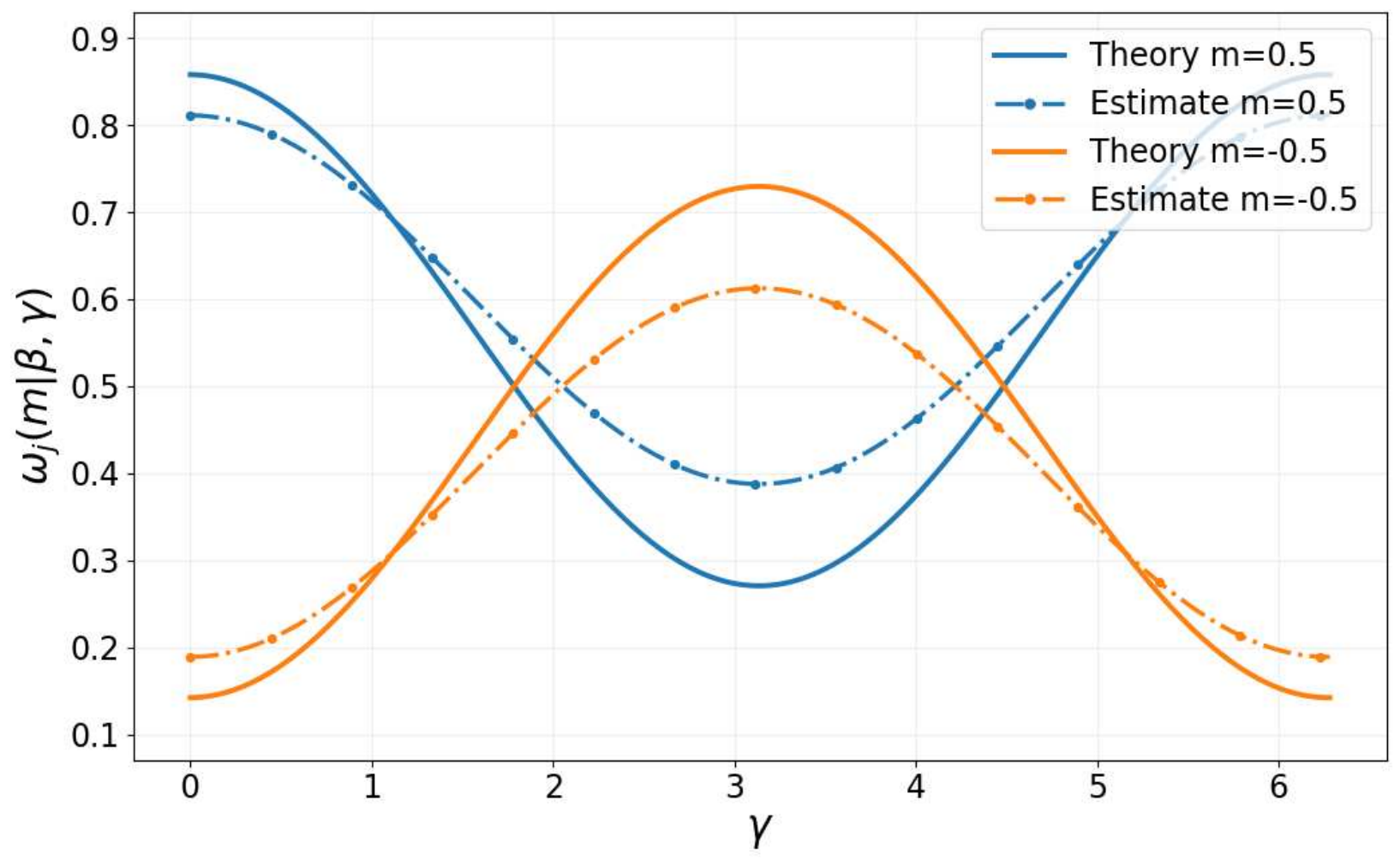}
        \caption{Fixed $\beta$}
        \label{fig:spin_jhalf_fixed_beta}
    \end{subfigure}
    \caption{Spin tomograms $\omega_{1/2}(m,\beta,\gamma)$ reconstructed using the KCFE method. a) fixed $\gamma=\pi/3$, $\beta\in[0,2\pi)$; b) fixed $\beta=\pi/3$,  $\gamma\in[0,2\pi)$. Solid lines show the theoretical predictions \eqref{1222}, while dash-dotted lines correspond to the KCFE based estimates obtained from experimental data.}
    \label{fig:4}
\end{figure}
\begin{figure}[t]
    \begin{subfigure}{0.49\linewidth}
        \centering
    \includegraphics[width=\linewidth]{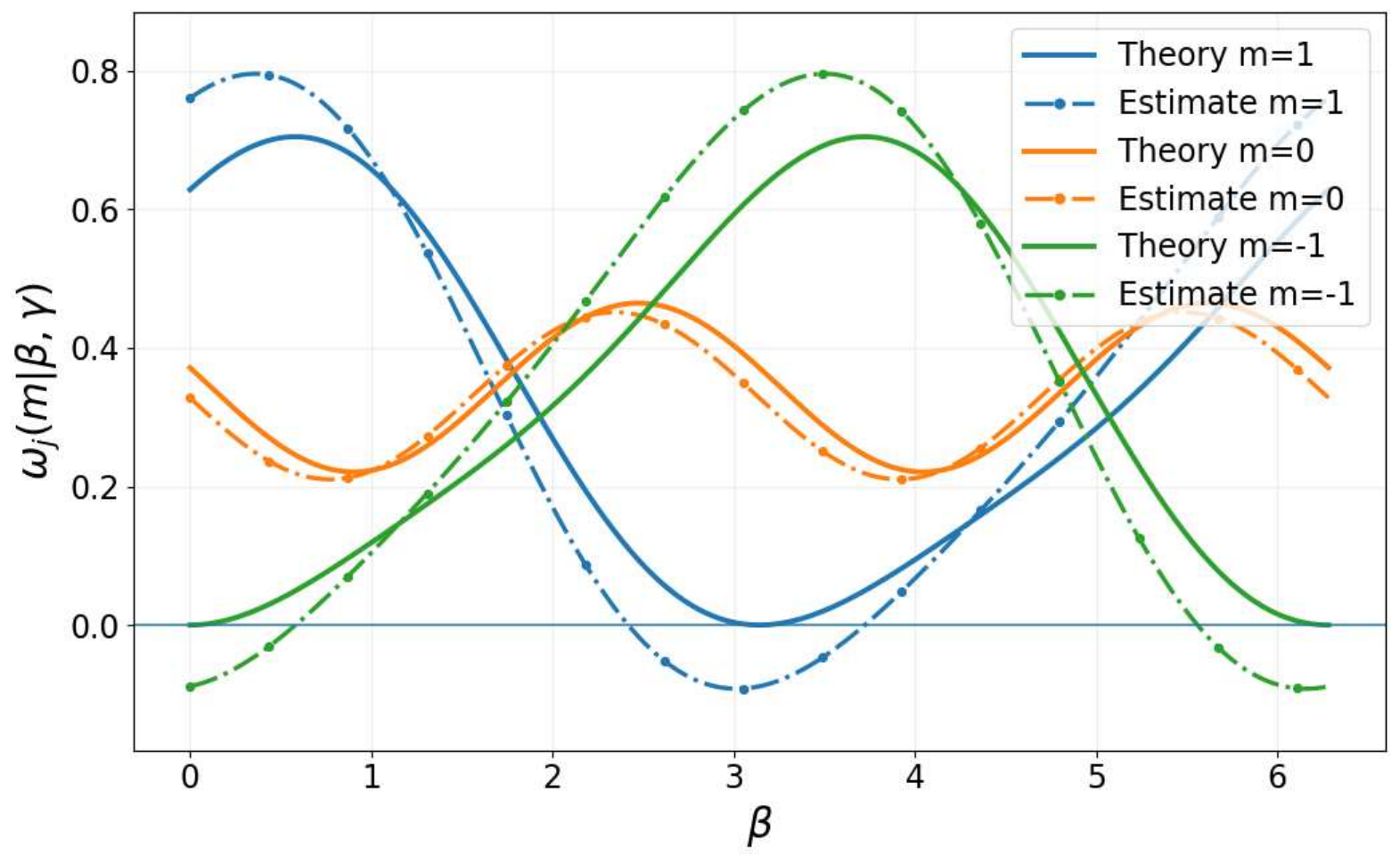}
    \caption{Fixed $\gamma$}
     \end{subfigure}\hfill
    \begin{subfigure}{0.49\linewidth}
        \centering
    \includegraphics[width=\linewidth]{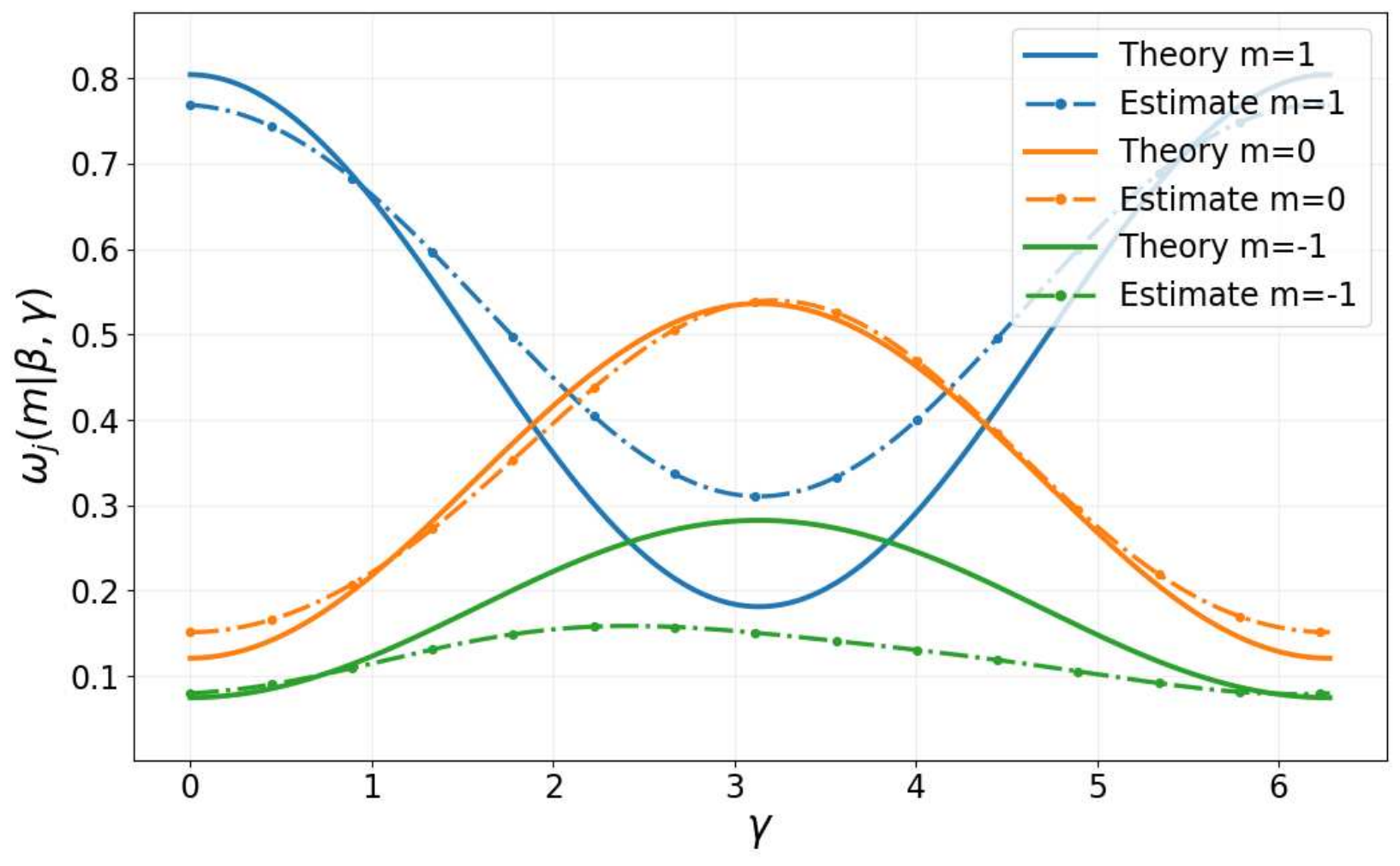}
\caption{Fixed $\beta$}
 \end{subfigure}
  \caption{Spin tomograms $\omega_{1}(m,\beta,\gamma)$ reconstructed using the KCFE method. a) fixed $\gamma=\pi/3$, $\beta\in[0,2\pi)$; b) fixed $\beta=\pi/3$, $\gamma\in[0,2\pi)$. Solid lines show the theoretical predictions \eqref{1223}, while dash-dotted lines correspond to the KCFE based estimates obtained from experimental data.}
    \label{fig:5}
\end{figure}
\par The experimentally measured homodyne data (quadrature samples) are distributed according to the symplectic tomogram~\eqref{1449}. We use the KQSE procedure introduced for estimating continuous-variable quantum states from raw homodyne measurements, described in detail in Ref.~\cite{markovich2025nonparametric}. The statistical properties of this estimator, including its behavior under experimental noise, comparisons with maximum-likelihood, Bayesian, convex optimization, and neural network approaches, and its mean squared error performance, have been analyzed in Ref.~\cite{markovich2025nonparametric}. In the present work, we therefore focus on the tomographic transformations themselves, while relying on these established results for the data-processing stage. In this paper we use the KQSEs intermediate step \textit{kernel CF estimation} (KCFE).
Without going into the details, we mention that KQSE uses the KCFE  to reconstruct the state, its trace characteristics and even the Wigner function achieving nearly optimal $\tilde{O}(1/T)$ rate of convergence in $L_2$ norm.  Here $T$ is the amount of homodyne measurements. In the present work, the estimated CFs are used as an input for the kernel transformations \eqref{1350_1} and \eqref{eq:W_to_omega_int_12}, which allow to reconstruct the corresponding photon-number tomograms 
$\mathsf{w}_r(\alpha)$ and spin tomograms $\omega_{j}(m,\beta,\gamma)$ directly, without performing an explicit quantum state reconstruction. Figure~\ref{fig:w1r} shows the comparison of the theoretical photon-number tomograms \eqref{1417} with those reconstructed from the experimental homodyne data, using the KCFE. 
{Although the KQSE estimates are quantitatively close to those obtained by maximum likelihood estimation in the comparison of Ref.~\cite{markovich2025nonparametric}, small regions of negativity may appear in the estimated photon-number tomogram (not in symplectic tomogram).} Unlike in the case of the Wigner function, such negativities do not indicate non-classicality. They arise because the KQSE procedure guarantees that the estimated symplectic tomogram is a valid probability density (i.e., non-negative), but it does not enforce the full set of necessary and sufficient conditions ensuring that the symplectic tomogram (and its characteristic function) corresponds to a physical quantum state (see Theorem~\ref{thm_3_0}). These conditions are highly sensitive, so even small deviations of the estimated photon-number tomogram from the exact one may lead to violations of Theorem~\ref{thm_2}, resulting in the appearance of negative regions. {This issue is not specific to the KQSE procedure: any estimation method that does not explicitly enforce the admissibility conditions of Theorems~\ref{thm_2} and \ref{thm_3_0} may exhibit the same effect within the present kernel transformation framework.} This issue has been analyzed in detail in Refs.~\cite{markovich2024not} and indicates a direction for further development of the KQSE method. However, the KQSE method does not require prior information about the underlying state and is fully data-driven, which makes it applicable to non-Gaussian states.  A detailed comparison with the MLE approach, whose performance strongly depends on the assumed prior model, is presented in Ref.~\cite{markovich2025nonparametric}. Figures~\ref{fig:4} and \ref{fig:5} show the estimated spin tomograms for spin-$1/2$ and $1$ systems.
\par Equations \eqref{1222}--\eqref{1223} are written in a fixed phase convention: the interference term is taken real and the angle $\gamma$ is measured from a chosen phase origin, i.e. $\Phi=0$. In the experiment, however, the absolute phase origin is not fixed a priori (the local--oscillator phase defines $\gamma$ only up to an additive constant), so the reconstructed off--diagonal element generally carries a phase $\rho_{m_1m_2}=|\rho_{m_1m_2}|e^{-i\Phi}$. Since
$A^{(j,\mathrm{HP})}_{\phi\to\omega}(m,\beta,\gamma,\mu,\nu)$ in the kernel transformation \eqref{eq:W_to_omega_int_12} depends on the Wigner $D$-function
$D^{(j)}(\beta,\gamma)$ (see~\ref{app:HP_symplectic_spin}) that contains exponential factors $e^{-im\gamma}$, this phase appears as a constant shift of the interference contribution, $\cos\gamma\to\cos(\gamma-\Phi)$. We determine $\Phi=3.134740$ from the most phase--sensitive configuration $j=\tfrac12$ at $\beta=\pi/2$.
\par
\begin{figure}[t]
\centering
    \begin{subfigure}{0.49\linewidth}
        \centering
        \includegraphics[width=\linewidth]{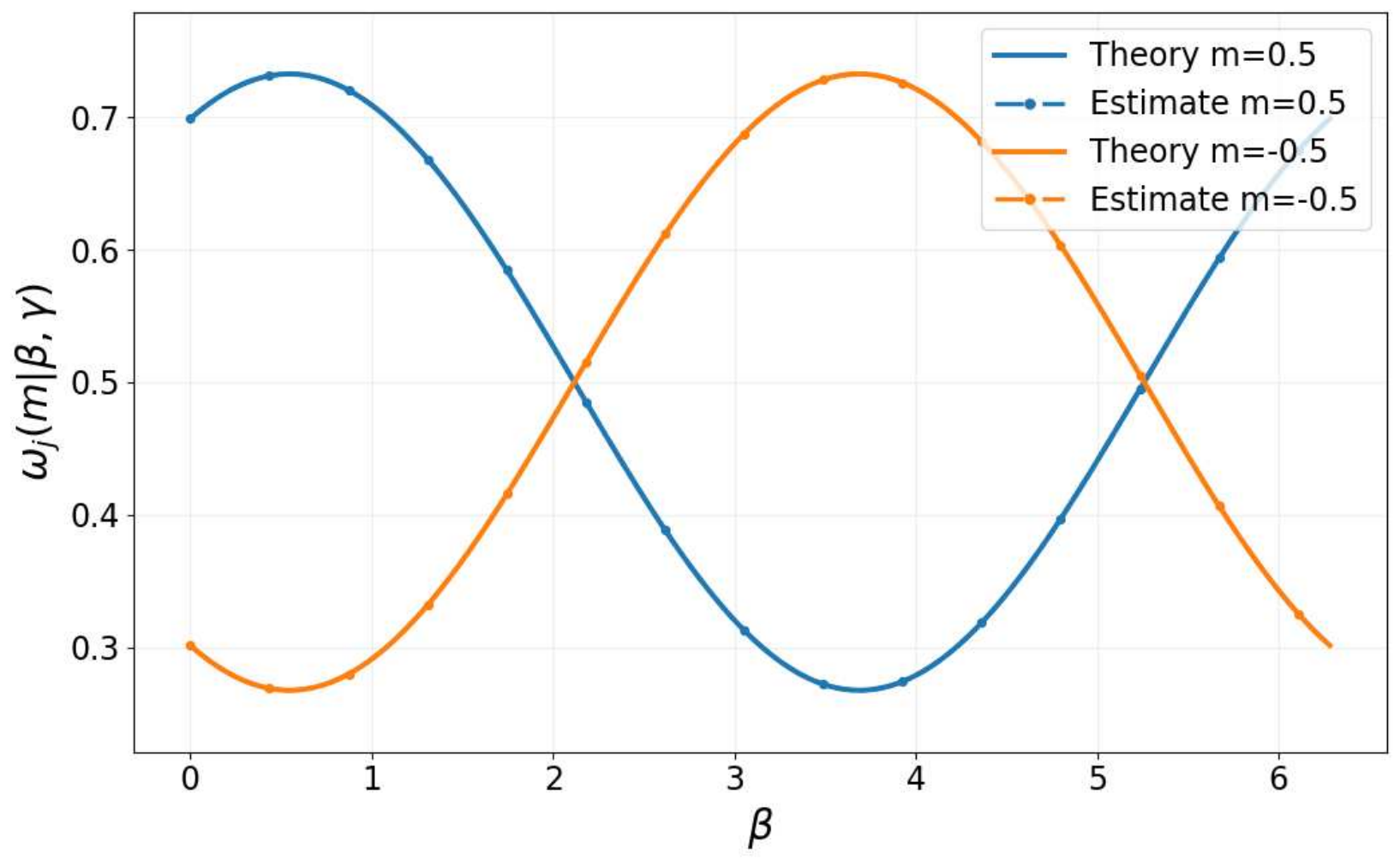}
        \caption{Fixed $\gamma$}
        \label{fig:spin_jhalf_fixed_gamma_1}
    \end{subfigure}\hfill
    \begin{subfigure}{0.49\linewidth}
        \centering
        \includegraphics[width=\linewidth]{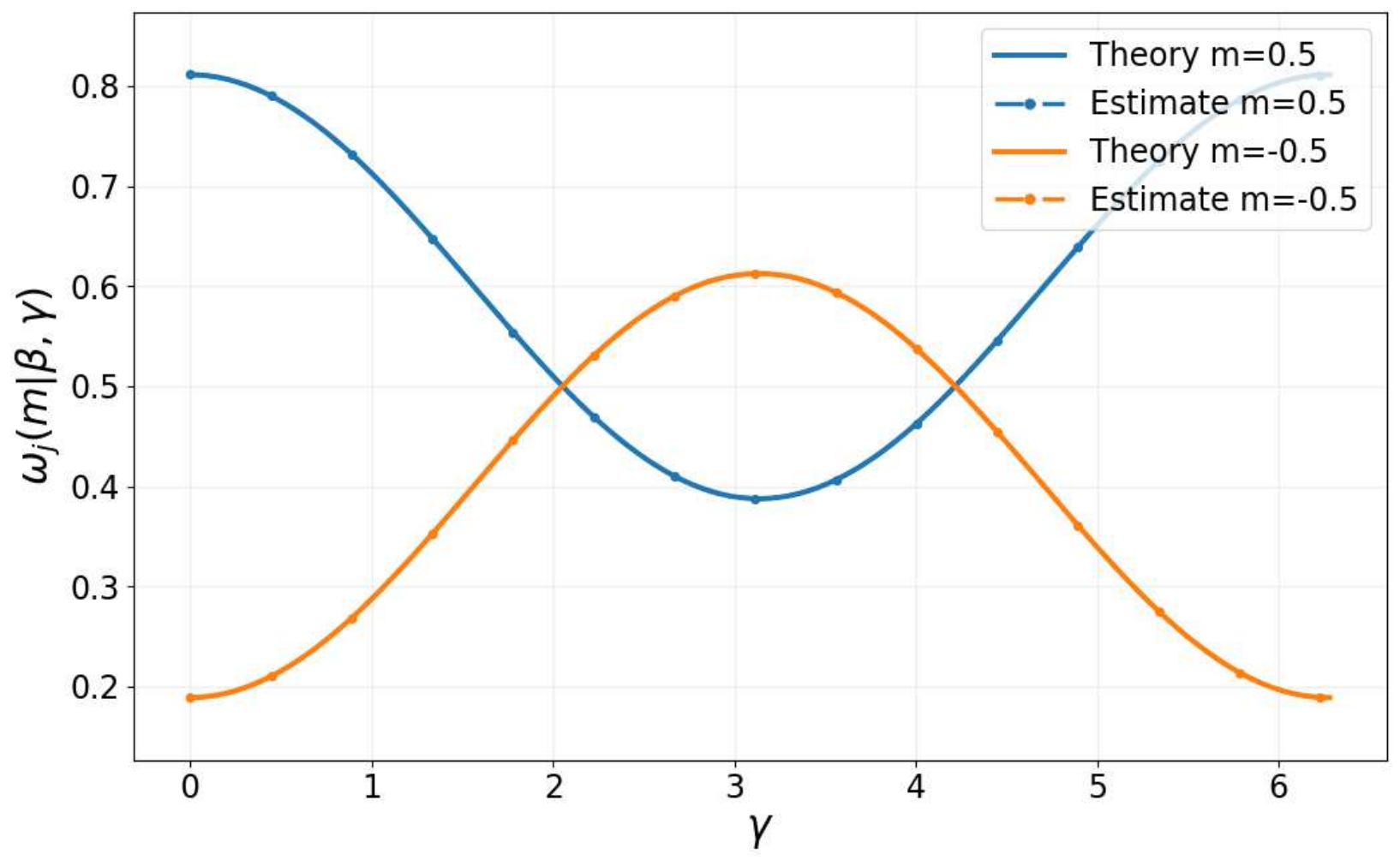}
        \caption{Fixed $\beta$}
        \label{fig:spin_jhalf_fixed_beta_1}
    \end{subfigure}
    \caption{Spin tomograms $\omega_{1/2}(m,\beta,\gamma)$ for the model \eqref{1207} reconstructed using the KCFE method. a) fixed $\gamma=\pi/3$, $\beta\in[0,2\pi)$; b) fixed $\beta=\pi/3$,  $\gamma\in[0,2\pi)$. Solid lines show the theoretical predictions \eqref{eq:w12_pdiag_pcoh_phi}, while dash-dotted lines correspond to the KCFE based estimates obtained from experimental data.}
\label{fig:spin_jhalf_slices_corr}
\end{figure}

\begin{figure}[t]
    \begin{subfigure}{0.49\linewidth}
        \centering
    \includegraphics[width=\linewidth]{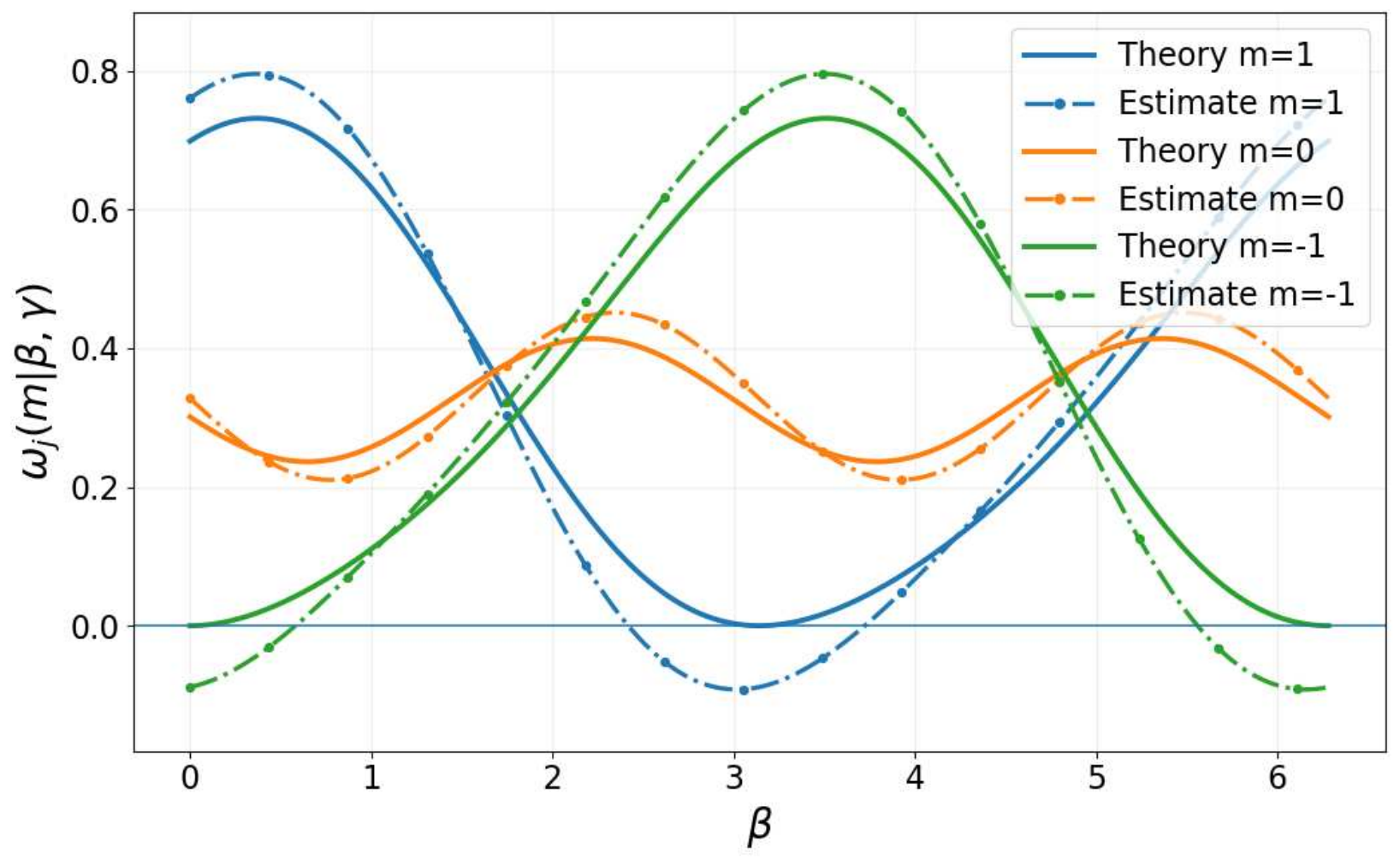}
    \caption{Fixed $\gamma$}
     \end{subfigure}\hfill
    \begin{subfigure}{0.49\linewidth}
        \centering
    \includegraphics[width=\linewidth]{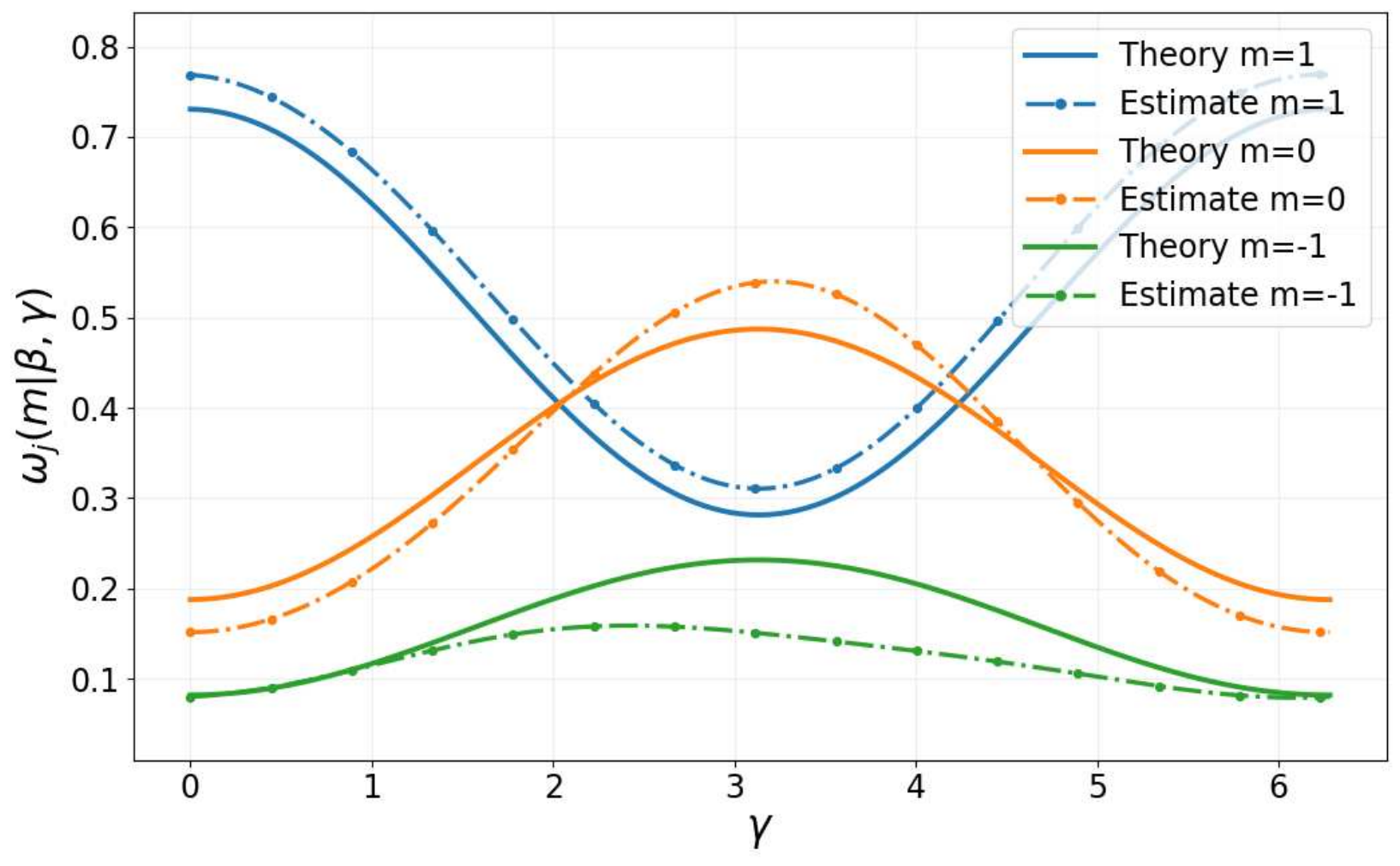}
\caption{Fixed $\beta$}
 \end{subfigure}
  \caption{Spin tomograms $\omega_{1}(m,\beta,\gamma)$ for the  model \eqref{1207} reconstructed using the KCFE method. a) fixed $\gamma=\pi/3$, $\beta\in[0,2\pi)$; b) fixed $\beta=\pi/3$, $\gamma\in[0,2\pi)$. Solid lines show the theoretical predictions \eqref{eq:w1_pdiag_pcoh_phi}, while dash-dotted lines correspond to the KCFE based estimates obtained from experimental data.}
    \label{fig:spin_j1_slices_corr}
\end{figure}

 One can see that the spin tomograms \eqref{1222}, \eqref{1223} and their estimates in Figures~\ref{fig:4}--\ref{fig:5} have similar shape, but are biased. The remaining discrepancies can be explained by an incorrectly chosen data description model \eqref{eq:rho_mixed}. A single parameter $p$ simultaneously parametrizes both the diagonal population and the coherence. 
In realistic experiments, phase noise and interferometric instability suppress off-diagonal coherences more strongly than populations. To account for this effect, we introduce two independent parameters: 
$p_{\mathrm{diag}}$, describing the weight of the diagonal single-photon contribution, and 
$p_{\mathrm{coh}}$, characterizing the residual coherence. This leads to an effective mixed state model where 
$p_{\mathrm{coh}}\leq p_{\mathrm{diag}}$ and
\begin{align}\label{1207}
    \rho&=(1-p_{diag})\ket{0}\bra{0}+p_{diag}(|c_0|^2\ket{0}\bra{0}+|c_1|^2\ket{1}\bra{1})\\\nonumber&+p_{coh}(c_0c_1^{\star}\ket{0}\bra{1}+c_0^{\star}c_1\ket{1}\bra{0}).
\end{align}
{The corresponding theoretical tomograms change accordingly. However, this does not affect the KCFE procedure, which is fully data-driven and does not use this model as an input.} The new theoretical spin-tomograms corresponding to the new model are
\begin{align}
\label{eq:w12_pdiag_pcoh_phi}
 &\omega_{1/2}(m|\beta,\gamma) = \frac12 +2m\Big[ \Big(\frac12-p_{\mathrm{diag}}\,c_1^2\Big)\cos\beta -\big(p_{\mathrm{coh}}\,c_0c_1\big)\sin\beta\,\cos(\gamma-\Phi) \Big], \nonumber \\
& m=\{-\tfrac12, \tfrac12\}
\end{align}
\begin{align}\label{eq:w1_pdiag_pcoh_phi} &\omega_{1}(m,\beta,\gamma)= \frac{1}{3} +\frac{1-p_{\mathrm{diag}}c_{1}^{2}}{2}\,m\cos\beta +\frac{1-3p_{\mathrm{diag}}c_{1}^{2}}{12}\,(3m^{2}-2)\,\bigl(3\cos^{2}\beta-1\bigr)\nonumber \\ &-\frac{p_{\mathrm{coh}}c_{0}c_{1}}{\sqrt2}\,\sin\beta\;\Bigl[m+(3m^{2}-2)\cos\beta\Bigr]\cos(\gamma-\Phi), \!\!\!\quad m\in\{-1,0,1\}. \end{align}
The parameters $p_{\mathrm{diag}}=0.5523$ and $p_{\mathrm{coh}}=0.4913$ are estimated  from the data. Then the KCFE based spin tomogram estimators  shown in Figures~\ref{fig:spin_jhalf_slices_corr} and \ref{fig:spin_j1_slices_corr} {show substantially improve agreement with the theoretical predictions \eqref{eq:w12_pdiag_pcoh_phi} and \eqref{eq:w1_pdiag_pcoh_phi}.} The remaining discrepancy is likely due to noise sources not included in the model \eqref{1207}, such as population leakage into the $n\ge 2$ Fock subspace. However, further adjustments are out of scope of our paper. 
\section{Discussion and Conclusion}\label{sec_5}
{We have established a sector-wise connection between tomographic representations of DV spin systems and restricted CV bosonic systems by employing the Jordan--Schwinger and Holstein--Primakoff maps along with the quantizer--dequantizer formalism.}
 This provides a practical framework for transforming measurement data across different physical platforms {without the intermediate step of full density matrix reconstruction, which can be numerically expensive and may accumulate errors in practical implementations. For example, one can start directly from the measured spin tomogram obtained in a DV setup and, without reconstructing the density matrix, obtain the tomogram of the corresponding restricted bosonic state within the associated fixed excitation sector.} The resulting data can then be used for state verification, error correction through the distance between the theoretical and reconstructed tomograms, recovery of trace characteristics of the state, and many other tasks.
Our analysis focuses on a fixed excitation subspace ($n=2j$), relevant for systems where superpositions of different total particle numbers are not of primary interest. {Within this regime, we establish a transparent sector-wise mapping between spin architectures and the corresponding restricted bosonic sectors, enabling hybrid processing protocols that combine the advantages of discrete and continuous variables. Importantly, this restriction has a clear physical interpretation rather than being merely a technical limitation.} In~\cite{descamps2024superselection,DescampsUnified2025,SaharyanMetrology2025} it is shown that the commonly used ``CV limit'' emerges as an approximation within a large fixed-$n$ sector, providing a controlled bridge between discrete and continuous descriptions. 
Looking ahead, extending the present method to multi-mode scenarios will enable the comparison and interfacing of more complex CV-DV architectures, including quantum memories and photonic communication links, within the same tomographic framework. A detailed analysis of the estimation errors arising at each step of our method, combined with the KQSE approach, will be the subject of future research.
\section*{Acknowledgements}
We thank A. Lvovsky for providing the experimental homodyne- tomographic data used in this work and for valuable discussions. We thank X. Liu for significant help in simulation study.
L.\,M.\ was supported by the Netherlands Organisation for Scientific Research (NWO/OCW), 
as part of the Quantum Software Consortium program (project number 024.003.037 / 3368).
\bibliographystyle{unsrtnat}
\bibliography{bib_corrected}
\appendix
\section{Dequantizer and Quantizer Derivation for the Spin Tomogram}\label{ap_1}
\begin{statement}
    The expansions of the dequantizer and quantizer in the irreducible-tensor operator (ITO) basis  $\hat{T}^{(j)}_{LM}$ are
\begin{align}
   &\hat{\mathcal{U}}(m,\Omega)
= \sum\limits_{L=0}^{2j} \sum\limits_{M=-L}^{L}C_{L,M}(m,\Omega) 
\hat{T}^{(j)}_{L\,M},\label{U_1}\\
  & \hat{\mathcal{D}}(m,\Omega)
= \sum\limits_{L=0}^{2j} \sum\limits_{M=-L}^{L}c_LC_{L,M}(m,\Omega) 
\hat{T}^{(j)}_{L\,M},\label{D_1}
\end{align}
where we used the notation
\begin{eqnarray*}
C_{L,M}(m,\Omega)=
(-1)^{j-m}\!
\bigl\langle j,m;\,j,-m\! \bigm|\! L,0\bigr\rangle
D^{(L)}_{M,0}(\Omega), \quad c_L=(2L+1)/8\pi^2,
\end{eqnarray*}
and $D^{(L)}_{M,0}(\Omega)$ is the Wigner D-matrix~\cite{Edmonds1957}.
\end{statement}
\begin{proof}We first deduce the form \eqref{U_1}. The dequantizer operator is the rotated projector:
\begin{eqnarray}\label{1101}
    \hat{\mathcal{U}}(m, \Omega)\equiv \hat{R}^{\dagger}(\Omega) |j,m\rangle\langle j,m| \hat{R}(\Omega),
\end{eqnarray}
and $\sum_m \hat{\mathcal{U}}(m, \Omega)=\hat{1}$, $\forall \Omega$.  The tomogram for rotating axes satisfies the normalization conditions:
\begin{eqnarray}
 \!\!\!   \sum_{m=-j}^{j} \omega_j(m,  \Omega) = 1, \quad 
\frac{2j+1}{8\pi^2} \int \omega_j(m,  \Omega) d\Omega= 1,
\end{eqnarray}
where $\int d\Omega\equiv\int_{0}^{2\pi} d\alpha \int_{0}^{\pi} d\beta \,\sin\beta \int_{0}^{2\pi} d\gamma $. Then it can be written as
\begin{eqnarray}\label{Spin_Deq_Wigner}
 \!\! \! \! \! \!    \!  \hat{\mathcal{U}}(m,\Omega)= \! \! \! \! \! \! \! \! \sum\limits_{m_1,m_2=-j}^j\!\!  \! \!   \! \!  D_{m,m_2}^{(j) *}(\Omega) D_{m,m_1}^{(j)}(\Omega) |j,m_2\rangle \langle j,m_1|,
\end{eqnarray}
where  the Wigner D-matrix is
\begin{align}\label{Dmatrix}
  D^{(j)}_{m,m^{\prime}}(\Omega)=e^{-im\alpha} d^{(j)}_{mm^{\prime}}(\beta) e^{-im^{\prime}\gamma}  .
\end{align}  
and its elements 
 form a set of orthogonal functions of the Euler angles:
 \begin{equation}\label{1635}\int d\Omega
D^{(j)*}_{m,k'}(\Omega)
D^{(j)}_{m, k}(\Omega)
=\frac{8\pi^{2}}{2j+1}
\delta_{k'k}.
\end{equation}
For $m+m'\geq 0$ the $d$-matrix has the following form
\begin{eqnarray}
d^{(j)}_{m,m'}(\beta)
&=&
\sqrt{\frac{(j+m')!\,(j-m')!}{(j+m)!\,(j-m)!}}
\Bigl(-\sin\tfrac\beta2\Bigr)^{m-m'}\\\nonumber
&\times&\Bigl(\cos\tfrac\beta2\Bigr)^{m+m'}
P_{\,j-m}^{\,(m-m',\,m+m')}\!\bigl(\cos\beta\bigr),
\end{eqnarray}
where $P^{(a,b)}_{n}(x)$ are Jacobi polynomials. Similar expression exists for the case of $m+m'<0$.
The matrix elements of \eqref{Spin_Deq_Wigner}  are
\begin{eqnarray}\label{1652}
    U_{m_2m_1}(m,\Omega)&\equiv&\bra{j,m_2} \hat{\mathcal{U}}(m,\Omega)\ket{j,m_1}= D_{m,m_2}^{(j) *}(\Omega) D_{m,m_1}^{(j)}(\Omega).
\end{eqnarray}
\par In practice, while any operator on a spin-$j$ Hilbert space can be expanded in the basis $\{\ket{j,m}\bra{j,m'}\}$, that representation is complicated when handling rotations. A more convenient choice is the basis of ITOs $\hat{T}^{(j)}_{LN}$, which carry definite rank $L\in[0,2j]$ and projection $N\in[-L,L]$:
\begin{equation}
\hat{T}^{(j)}_{LN}
=  \sum_{m,m'=-j}^{j} 
(-1)^{j-m'}\,
\langle j,m;\,j,-m' \,\vert\,L,N\rangle\;
\ket{j,m}\bra{j,m'}\,.
\end{equation}
The set $\{\hat{T}^{(j)}_{LN}\}$ is complete and orthogonal under the trace inner product in the $\mathfrak{su}(2)$ irreducible representation, satisfying \begin{eqnarray}\label{1752}
\tr{\hat{T}^{(j)\dagger}_{L'N'}\hat{T}^{(j)}_{LN}}=\delta_{LL'}\delta_{NN'}.\end{eqnarray}
In this basis the expansion is  
\begin{equation}\label{eq:ito_basis_change}
\ket{j,m}\!\bra{j,m'} 
= \sum_{L=0}^{2j} \sum_{N=-L}^{L} 
(-1)^{\,j-m'}
\langle j,m;\,j,-m'\,\vert\,L,N\rangle\;
\hat{T}^{(j)}_{LN},
\end{equation}  
where $\langle j,m;\,j,-m'\,\vert\,L,N\rangle$ is the Clebsch--Gordan coefficient.
Under rotation $\hat{R}(\Omega)$ each $\hat{T}^{(j)}_{LN}$ is carried into a linear combination of the same family of operators, mixing its original component $N$ into new components $M\in[-L,L]$:
\begin{eqnarray}
R(\Omega)\,\hat T^{(j)}_{L\,N}\,R^\dagger(\Omega)
&= \sum\limits_{M=-L}^{L}
D^{(L)}_{M,N}(\Omega)\;\hat T^{(j)}_{L\,M},
\end{eqnarray}
where the $D_{M,N}^{(L)}$ are the Wigner $D$-matrices of spin $L$.
Taking $m=m'$ in \eqref{eq:ito_basis_change}, we get
\begin{equation}\label{1713_1}
\ket{j,m}\!\bra{j,m} 
\!= \!\sum_{L=0}^{2j} 
(-1)^{\,j-m}\!
\langle j,m;\,j,-m\,\vert\,L,0\rangle\;
\hat{T}^{(j)}_{L0},
\end{equation} 
where the Clebsch--Gordan coefficient enforces the selection rule on total projection $ \bigl\langle j,m;\,j,-m \bigm| L,N\bigr\rangle\neq 0$,
so the only surviving term is $N=0$.
Then the expansion of the dequantizer in the irreducible-tensor basis is
given by \eqref{U_1}. 
Next, we seek the quantizer in the form \eqref{D_1}, 
where $c_L$ is the unknown coefficient. We expand
\begin{eqnarray}
  \!\!\!\! &&\!\!\! \tr{\hat{\mathcal{U}}(m,\Omega)\hat{\mathcal{D}}(m',\Omega')}=
   \sum\limits_{L,L'=0}^{2j} \sum\limits_{M=-L}^{L} \sum\limits_{M'=-L'}^{L'} (-1)^{2j-m-m'+M+M'} c_{L'}\\\nonumber
   \!\!\!\!&\times&\!\!\!\!\bigl\langle j,m;\,j,-m\! \bigm|\! L,0\bigr\rangle \bigl\langle j,m';\,j,-m'\! \bigm|\! L',0\bigr\rangle D^{(L)}_{0,-M}(\Omega)D^{(L')}_{0,-M'}(\Omega')\tr{\hat{T}^{(j)}_{L\,M}\hat{T}^{(j)}_{L'\,M'}},
\end{eqnarray}
such that the condition \eqref{1727} in the form
\begin{eqnarray}
\tr{\hat{\mathcal{U}}(m,\Omega)\hat{\mathcal{D}}(m',\Omega')}=\delta_{m,m'}\delta(\Omega-\Omega'),
\end{eqnarray}
holds. We use \eqref{1752} with $(\hat{T}^{(j)}_{L,M})^{\dagger}=(-1)^M\hat{T}^{(j)}_{L,-M}$ and
 the Haar orthogonality to get $c_L=(2L+1)/8\pi^2$.\end{proof}
\section{Symplectic to Spin Tomogram Transformation via the Jordan--Schwinger Map}\label{app_2}
We start with deduction of the integral transformation \eqref{eq:omega_from_W} with the kernel \eqref{952}.
The  transition from a symplectic tomogram to a spin tomogram is:
\begin{eqnarray}\label{1248}
\omega_j(m,\Omega) = 
\frac{\tr{\mathcal{E}_j \hat{\Pi}_{2j}\hat{\rho}_{(2)}\hat{\Pi}_{2j} \mathcal{E}_j^{\dagger} \hat{\mathcal{U}}(m,\Omega)}}{\tr{\hat{\Pi}_{2j}\hat{\rho}_{(2)}}}.
\end{eqnarray}
We use the relation between the density matrix operator of the two-mode state with the symplectic tomogram to write:
\begin{eqnarray}\label{1055}
\!\!\!&&\!\!\!\tr{\mathcal{E}_j \hat{\Pi}_{2j}\hat{\rho}_{(2)}\hat{\Pi}_{2j} \mathcal{E}_j^{\dagger} \hat{\mathcal{U}}(m,\Omega)}
\\\nonumber&=&\!\!\!\int_{\mathbb{R}^6}\!\!\!d^2\boldsymbol{x}
d^2\boldsymbol{\mu}
 d^2\boldsymbol{\nu}\mathcal{W}(\boldsymbol{x}|\boldsymbol{\mu},\boldsymbol{\nu}) \tr{\hat{\Pi}_{2j}\hat{\mathcal{D}}(\boldsymbol{x}, \boldsymbol{\mu}, \boldsymbol{\nu})\hat{\Pi}_{2j}\hat{\mathcal{U}}^{\mathrm{JS}}_j(m,\Omega)},\\\nonumber
\!\!\!&&\!\!\!\!\!\!
 \tr{\hat{\Pi}_{2j}\hat{\rho}_{(2)}}\!=\!\!\!\displaystyle  \int_{\mathbb{R}^6}\!\!\!d^2\boldsymbol{x}
d^2\boldsymbol{\mu}
 d^2\boldsymbol{\nu}\mathcal{W}(\boldsymbol{x}|\boldsymbol{\mu},\boldsymbol{\nu})\tr{\hat{\Pi}_{2j}\hat{\mathcal{D}}(\boldsymbol{x}, \boldsymbol{\mu}, \boldsymbol{\nu})},
\end{eqnarray}
where we use the cyclicity of trace and the lifted spin dequantizer \eqref{lifted_spin_dequantizer} on Fock space,
that acts as $\hat{\mathcal{U}}(m,\Omega)$ on $S_{2j}$ and zero otherwise. 
\par Let us take a generic two-mode bosonic Fock state $\ket{\psi}=\ket{\psi_{2j}}+\ket{\psi_{\perp}}$, where $\ket{\psi_{2j}}\in S_{2j}$ and $\ket{\psi_{\perp}}$ lies in all other sectors. Then
\begin{eqnarray}
\hat{\Pi}_{2j}\hat{\mathcal{U}}^{\mathrm{JS}}_j(m,\Omega)\hat{\Pi}_{2j}\ket{\psi}=
    \hat{\Pi}_{2j}\hat{\mathcal{U}}^{\mathrm{JS}}_j(m,\Omega)\ket{\psi_{2j}}.
\end{eqnarray}
Since $ \hat{\mathcal{U}}^{\mathrm{JS}}_j(m,\Omega)$ acts nontrivially only inside the two-mode subspace with total number $2j$, we can  write
\begin{eqnarray}
\hat{\Pi}_{2j}\hat{\mathcal{U}}^{\mathrm{JS}}_j(m,\Omega)\ket{\psi_{2j}}=\hat{\mathcal{U}}^{\mathrm{JS}}_j(m,\Omega)\ket{\psi_{2j}}.
\end{eqnarray}
Using the latter conclusions we immediately get the integral transformation \eqref{eq:omega_from_W}. 
  The transition  from the restricted symplectic tomogram to a spin tomogram is derived in the similar way:
\begin{eqnarray*}
\omega_j(m,\Omega)=\int_{\mathbb{R}^6}d^2\boldsymbol{x}
d^2\boldsymbol{\mu}
 d^2\boldsymbol{\nu}\mathcal{W}^{(2j)}_{\mathrm{JS}}(\boldsymbol{x}|\boldsymbol{\mu},\boldsymbol{\nu}) K^{(j, \mathrm{JS})}_{\mathcal{W}\to\omega}(m,\Omega; \boldsymbol{x}, \boldsymbol{\mu}, \boldsymbol{\nu}).
\end{eqnarray*}
Following similar steps, one can deduce the reversed transformation \eqref{eq:W_from_omega} from $\omega_j(m,\Omega)$ to the restricted tomogram.
\begin{statement}
   An explicit view of the kernels \eqref{952} and  \eqref{953} are:
\begin{align}
    \label{1325}
    &K^{(j, \mathrm{JS})}_{\mathcal{W}\to\omega}(m,\Omega; \boldsymbol{x}, \boldsymbol{\mu}, \boldsymbol{\nu})
=\frac{1}{(2\pi)^2}e^{i(x_1+x_2)}e^{-\tfrac{1}{2}(|\xi_{1}|^2+|\xi_{2}|^2)} 
\\\nonumber&\times \sum\limits_ {m_1,m_2=-j}^jD_{m,m_2}^{(j) *}(\Omega) D_{m,m_1}^{(j)}(\Omega) \mathcal{M}^{(j)}_{m_1,m_2}(\xi_1,\xi_2),\\
\label{1233_2}
&K^{(j, \mathrm{JS})}_{\omega\to \mathcal{W}}(m,\Omega; \boldsymbol{x}, \boldsymbol{\mu}, \boldsymbol{\nu})=\frac{1}{\pi }e^{-\tfrac{1}{2}\left(\tfrac{x_1^2}{|\xi_1|^2}+\tfrac{x_2^2}{|\xi_2|^2}\right)} \\\nonumber
&\times \sum\limits_ {m_1,m_2=-j}^j \frac{\mathcal{P}^{(j)}_{m_1,m_2}(x_1,\xi_1)\mathcal{P}^{(j)}_{m_1,m_2}(x_2,\xi_2)}{(2|\xi_1\xi_2|)^{|m_1-m_2|+1}}  \mathcal{D}_{m_2 m_1}^{(j)}(m,\Omega)\mathcal{K}^{(j)}_{m_1,m_2}(\xi_1,\xi_2).
  \end{align} 
where $D^{(j)}_{m,l}(\Omega)$ is the Wigner D-matrix and  we used the notations $\xi_{k}=-\frac{i}{\sqrt{2}}(\mu_{k}+i\nu_{k})$, $k=1,2$ and 
\begin{align}
\mathcal{M}^{(j)}_{m_1,m_2}(\xi_1,\xi_2)
&\equiv(-1)^{|m_1-m_2|} \mathcal{K}^{(j)}_{m_1,m_2}(\xi_1,\xi_2) \\\nonumber
&\times L_{j+\min\{m_1,m_2\}}^{(|m_1-m_2|)}(|\xi_1|^2)\,
L_{j-\max\{m_1,m_2\}}^{(|m_1-m_2|)}(|\xi_2|^2),
\end{align}
\begin{align}
    \mathcal{K}^{(j)}_{m_1,m_2}(\xi_1,\xi_2)&\equiv \sqrt{\frac{(j+\min\{m_1,m_2\})!\,(j-\max\{m_1,m_2\})!}
{(j+\max\{m_1,m_2\})!\,(j-\min\{m_1,m_2\})!}}\\\nonumber
&\times(\xi_1^{*}\xi_2)^{\max(0,m_2-m_1)}\,(\xi_1\xi_2^{*})^{\max(0,m_1-m_2)},
\end{align}
\begin{align}\label{1426}
 \mathcal{P}^{(j)}_{m_1,m_2}(x_k,\xi_k)&\equiv   \sum_{s=0}^{n_k}
\binom{n_k+|m_1-m_2|}{\,n_k-s\,}\,
\frac{1}{s!2^{s}}\,
H_{|m_1-m_2|+2s}\!\Big(\tfrac{\,x_k}{\sqrt{2}|\xi_k|}\Big),\\\nonumber
& n_1=j+\min{\{m_1,m_2\}},\quad 
n_2=j-\max{\{m_1,m_2\}},
\end{align}
\begin{eqnarray}\label{1250}
\mathcal{D}_{m_1 m_2}^{(j)}(m,\Omega) =\sum_{L=0}^{2j}c_L\!\!\!\!\sum_{M=-L}^{L} C_{L,M}(m,\Omega) (-1)^{j-m_2} \langle j,m_1; j,-m_2 | L,M \rangle.
\end{eqnarray}
\end{statement}
\begin{proof} We first deduce the form \eqref{1325}. Inserting the resolution of identity on the $S_{2j}$, one can write
\begin{eqnarray}\label{resolution}
\hat{\mathcal{U}}^{\mathrm{JS}}_j(m,\Omega)=\! \! \! \! \! \sum\limits_ {m_1,m_2=-j}^j \! \! \! \! \! \ket{j+m_2,j-m_2}_2\bra{j+m_1,j-m_1}_2U_{m_2m_1}(m,\Omega),
\end{eqnarray}
where the matrix elements $U_{m_2m_1}(m,\Omega)$ can be written in terms of the Wigner D-matrix elements as in \eqref{1652}. 
The symplectic quantizer \eqref{symplectic_quantizer} corresponding to the tomogram of the two-mode Fock state can be written as a direct product of the Weyl displacement $\hat{D}(\xi)$ operators:
\begin{eqnarray}\label{1141}
    \hat{\mathcal{D}}(\boldsymbol{x}, \boldsymbol{\mu}, \boldsymbol{\nu})=(2\pi)^{-2}e^{i(x_1+x_2)}\hat{D}(\xi_1)\otimes\hat{D}(\xi_2).
\end{eqnarray}
Since the operator factorizes over two modes, the matrix element factorizes too, using a standard property of tensor product: 
\begin{eqnarray}\label{1147}
D_{m_1m_2}^{(j)}(\boldsymbol{x}, \boldsymbol{\mu}, \boldsymbol{\nu})\equiv%
\frac{e^{i(x_1+x_2)}}{(2\pi)^2}\bra{j+m_1}\hat{D}(\xi_1)\ket{j+m_2}\bra{j-m_1}\hat{D}(\xi_2)\ket{j-m_2}.
\end{eqnarray}
 The matrix elements of $\hat{D}(\xi)$ in the Fock state basis $\{|n\rangle\}$ are given by~\cite{cahill1969ordered}:
\begin{equation} \label{eq:displacement_matrix_element}
 \!\!\! \!\!\! \langle n' | \hat{D}(\xi) | n \rangle = \left(n!/n'!\right)^{1/2} \xi^{n'-n} e^{-|\xi|^2/2} L_n^{(n'-n)}(|\xi|^2),
\end{equation}
for $ n' \ge n$, 
where $L_q^{(p)}(z)$ is an associated (generalized) Laguerre polynomial. For the case $n' < n$, one can use the relation $\langle n' | D(\xi) | n \rangle = (\langle n | D(-\xi) | n' \rangle)^*$. Substituting \eqref{1652} and \eqref{1147}  in
\begin{align}
     K^{(j, \mathrm{JS})}_{\mathcal{W}\to\omega}(m,\Omega; \boldsymbol{x}, \boldsymbol{\mu}, \boldsymbol{\nu})
    =\sum\limits_ {m_1,m_2=-j}^j\! \! \! \!  U_{m_2m_1}(m,\Omega)D_{m_1m_2}^{(j)}(\boldsymbol{x}, \boldsymbol{\mu}, \boldsymbol{\nu}),
\end{align}
we get \eqref{1325}. 
\par To deduce the inverse transformation kernel \eqref{953} in form \eqref{1233_2} we write the symplectic dequantizer as
\begin{eqnarray}
    \hat{\mathcal{U}}(\boldsymbol{x}, \boldsymbol{\mu}, \boldsymbol{\nu})
    = \frac{1}{(2\pi)^2} \int\limits_{-\infty}^{\infty} dk_1dk_2  e^{i(k_1x_1\hat{1}_1+k_2x_2\hat{1}_2)} \hat{D}(k_1\xi_1)\otimes\hat{D}(k_2\xi_2),
\end{eqnarray}
Then the kernel can be written as follows
\begin{eqnarray}\label{1233}
&&\!\!\!\!\!\!\!\!K^{(j, \mathrm{JS})}_{\omega\to \mathcal{W}}(m,\Omega; \boldsymbol{x}, \boldsymbol{\mu}, \boldsymbol{\nu})=\frac{1}{(2\pi)^2} \int\limits_{-\infty}^{\infty} dk_1dk_2  e^{i(k_1x_1+k_2x_2)} e^{-\tfrac{1}{2}(|k_1\xi_1|^2+|k_2\xi_2|^2)}\nonumber\\
&\times& \sum\limits_ {m_1,m_2=-j}^j \mathcal{D}_{m_2 m_1}^{(j)}(m,\Omega)\mathcal{M}^{(j)}_{m_1,m_2}(k_1\xi_1,k_2\xi_2), 
  \end{eqnarray}
where the quantizer matrix elements are equal to \eqref{1250}.
To calculate the integrals in the latter relation we use the series expansion of the associated Laguerre polynomial:
\begin{eqnarray}
L_q^{(p)}(z) = \sum_{l=0}^q (-1)^l \binom{q+p}{q-l} \frac{z^l}{l!},
\end{eqnarray}
and the known integral
\begin{eqnarray*}
    \int\limits_{-\infty}^{\infty}x^n e^{-px^2-qx}dx=\left(\frac{i}{2}\right)^n\sqrt{\pi}p^{-(n+1)/2}e^{\frac{q^2}{4p}}H_n\left(\frac{iq}{2\sqrt{p}}\right).
\end{eqnarray*}
We introduce the notations $\Delta=|m_1-m_2|$, $n_1=j+\min{\{m_1,m_2\}}$,
$n_2=j-\max{\{m_1,m_2\}}$ %
and get
\begin{eqnarray}
    && \int\limits_{-\infty}^{\infty} dk_r e^{ik_rx_r-k_r^2\frac{|\xi_r|^2}{2}}k_r^{\max(0,m_2-m_1)}k_r^{\max(0,m_1-m_2)}L_{n_r}^{(\Delta)}(k_r^2|\xi_r|^2)\\\nonumber
     &=& \frac{\sqrt{2\pi}}{|\xi_r|}\,
\exp\!\Big[-\tfrac{x_r^2}{2|\xi_r|^2}\Big]\,
\Big(\tfrac{i}{\sqrt{2}|\xi_r|}\Big)^{\Delta}
\left[\sum_{s_r=0}^{n_r}
\binom{n_r+\Delta}{\,n_r-s_r\,}\,
\frac{1}{s_r!2^{s_r}}\,
H_{\Delta+2s_r}\!\Big(\tfrac{x_r}{\sqrt{2}|\xi_r|}\Big)\right].
  \end{eqnarray}
Using this result, we can rewrite  \eqref{1233} into the final form \eqref{1233_2}.\end{proof}
\begin{statement}
The explicit form of the denominator of \eqref{eq:omega_from_W} is
    \begin{eqnarray}\label{eq:trace_Pi2j_D_simplified}
\tr{\hat{\Pi}_{2j}\hat{\mathcal{D}}(\boldsymbol{x}, \boldsymbol{\mu}, \boldsymbol{\nu})} =\frac{1}{(2\pi)^2}e^{i(x_1+x_2)}\,
e^{-\tfrac{1}{2}\big(|\xi_{1}|^2+|\xi_{2}|^2\big)}
L_{2j}^{(1)}\!\big(|\xi_1|^2+|\xi_2|^2\big).
\end{eqnarray}
\end{statement} 
\begin{proof}Since \eqref{1141} holds, we use \eqref{1147} to write  
\begin{eqnarray}
 \!\!\!   &&\tr{\hat{\Pi}_{2j}\hat{\mathcal{D}}(\boldsymbol{x}, \boldsymbol{\mu}, \boldsymbol{\nu})}=\sum\limits_{m=-j}^jD_{mm}^{(j)}(\boldsymbol{x}, \boldsymbol{\mu}, \boldsymbol{\nu})\\
     &=&\frac{e^{i(x_1+x_2)}}{(2\pi)^2} 
e^{-\tfrac{1}{2}(|\xi_{1}|^2+|\xi_{2}|^2)}\sum\limits_ {m=-j}^jL_{j+m}(|\xi_1|^2)L_{j-m}(|\xi_2|^2).\nonumber
\end{eqnarray}
\par Using the finite convolution identity for Laguerre polynomials, which follows directly from the generating function,
\begin{equation}\label{eq:Laguerre_convolution}
\sum_{n=0}^{N} L_n(a)\,L_{N-n}(b)=L_{N}^{(1)}(a+b),
\end{equation}
and the change of summation index $n=j+m$ (so that $n=0,1,\dots,2j$ and $j-m=2j-n$), we obtain \eqref{eq:trace_Pi2j_D_simplified}.\end{proof}
\section{ Spin to Photon-Number Tomogram Transformations via the Jordan--Schwinger Map}\label{app_3}
\begin{statement}
   The transition kernel  \eqref{1554_1} and \eqref{1554_2} of the integral transformations between the spin tomograms and the photon-number tomogram \eqref{eq:photon-to-omega} and \eqref{eq:omega-to-photon} have the form
\begin{align}\label{spin_to_photon}
  &\! \! K^{(j, \mathrm{JS})}_{\omega\rightarrow \mathsf{w}}(m,\Omega;\vec r, \boldsymbol{\alpha})= \sum\limits_ {m_1,m_2=-j}^j \mathcal{D}_{m_2 m_1}^{(j)}(m,\Omega)  \nonumber \\&\times
   M(j+m_1, r_1; -\alpha_1)  M(r_1, j+m_2; \alpha_1)   M(j-m_1, r_2; -\alpha_2)M(r_2, j-m_2; \alpha_2),\\\nonumber
     &\! \! K^{(j, \mathrm{JS})}_{\mathsf{w}\rightarrow \omega}(m,\Omega;\vec r, \boldsymbol{\alpha})=\!\frac{16}{\pi^{2} (1-s^2)^{2}}  \sum\limits_ {m_1,m_2=-j}^j  \!\!\!D_{m,m_2}^{(j) *}(\Omega) D_{m,m_1}^{(j)}\!(\Omega)   \sum_{l_1,l_2=0}^{\infty}  g(s)^{l_1-r_1 + l_2-r_2} \\\nonumber&\times
   M(j+m_1, l_1; -\alpha_1)  M(l_1, j+m_2; \alpha_1)  M(j-m_1, l_2; -\alpha_2)  M(l_2, j-m_2; \alpha_2),
\end{align}
where we used the notations
\begin{equation}\label{1176}
M(a, b; \beta) = 
\begin{cases}
\sqrt{\dfrac{b!}{a!}} \beta^{a-b} e^{-|\beta|^2/2} L_b^{(a-b)}(|\beta|^2), & a \geq b \\
\sqrt{\dfrac{a!}{b!}} (-\beta^*)^{b-a} e^{-|\beta|^2/2} L_a^{(b-a)}(|\beta|^2), & a < b,
\end{cases}
\end{equation}
and $\mathcal{D}_{m_2 m_1}^{(j)}(m,\Omega)$ is given by \eqref{1250}.
\end{statement}
\begin{proof}
With resolution~\eqref{resolution} kernels \eqref{1554_1} and \eqref{1554_2} can be rewritten as
\begin{eqnarray}\label{photon_kernels}
     &&\!\!\!\!\!\!\!\!\!K^{(j, \mathrm{JS})}_{\mathsf{w}\rightarrow \omega}(m,\Omega; \vec r, \boldsymbol{\alpha})=\!\!\!\!\!\!\!\!\sum\limits_ {m_1,m_2=-j}^j\!\!\!\!\!\!\!D_{m_1 m_2}^{(j)}(\vec r, \boldsymbol{\alpha}) U^{(j)}_{m_2 m_1}(m,\Omega),\\\nonumber
  &&\!\!\!\!\!\!\!\!\!K^{(j, \mathrm{JS})}_{\omega\rightarrow \mathsf{w}}(m,\Omega;\vec r, \boldsymbol{\alpha})=\!\!\!\!\!\!\!\!\sum\limits_ {m_1,m_2=-j}^j \!\!\!\!\!\! U_{m_1 m_2}^{(j)}(\vec r, \boldsymbol{\alpha})\mathcal{D}^{(j)}_{m_2 m_1}(m,\Omega),
   \end{eqnarray}
   where we denote
   \begin{eqnarray*} 
    &&D_{m_1 m_2}^{(j)}(\vec r, \boldsymbol{\alpha}) \!\equiv \! \bra{j+m_1,j-m_1}_2 \hat{\mathcal{D}}_{\vec r}(\boldsymbol{\alpha}) \!\ket{j+m_2,j-m_2}_2,\\\nonumber
    &&U_{m_1,m_2}^{(j)}(\vec r, \boldsymbol{\alpha}) \!\equiv \! \bra{j+m_1,j-m_1}_2 \hat{\mathcal{U}}_{\vec r}(\boldsymbol{\alpha}) \!\ket{j+m_2,j-m_2}_2.
\end{eqnarray*}
\par Now we obtain an explicit form of the transition kernels from spin to photon-number tomograms and vice versa, using the definition of the photon-number tomogram \eqref{dequantizer-ph-number}. The photon-number dequantizer can be written as:
\begin{eqnarray}
    &&\!\!\!\!\!\!
   U_{m_1 m_2}^{(j)}(\vec r, \boldsymbol{\alpha})
   = M(j+m_1, r_1; -\alpha_1)  M(r_1, j+m_2; \alpha_1) \nonumber \\&\times&  M(j-m_1, r_2; -\alpha_2)  M(r_2, j-m_2; \alpha_2),
\end{eqnarray}
where we denote $M(a, b; \beta)=\bra{a}  \hat{D}(\beta)\, |b\rangle$. We use \eqref{eq:displacement_matrix_element} to deduce \eqref{1176}.
The expression for the quantizer \eqref{quantizer-ph-number} is not so straightforward. Here the operator power is understood in the spectral sense for the number operator
$\hat n=\hat a^\dagger\hat a$. We define
$g(s)^{\hat{a}^\dagger\hat{a}-r}$ by its action in the Fock basis and use the unitary equivalence of operators:
\begin{eqnarray}
f\left( (\hat{a}^\dagger+\alpha^{*})(\hat{a}+\alpha) \right)
= \hat{D}(-\alpha)\, f(\hat{a}^\dagger\hat{a})\, \hat{D}^\dagger(-\alpha),
\end{eqnarray}
where we take as the function $f(x)=g(s)^{x-r}$:
\begin{eqnarray}
g(s)^{\left[(\hat{a}^\dagger+\alpha^{*})(\hat{a}+\alpha)-r\right]}
= \hat{D}(-\alpha)\, g(s)^{\hat{a}^\dagger\hat{a}-r}\, \hat{D}^\dagger(-\alpha).
\end{eqnarray}
Operator $g(s)^{\hat{a}^\dagger\hat{a}-r}$ in the Fock basis,
\begin{eqnarray}
g(s)^{\hat{a}^\dagger\hat{a}-r} = \sum_{l=0}^{\infty} g(s)^{l-r} |l\rangle\langle l|.
\end{eqnarray}
For $-1<s<1$ one has $g(s)<0$, and since $l-r\in\mathbb{Z}$ the powers $g(s)^{l-r}$ are unambiguous.
Then the quantizer \eqref{quantizer-ph-number} for one mode can be rewritten as follows
\begin{eqnarray}\label{1348}
\!\!\!\!\!\!\!\!\!\hat{\mathcal{D}}_r(\alpha) =\! \frac{4  \hat{D}(-\alpha)}{\pi (1-s^2)}\!\!\left[ \sum_{l=0}^{\infty} g(s)^{l-r} |l\rangle\langle l| \right]\! \hat{D}^\dagger(-\alpha).
\end{eqnarray}
Using this, the matrix element of the two-mode quantizer has the form:
\begin{align}
\!\!\!&D_{m_1 m_2}^{(j)}(\vec r, \boldsymbol{\alpha}) = \frac{16}{\pi^2 (1-s^2)^2}\sum_{l_1,l_2=0}^{\infty} g(s)^{l_1-r_1 + l_2-r_2}\\\nonumber
&\times M(j+m_1,l_1;-\alpha_1)M(l_1,j+m_2;\alpha_1) M(j-m_1,l_2;-\alpha_2)M(l_2,j-m_2;\alpha_2).
\end{align}
Thereby, we can immediately write the transition kernel from the spin tomograms to the photon-number tomogram in the form \eqref{spin_to_photon}.\end{proof}
\begin{statement}
The denominator of \eqref{1055_2} has the form:
    \begin{align}\label{1356}
\tr{\hat{\Pi}_{2j}\,\hat{\mathcal{D}}_{\vec r}(\boldsymbol{\alpha})}
&= \frac{16}{\pi^2(1-s^2)^2}\,
g(s)^{2j-r_1-r_2}\,
e^{-(1-g(s))(|\alpha_1|^2+|\alpha_2|^2)} \\
&\quad\times
L_{2j}^{(1)}\!\left(-\frac{(1-g(s))^2}{g(s)}\big(|\alpha_1|^2+|\alpha_2|^2\big)\right).
\end{align}
\end{statement}
\begin{proof}
 We use \eqref{1348}, to calculate the denominator of \eqref{1055_2}:  \begin{eqnarray}
 &&\! \! \!\! \! \! \tr{\hat{\Pi}_{2j}\,\hat{\mathcal{D}}_{\vec r}(\boldsymbol{\alpha})}
=\sum_{m=-j}^{j} D_{m m}^{(j)}(\vec r, \boldsymbol{\alpha})
\end{eqnarray}
The matrix diagonal element of the two-mode quantizer has the form:
\begin{eqnarray}
&&\! \! \!\! \! \!  D_{m m}^{(j)}(\vec{r}, \boldsymbol{\alpha}) = \frac{16}{\pi^2 (1-s^2)^2} \sum_{l_1,l_2=0}^{\infty}  g(s)^{l_1 - r_1 + l_2 - r_2}\\ \nonumber 
&\times& M(j+m, l_1; -\alpha_1) M(l_1, j+m; \alpha_1)  M(j-m, l_2; -\alpha_2) M(l_2, j-m; \alpha_2).
\end{eqnarray}
To simplify, we can use the property $M(j+m, l_1; -\alpha_1) M(l_1, j+m; \alpha_1) = |M(j+m, l_1; -\alpha_1)|^2$
that follows from the fact that $\hat{D}^{\dagger}(\alpha) = \hat{D}(-\alpha) $. In addition, the generating function for the squares of generalized Laguerre polynomials plays a key role:
\begin{eqnarray}
\sum_{l_1=0}^{\infty} z^{l_1} |\langle l_2 | \hat{D}(\alpha) | l_1 \rangle|^2=z^{l_2} e^{-(1-z)|\alpha|^2}L_{l_2}\left(\!-\frac{(1-z)^2|\alpha|^2}{z}\right). \nonumber
\end{eqnarray}
Then the explicit form of the trace is the following:
\begin{eqnarray}
&&\! \! \!\! \! \! \tr{\hat{\Pi}_{2j}\,\hat{\mathcal{D}}_{\vec r}(\boldsymbol{\alpha})} = \frac{16}{\pi^2(1-s^2)^2} g(s)^{2j - r_1 - r_2}  e^{-(1-g(s))(|\alpha_1|^2 + |\alpha_2|^2)}  \nonumber \\
&\times& \sum_{m=-j}^{j} L_{j+m} ( -\frac{(1-g(s))^2}{g(s)} \, |\alpha_1|^2 )  L_{j-m}( -\frac{(1-g(s))^2}{g(s)} \, |\alpha_2|^2 ).
\end{eqnarray}
Using the property of Laguerre polynomials \eqref{eq:Laguerre_convolution}, we obtain \eqref{1356}. 
\end{proof}
Collecting these results, one gets \eqref{1055_2} written in terms of the Laguerre polynomials.
\section{Spin Tomogram to Wigner Function Transformations via the Jordan--Schwinger Map}\label{app_4}
\begin{statement}
    The transition kernel  \eqref{1713}  from the Wigner function to the spin tomogram has the form
\begin{align}
\label{Wigner_to_spin}
&\! \! \!\! \! \! 
K^{(j, \mathrm{JS})}_{\mathbb{W}\rightarrow \omega}(m,\Omega; \boldsymbol{\alpha})=  4 (-1)^{2j}\sum\limits_ {m_1,m_2=-j}^j  D_{m,m_2}^{(j) *}(\Omega) D_{m,m_1}^{(j)}(\Omega) \\ \nonumber
&\times M(j+m_1, j+m_2; 2\alpha_1)   M(j-m_1, j-m_2; 2\alpha_2),\\
&\! \! \!\! \! \! \label{spin_to_Wigner}
  K^{(j, \mathrm{JS})}_{\omega\rightarrow \mathbb{W}}(m,\Omega; \boldsymbol{\alpha})=\frac{ (-1)^{2j}}{\pi^{2}} \sum\limits_ {m_1,m_2=-j}^j \mathcal{D}_{m_1 m_2}^{(j)}(m,\Omega)  \\\nonumber
  &\times  M(j+m_1, j+m_2; 2\alpha_1)   M(j-m_1, j-m_2; 2\alpha_2),
\end{align}
where $\mathcal{D}_{m_2 m_1}^{(j)}(m,\Omega)$ is given by \eqref{1250}.
\end{statement}
\begin{proof}
Using \eqref{resolution}, the transition kernels \eqref{1713}   can be rewritten as
\begin{eqnarray}\label{wigner_kernels}
   &&\! \! \!\! \! \!K^{(j, \mathrm{JS})}_{\mathbb{W}\rightarrow \omega}(m,\Omega;  \boldsymbol{\alpha})=\!\!\!\!\sum\limits_ {m_1,m_2=-j}^j \mathbb{D}_{m_1 m_2}^{(j)}( \boldsymbol{\alpha}) U^{(j)}_{m_2 m_1}(m,\Omega),\nonumber\\
     &&\! \! \!\! \! \!K^{(j, \mathrm{JS})}_{\omega\rightarrow \mathbb{W}}(m,\Omega; \boldsymbol{\alpha})=\!\!\!\!\sum\limits_ {m_1,m_2=-j}^j \!\!\!\!\mathbb{U}_{m_1 m_2}^{(j)}( \boldsymbol{\alpha})\mathcal{D}^{(j)}_{m_2 m_1}(m,\Omega),
 \end{eqnarray}   
where we use the notations
\begin{eqnarray*}
   &&\mathbb{D}_{m_1 m_2}^{(j)}( \boldsymbol{\alpha})\equiv \bra{j+m_1,j-m_1}_2 \hat{\mathbb{D}}(\boldsymbol{\alpha})\ket{j+m_2,j-m_2}_2,\\\nonumber
     &&\mathbb{U}_{m_1 m_2}^{(j)}( \boldsymbol{\alpha})\equiv \bra{j+m_1,j-m_1}_2 \hat{\mathbb{U}}(\boldsymbol{\alpha})\ket{j+m_2,j-m_2}_2.
\end{eqnarray*}
 Using Eq.~\eqref{quant_and_dequant_wigner}, the expression for the matrix elements of the Wigner quantizer can be rewritten as
\begin{eqnarray}\label{1270}
&&\! \! \!\! \! \! 
\mathbb{D}_{m_1 m_2}^{(j)}( \boldsymbol{\alpha})\ = 4 \sum_{n_1,n_2=0}^{\infty}   \bra{j+m_1} \hat{D}(2\alpha_1)\ket{n_1} \\
&\times& \bra{n_1} \hat{P} \ket{j+m_2}     \bra{j-m_1} \hat{D}(2\alpha_2)\ket{n_2}\bra{n_2} \hat{P} \ket{j-m_2}. \nonumber
\end{eqnarray}
The formula for the matrix elements of the dequantizer is obtained in a similar way as for the quantizer:
\begin{eqnarray}
&&\! \! \!\! \! \! 
\mathbb{U}_{m_1 m_2}^{(j)}( \boldsymbol{\alpha})\ = \frac{1}{\pi^2} \sum_{n_1,n_2=0}^{\infty} \bra{j+m_1} \hat{D}(2\alpha_1)\ket{n_1} \\
&\times& \bra{n_1} \hat{P} \ket{j+m_2}     \bra{j-m_1} \hat{D}(2\alpha_2)\ket{n_2}\bra{n_2} \hat{P} \ket{j-m_2}. \nonumber
\end{eqnarray}
We use the fact that the matrix element of the parity operator is given by $    \bra{m}\hat{P}\ket{n}=(-1)^{n}\delta_{n, m}$.
Using this relation, the transition kernel from the Wigner function to the spin tomogram can be written as 
\begin{eqnarray}
&&\! \! \!\! \! \! 
K^{(j, \mathrm{JS})}_{\mathbb{W}\rightarrow \omega}(m,\Omega; \boldsymbol{\alpha})=4\sum\limits_ {m_1,m_2=-j}^j D_{m,m_2}^{(j) *}(\Omega) D_{m,m_1}^{(j)}(\Omega)  \\\nonumber
&\times& \!\!\sum_{n_1,n_2=0}^{\infty}  (-1)^{n_1+n_2} \delta_{j+m_2,n_1} \delta_{j-m_2,n_2}  M(j+m_1, n_1; 2\alpha_1)  M(j-m_1, n_2; 2\alpha_2),
\end{eqnarray}
where the Kronecker delta functions have been used to eliminate the summation over  $n_1 $ and $n_2 $ to get \eqref{Wigner_to_spin}. The inverse transformation kernel \eqref{spin_to_Wigner} is obtained in an analogous way.
\end{proof}
\begin{statement}
    The denominator of the transformation~\eqref{1248_4} is
    \begin{eqnarray}\label{1400}
\tr{\hat{\Pi}_{2j} \hat{\mathbb{D}}(\boldsymbol{\alpha})} =  4 (-1)^{2j} e^{-2 (|\alpha_1|^2 + |\alpha_2|^2)} L_{2j}^{(1)} \left( 4 (|\alpha_1|^2 + |\alpha_2|^2) \right).
\end{eqnarray}
\end{statement}
\begin{proof}
We use Eq.~\eqref{1270} to calculate the denominator of Eq.~\eqref{1248_4}:
\begin{eqnarray}
\tr{\hat{\Pi}_{2j}\,\hat{\mathbb{D}}(\boldsymbol{\alpha})} %
=4 (-1)^{2j}  e^{-2(|\alpha_1|^2 + |\alpha_2|^2)} \sum_{k=0}^{2j} L_k(4|\alpha_1|^2) L_{2j-k}(4|\alpha_2|^2). %
\end{eqnarray}
Applying the identity for Laguerre polynomials,
\(\sum_{k=0}^{n} L_k (x) L_{n-k} (y) = L_n^{(1)} (x + y)\),
the result simplifies to \eqref{1400}.
\end{proof}
Collecting these results, the Eq.~\eqref{1248_4} can be written in terms of the Laguerre polynomials. 
\section{Transition Kernels for the Integral
Transformation Between Symplectic and Photon-Number Tomograms}\label{app_5}
\begin{statement}
    The explicit form of the symplectic-to--photon-number transition kernels \eqref{636} and \eqref{636_1}: 
\begin{align}\label{1406}
K_{\mathcal{W}\to\mathsf{w}}(x,\mu,\nu,r,\alpha)
&=\frac{1}{2\pi} e^{-\frac{\nu^2+\mu^2}{4}}
e^{ ix + \frac{\nu - i\mu}{\sqrt{2}}\alpha^* - \frac{\nu + i\mu}{\sqrt{2}}\alpha}
L_r\!\left(\frac{\nu^2+\mu^2}{2}\right),
\\
K_{\mathsf{w}\to\mathcal{W}}(x,\mu,\nu,r,\alpha)
&=\frac{2}{\pi^{3/2}(1-s)\sqrt{s(\nu^2+\mu^2)}}
\left(\frac{s-1}{s+1}\right)^{-r} \nonumber\\
&\quad\times \exp\!\left(
-\frac{\left( x + \sqrt{2}(\mu\,\operatorname{Re}\alpha + \nu\,\operatorname{Im}\alpha)\right)^2}
{s(\nu^2+\mu^2)}
\right). \label{photon_to_sympl_corrected}
\end{align}
\end{statement}
\begin{proof}
The transition kernels \eqref{636} and \eqref{636_1} can be rewritten as
\begin{align}
&K_{\mathcal{W}\to\mathsf{w}}(x, \mu, \nu, r, \alpha) = \frac{1}{2\pi} \langle r | \hat{D}(\alpha)e^{i(x\hat{1} - \mu \hat{q} - \nu \hat{p})}\hat{D}^{\dagger}(\alpha) | r \rangle, \nonumber \\
&K_{\mathsf{w}\to\mathcal{W}}(x, \mu, \nu, r, \alpha) = \tr{\frac{4}{\pi (1-s^2)} \left( \frac{s-1}{s+1} \right)^{ (\hat{a}^\dagger + \alpha^*)(\hat{a} + \alpha) - r }\!\!\!\! \delta(x \hat{1} - \mu \hat{q} - \nu \hat{p})}, \nonumber
\end{align}
that establishes a direct connection between the symplectic tomogram and the photon-number tomogram. The explicit form of the symplectic-to--photon-number transition kernel can be immediately deduced to be equal to \eqref{1406}. Using the integral representation of the delta function, the symplectic dequantizer can be expressed as
\begin{eqnarray}
\hat{\mathcal{U}}(x, \mu, \nu) = \frac{1}{2\pi} \int_{-\infty}^{\infty} dk \, e^{i k x} \hat{D}\left( \frac{k (\nu - i \mu)}{\sqrt{2}} \right).
\end{eqnarray}
Substituting this into the previous expression, the photon-number to symplectic kernel becomes
\begin{eqnarray}
&&\! \! \!\! \! \!
K_{\mathsf{w}\to\mathcal{W}}(x, \mu, \nu, r, \alpha) = \frac{2}{\pi^2 (1-s^2)} \int_{-\infty}^{\infty} dk \, e^{i k x}  \\
&\times& \tr{\left( \frac{s-1}{s+1} \right)^{(\hat{a}^\dagger + \alpha^{*})(\hat{a} + \alpha) - r} \hat{D}\left(\frac{k (\nu - i \mu)}{\sqrt{2}}\right)}. \nonumber
\end{eqnarray}
To evaluate the trace, we use the generating function identity for Laguerre polynomials,
\begin{eqnarray}
\sum_{m=0}^{\infty} \left( \frac{s-1}{s+1} \right)^m L_m(z) = \frac{s+1}{2} \exp\left( -\frac{s-1}{2} z \right).
\end{eqnarray}
This allows us to simplify the trace as follows
\begin{eqnarray}
&&\! \! \!\! \! \!
\tr{\left( \frac{s-1}{s+1} \right)^{(\hat{a}^\dagger + \alpha^{*})(\hat{a} + \alpha) - r} \hat{D}\left(\frac{k (\nu - i \mu)}{\sqrt{2}}\right)} =  \\
&=& \sum_{m=0}^{\infty} \left( \frac{s-1}{s+1} \right)^{m - r} e^{i k \sqrt{2}( \mu \operatorname{Re}(\alpha) + \nu \operatorname{Im}(\alpha) )}  e^{-k^2 \frac{\nu^2 + \mu^2}{4}}   \nonumber \\
&\times& L_m\left( k^2 \frac{\nu^2 + \mu^2}{2} \right) = \frac{s+1}{2} \left(\frac{s-1}{s+1} \right)^{- r} e^{i k \sqrt{2}( \mu \operatorname{Re}(\alpha) + \nu \operatorname{Im}(\alpha) )}  e^{-s k^2 \frac{\nu^2 + \mu^2}{4}}. \nonumber
\end{eqnarray}
Performing the integral over  $k$  and using the well known identity
\begin{eqnarray}
\int_{-\infty}^{\infty} dk \, e^{i a k} e^{-b k^2} = \sqrt{\frac{\pi}{b}} \exp\left( -\frac{a^2}{4b} \right),
\end{eqnarray}
with $ a = x + \sqrt{2} ( \mu \operatorname{Re}(\alpha) + \nu \operatorname{Im}(\alpha) )$ and $b = s \frac{\nu^2 + \mu^2}{4}$,
we obtain the explicit expression for the transition kernel \eqref{photon_to_sympl_corrected}.
For $\mathrm{Re}\,s>0$ the series and the Gaussian integral converge and the kernel is obtained as an ordinary function. For other values of $s$ the expression can be understood via analytic continuation.
\end{proof}
\section{Symplectic to Spin Tomogram Transformations via the Holstein--Primakoff map}
\label{app:HP_symplectic_spin}
\begin{statement}
    The explicit forms of the kernels
$K^{(j,\mathrm{HP})}_{\mathcal W\to\omega}$ and
$K^{(j,\mathrm{HP})}_{\omega\to\mathcal W}$ defined in
Eqs.~\eqref{eq:kernel-W-to-omega-HP}--\eqref{eq:kernel-omega-to-W-HP} are
\begin{align}
K^{(j,\mathrm{HP})}_{\mathcal W\to\omega}(m,\Omega; x,\mu,\nu)
&=\frac{e^{ix}}{2\pi}A^{(j,\mathrm{HP})}_{\phi\to\omega}(m,\Omega,\mu,\nu)
\label{eq:K_forward_HP_sympl}\\
A^{(j,\mathrm{HP})}_{\phi\to\omega}(m,\Omega,\mu,\nu)&\equiv
\sum_{m_1,m_2=-j}^{j}
D^{(j)*}_{m,m_2}(\Omega)\,
D^{(j)}_{m,m_1}(\Omega)\,
M(j-m_1,\,j-m_2;\xi),
\nonumber\\
K^{(j,\mathrm{HP})}_{\omega\to\mathcal W}(x,\mu,\nu; m,\Omega)
&=\frac{e^{-\frac{x^2}{2|\xi|^2}}}{\sqrt{2\pi}\,|\xi|}
\sum_{m_1,m_2=-j}^{j}
\mathcal D^{(j)}_{m_2m_1}(m,\Omega)\,
\mathcal{P}^{(j)}_{m_1,m_2}(x,\xi)
\nonumber\\
&\quad\times
\sqrt{\frac{n!}{(n+|m_1-m_2|)!}}
\left(\frac{i\,\xi'}{\sqrt{2}|\xi|}\right)^{|m_1-m_2|},
\label{eq:K_inverse_HP_sympl}
\end{align}
where $\xi=(\nu-i\mu)/\sqrt2$ and 
\begin{equation}
\xi'\equiv 
\begin{cases}
\xi, & m_1\le m_2,\\
-\xi^{*}, & m_1> m_2.
\end{cases}
\end{equation}
We used that 
$D^{(j)}_{m,m'}(\Omega)$ is the Wigner $D$-matrix, $M(j-m_1,\,j-m_2;\xi)$ is given in \eqref{1176}, $\mathcal{P}^{(j)}_{m_1,m_2}(x,\xi)$
is defined in \eqref{1426},  $\mathcal{D}_{m_2 m_1}^{(j)}(m,\Omega)$ is given by \eqref{1250}.
\end{statement}
\begin{proof}
    We work on the truncated bosonic subspace $S_{2j}=\mathrm{span}\{\ket{k}\}_{k=0}^{2j}$
associated with the Holstein--Primakoff map, where $k=j-m$.
Inserting the resolution of identity on $S_{2j}$ into the traces in
Eqs.~\eqref{eq:kernel-W-to-omega-HP}--\eqref{eq:kernel-omega-to-W-HP}, we obtain finite sums
\begin{align}
K^{(j,\mathrm{HP})}_{\mathcal W\to\omega}(m,\Omega; x,\mu,\nu)
&=\sum_{m_1,m_2=-j}^{j}
\mathcal D^{(j,\mathrm{HP})}_{m_1m_2}(x,\mu,\nu)\,
U^{(j)}_{m_2m_1}(m,\Omega),
\label{eq:HP_sympl_kernels_1}\\
K^{(j,\mathrm{HP})}_{\omega\to\mathcal W}(x,\mu,\nu; m,\Omega)
&=\sum_{m_1,m_2=-j}^{j}
\mathcal U^{(j,\mathrm{HP})}_{m_1m_2}(x,\mu,\nu)\,
\mathcal D^{(j)}_{m_2m_1}(m,\Omega),
\label{eq:HP_sympl_kernels_2}
\end{align}
where $U^{(j)}_{m_2m_1}(m,\Omega)$ are matrix elements of the spin dequantizer
(see Eq.~\eqref{1652}), while $\mathcal D^{(j)}_{m_2m_1}(m,\Omega)$ are matrix elements
of the spin quantizer (see~\ref{ap_1}). The HP--projected bosonic matrix elements are
\begin{align}
\mathcal D^{(j,\mathrm{HP})}_{m_1m_2}(x,\mu,\nu)
&\equiv \bra{j-m_1}\,\hat{\mathcal D}(x,\mu,\nu)\,\ket{j-m_2},
\label{eq:D_HP_def}\\
\mathcal U^{(j,\mathrm{HP})}_{m_1m_2}(x,\mu,\nu)
&\equiv \bra{j-m_1}\,\hat{\mathcal U}(x,\mu,\nu)\,\ket{j-m_2}.
\label{eq:U_HP_def}
\end{align}
Using the symplectic quantizer \eqref{symplectic_quantizer}, we write it through the Weyl
displacement operator $\hat D(\xi)=\exp(\xi\hat a^\dagger-\xi^*\hat a)$ as
\begin{equation}
\hat{\mathcal D}(x,\mu,\nu)=\frac{1}{2\pi}\exp\!\bigl(i x\hat 1-i\nu\hat p-i\mu\hat q\bigr)
=\frac{e^{ix}}{2\pi}\,\hat D(\xi),
\label{eq:sympl_quantizer_single}
\end{equation}
so that $|\xi|^2=(\mu^2+\nu^2)/2$ and equivalently $\xi=(\nu-i\mu)/\sqrt2$.
Introduce $M(a,b;\beta)\equiv\langle a|\hat D(\beta)|b\rangle$ (explicit form is given in
Eq.~\eqref{eq:displacement_matrix_element}). Then
\begin{equation}
\mathcal D^{(j,\mathrm{HP})}_{m_1m_2}(x,\mu,\nu)=\frac{e^{ix}}{2\pi}\,
M(j-m_1,\,j-m_2;\xi),
\label{eq:D_HP_in_M}
\end{equation}
and we get the final form \eqref{eq:K_forward_HP_sympl}.
\par  For the dequantizer we use Eq.~\eqref{eq:deq-simpl}:
\begin{equation}
\hat{\mathcal U}(x,\mu,\nu)=\delta(x\hat 1-\mu\hat q-\nu\hat p)
=\frac{1}{2\pi}\int_{-\infty}^{\infty}dk\;e^{ikx}\,
\exp\!\bigl(-ik\nu\hat p-ik\mu\hat q\bigr),
\label{eq:sympl_dequantizer_single_a}
\end{equation}
which yields the one-mode representation
\begin{equation}
\hat{\mathcal U}(x,\mu,\nu)=\frac{1}{2\pi}\int_{-\infty}^{\infty}dk\;e^{ikx}\,\hat D(k\xi).
\label{eq:sympl_dequantizer_single}
\end{equation}
Hence
\begin{equation}
\mathcal U^{(j,\mathrm{HP})}_{m_1m_2}(x,\mu,\nu)
=\frac{1}{2\pi}\int_{-\infty}^{\infty}dk\;e^{ikx}\,
M(j-m_1,\,j-m_2; k\xi).
\label{eq:U_HP_integral_M}
\end{equation}
Evaluating the integral by the same steps as in~\ref{app_2} (one-mode case),
we obtain a closed form. Let $\Delta\equiv|m_1-m_2|$ and $n\equiv j-\max\{m_1,m_2\}$.
Then, for $(\mu,\nu)\neq(0,0)$ we get
\begin{align}
\mathcal U^{(j,\mathrm{HP})}_{m_1m_2}(x,\mu,\nu)
&=
\frac{1}{\sqrt{2\pi}\,|\xi|}\exp\!\left[-\frac{x^2}{2|\xi|^2}\right]\,
\sqrt{\frac{n!}{(n+\Delta)!}}\,
\left(\frac{i}{\sqrt2|\xi|}\right)^{\Delta}(\xi')^{\Delta}
\nonumber\\
&\quad\times
\sum_{s=0}^{n}\binom{n+\Delta}{n-s}\frac{1}{s!\,2^{s}}\,
H_{\Delta+2s}\!\left(\frac{x}{\sqrt2|\xi|}\right).
\label{eq:U_HP_closed_sympl}
\end{align}
Substitution of Eq.~\eqref{eq:U_HP_closed_sympl} into Eq.~\eqref{eq:HP_sympl_kernels_2}
gives \eqref{eq:K_inverse_HP_sympl}.
\end{proof}
The normalization trace entering the denominator of Eq.~\eqref{eq:W_to_omega_int} reads
\begin{align}
\tr\!\left\{\hat{\mathcal D}(x,\mu,\nu)\hat P_{2j}\right\}
&=\sum_{k=0}^{2j}\bra{k}\,\hat{\mathcal D}(x,\mu,\nu)\,\ket{k}
=\frac{e^{ix}}{2\pi}\,e^{-|\xi|^2/2}\,L_{2j}^{(1)}(|\xi|^2),
\label{eq:trace_D_P2j_sympl}
\end{align}
where we used $\sum_{k=0}^{N}L_k(z)=L_N^{(1)}(z)$.
\section{Photon-Number to Spin Tomogram Transformations via the Holstein--Primakoff Map}\label{app_7}
\begin{statement}
\begin{align}
K^{(j,\text{HP})}_{\omega\rightarrow \mathsf{w}}(m,\Omega; r, \alpha)
&= \sum_{m_1,m_2=-j}^j
\mathcal{D}_{m_2 m_1}^{(j)}(m, \Omega)\,
M(j-m_1, r; -\alpha)\,
M(r, j-m_2; \alpha),
\label{eq:1444}
\\
K^{(j,\text{HP})}_{\mathsf{w}\rightarrow \omega}(m, \Omega; r, \alpha)
&= \frac{4}{\pi(1-s^2)}
\sum_{m_1,m_2=-j}^j
U^{(j)}_{m_2 m_1}(m, \Omega)
\sum_{l=0}^\infty g(s)^{\,l-r}
\nonumber\\
&\quad\times
M(j-m_1, l; -\alpha)\,
M(l, j-m_2; \alpha),
\label{eq:1443}
\end{align}
where  $M(k,l,\alpha)$ is given in \eqref{1176}, $U^{(j)}_{m_2 m_1}(m, \Omega)$ is given in Eq.~\eqref{1652} and  $\mathcal{D}_{m_2 m_1}^{(j)}(m,\Omega)$ is given by Eq.~\eqref{1250}.
\end{statement}
\begin{proof}
Using the resolution of identity on the restricted subspace $S_{2j}$, $\hat{1}_{2j} = \sum_{m=-j}^j |j-m\rangle\langle j-m|$, the transition kernels between spin and photon-number tomograms (Eqs.~\eqref{eq:kernel-w-to-omega-HP} and \eqref{eq:kernel-omega-to-w-HP}) can be written in a form analogous to the JS case, but restricted to a single mode:
\begin{eqnarray}\label{hp_photon_kernels}
     K^{(j,\text{HP})}_{\mathsf{w}\rightarrow \omega}(m,\Omega; r, \alpha) &=& \sum_{m_1,m_2=-j}^j D_{m_1 m_2}^{(j,\text{HP})}(r, \alpha) U^{(j)}_{m_2 m_1}(m,\Omega), \\
     K^{(j,\text{HP})}_{\omega\rightarrow \mathsf{w}}(m,\Omega; r, \alpha) &=& \sum_{m_1,m_2=-j}^j U_{m_1 m_2}^{(j,\text{HP})}(r, \alpha) \mathcal{D}^{(j)}_{m_2 m_1}(m,\Omega),
\end{eqnarray}
where $U^{(j)}_{m_2 m_1}(m,\Omega)$ and $\mathcal{D}^{(j)}_{m_2 m_1}(m,\Omega)$ are the matrix elements of the spin dequantizer and quantizer. The bosonic matrix elements are defined as projections of the single-mode photon-number operators onto the HP basis states:
\begin{eqnarray}
    D_{m_1 m_2}^{(j,\text{HP})}(r, \alpha) &\equiv& \langle j-m_1 | \hat{\mathcal{D}}_{r}(\alpha) | j-m_2 \rangle, \label{hp_D_matrix}\\
    U_{m_1 m_2}^{(j,\text{HP})}(r, \alpha) &\equiv& \langle j-m_1 | \hat{\mathcal{U}}_{r}(\alpha) | j-m_2 \rangle. \label{hp_U_matrix}
\end{eqnarray}
Using the explicit form of the single-mode quantizer from Eq.~\eqref{1348}, we can express the matrix element (Eq.~\eqref{hp_D_matrix}) as a series:
\begin{equation}
    D_{m_1 m_2}^{(j,\text{HP})}(r, \alpha) = \frac{4}{\pi(1-s^2)} \sum_{l=0}^\infty g(s)^{l-r} M(j-m_1, l; -\alpha) M(l, j-m_2; \alpha),
\end{equation}
where $M(a, b; \beta) = \langle a | \hat{D}(\beta) | b \rangle$ are the matrix elements of the displacement operator expressed through generalized Laguerre polynomials in Eq.~\eqref{1176}.
Similarly, using $\hat{\mathcal{U}}_r(\alpha) = \hat{D}^\dagger(\alpha)|r\rangle\langle r|\hat{D}(\alpha)$, the dequantizer matrix element (Eq.~\eqref{hp_U_matrix}) takes a closed form without infinite summation:
\begin{equation}
     U_{m_1 m_2}^{(j,\text{HP})}(r, \alpha) = M(j-m_1, r; -\alpha)\, M(r, j-m_2; \alpha).
\end{equation}
Thereby, we can immediately write the transition kernels connecting the spin and photon-number tomograms via the HP map \eqref{eq:1444}.  Note that this kernel is a finite sum, reflecting the finite dimensionality of the spin reconstruction space. The inverse transition \eqref{eq:1443} from photon-number to spin tomogram involves the infinite series inherent in the photon-number quantizer.
\end{proof}
\begin{statement}
 The denominator of Eq.~\eqref{eq:photon-to-omega} is
    \begin{eqnarray}\label{1451}
  \!\!\!  \tr{ \hat{P}_{2j} \hat{\mathcal{D}}_r(\alpha) }= \frac{4 e^{-(1-g(s))|\alpha|^2}}{\pi(1-s^2)} \sum_{m=-j}^j g(s)^{j-m-r} L_{j-m} \left( - \frac{(1-g(s))^2}{g(s)} |\alpha|^2 \right).
\end{eqnarray}
\end{statement}
\begin{proof}
The denominator for Eq.~\eqref{eq:photon-to-omega}, which corresponds to the trace of the restricted photon-number quantizer:
\begin{eqnarray}
    \tr\left\{ \hat{P}_{2j} \hat{\mathcal{D}}_r(\alpha) \right\} &=& \sum_{m=-j}^j D_{mm}^{(j,\text{HP})}(r, \alpha) \nonumber \\
    &=& \frac{4}{\pi(1-s^2)} \sum_{m=-j}^j \sum_{l=0}^\infty g(s)^{l-r} \left| M(j-m, l; -\alpha) \right|^2.
\end{eqnarray}
Using the property $|M(a, b; -\alpha)|^2 = |\langle a | \hat{D}(-\alpha) | b \rangle|^2$, this expression can be computed numerically for any given truncated subspace; however, it also admits a closed form derived below.
 Using the single-mode definition from \ref{app_3}, the diagonal matrix element in the Fock basis $|k\rangle$ (where $k=j-m$) is:
\begin{equation}
    D_{mm}^{(j,\text{HP})}(r, \alpha) = \frac{4}{\pi(1-s^2)} g(s)^{-r} \sum_{l=0}^\infty g(s)^l \left| \langle l | \hat{D}(\alpha) | k \rangle \right|^2.
\end{equation}
Here we used the property $\hat{D}^\dagger(-\alpha) = \hat{D}(\alpha)$. The term $|\langle l | \hat{D}(\alpha) | k \rangle|^2$ is the probability of detecting $l$ photons in a displaced Fock state $|k\rangle$. We can now apply the generating function identity for the squares of generalized Laguerre polynomials from \ref{app_3}:
\begin{equation}
    \sum_{l=0}^\infty z^l \left| \langle l | \hat{D}(\alpha) | k \rangle \right|^2 = z^k e^{-(1-z)|\alpha|^2} L_k \left( - \frac{(1-z)^2}{z} |\alpha|^2 \right).
\end{equation}
Identifying $z = g(s)$, we obtain a closed-form expression for the diagonal element without any infinite summation:
\begin{equation}
    D_{mm}^{(j,\text{HP})}(r, \alpha) = \frac{4}{\pi(1-s^2)} e^{-(1-g(s))|\alpha|^2} g(s)^{k-r} L_k \left( - \frac{(1-g(s))^2}{g(s)} |\alpha|^2 \right).
\end{equation}
Finally, the trace over the spin-$j$ subspace is simply the finite sum of these diagonal elements for $m$ running from $-j$ to $j$ (where $k=j-m$) given by \eqref{1451}.
\end{proof}
\section{Spin Tomogram to Wigner Function Transformations via the Holstein--Primakoff Map}\label{app_8}
\begin{statement}
The transition kernels~\eqref{eq:Wigner-to-omega} and \eqref{eq:omega-to-Wigner} can be written as
\begin{align}
K^{(j,\text{HP})}_{\mathbb{W}\rightarrow \omega}(m,\Omega; \alpha)
&= 2 \sum_{m_1,m_2=-j}^j
(-1)^{j-m_2}
D_{m,m_2}^{(j) *}(\Omega)\,
D_{m,m_1}^{(j)}(\Omega)
\nonumber\\
&\quad\times
M(j-m_1, j-m_2; 2\alpha),
\label{hp_Wigner_to_spin}
\\
K^{(j,\text{HP})}_{\omega\rightarrow \mathbb{W}}(m,\Omega; \alpha)
&= \frac{1}{\pi}
\sum_{m_1,m_2=-j}^j
(-1)^{j-m_2}
\mathcal{D}_{m_2 m_1}^{(j)}(m,\Omega)
\nonumber\\
&\quad\times
M(j-m_1, j-m_2; 2\alpha),
\label{hp_spin_to_Wigner}
\end{align}
where  $M(k,l,\alpha)$ is given in \eqref{1176} and  $\mathcal{D}_{m_2 m_1}^{(j)}(m,\Omega)$ is given by Eq.~\eqref{1250}.
\end{statement}
\begin{proof}
We derive the kernels linking the spin tomogram and the Wigner function under the HP isomorphism. Using the resolution of identity on the restricted subspace $S_{2j}$, the transition kernels defined in Eqs.~\eqref{eq:Wigner-to-omega} and \eqref{eq:omega-to-Wigner} can be written as:
\begin{eqnarray}\label{hp_wigner_kernels}
   K^{(j,\text{HP})}_{\mathbb{W}\rightarrow \omega}(m,\Omega; \alpha)&=&\sum_{m_1,m_2=-j}^j \mathbb{D}_{m_1 m_2}^{(j,\text{HP})}(\alpha) U^{(j)}_{m_2 m_1}(m,\Omega),\\
   K^{(j,\text{HP})}_{\omega\rightarrow \mathbb{W}}(m,\Omega; \alpha)&=&\sum_{m_1,m_2=-j}^j \mathbb{U}^{(j,\text{HP})}_{m_1 m_2}(\alpha)\mathcal{D}^{(j)}_{m_2 m_1}(m,\Omega),
 \end{eqnarray}   
where the matrix elements of the Wigner quantizer and dequantizer in the HP basis are:
\begin{eqnarray*}
   \mathbb{D}_{m_1 m_2}^{(j,\text{HP})}(\alpha) &\equiv& \langle j-m_1 | \hat{\mathbb{D}}(\alpha) | j-m_2 \rangle, \\
   \mathbb{U}_{m_1 m_2}^{(j,\text{HP})}(\alpha) &\equiv& \langle j-m_1 | \hat{\mathbb{U}}(\alpha) | j-m_2 \rangle.
\end{eqnarray*}
Here $|k\rangle$ denotes a single-mode Fock state; in the HP basis one has $k=j-m$. Recall that the single-mode Wigner quantizer is given by $\hat{\mathbb{D}}(\alpha) = 2 \hat{D}(2\alpha)\hat{P}$, where $\hat{P}$ is the parity operator. Its action on a Fock state is $\hat{P}|n\rangle = (-1)^n|n\rangle$.
Substituting this into the matrix element definition and using the displacement matrix elements $M(a,b; \beta)$, we obtain a compact expression without infinite summation:
\begin{eqnarray}\label{hp_wigner_quant_element}
\mathbb{D}_{m_1 m_2}^{(j,\text{HP})}(\alpha) &=& 2 \langle j-m_1 | \hat{D}(2\alpha) \hat{P} | j-m_2 \rangle \nonumber \\
&=& 2 (-1)^{j-m_2} \langle j-m_1 | \hat{D}(2\alpha) | j-m_2 \rangle \nonumber \\
&=& 2 (-1)^{j-m_2} M(j-m_1, j-m_2; 2\alpha).
\end{eqnarray}
Similarly, using the relation $\hat{\mathbb{U}}(\alpha) = (2\pi)^{-1} \hat{\mathbb{D}}(\alpha)$ (for the standard Wigner normalization), the dequantizer matrix elements are:
\begin{equation}
\mathbb{U}_{m_1 m_2}^{(j,\text{HP})}(\alpha) = \frac{1}{\pi} (-1)^{j-m_2} M(j-m_1, j-m_2; 2\alpha).
\end{equation}
Substituting these into Eq.~\eqref{hp_wigner_kernels}, we arrive at the explicit forms of the transition kernels \eqref{hp_Wigner_to_spin} and \eqref{hp_spin_to_Wigner}.
\end{proof}
Finally, we calculate the normalization denominator for Eq.~\eqref{eq:Wigner-to-omega}. Unlike the two-mode case, where the trace reduces to a Laguerre polynomial via convolution, the single-mode HP trace involves an alternating sum of Laguerre polynomials:
\begin{eqnarray}
\tr\left\{\hat{P}_{2j}\,\hat{\mathbb{D}}(\alpha)\right\} &=& \sum_{m=-j}^j \mathbb{D}_{m m}^{(j,\text{HP})}(\alpha) =2 e^{-2|\alpha|^2} \sum_{k=0}^{2j} (-1)^k L_k(4|\alpha|^2).
\end{eqnarray}
This finite sum represents the exact projection of the Wigner quantizer onto the spin-$j$ subspace.
\end{document}